\documentclass[aps,prd,citeautoscript,longbibliography,superscriptaddress,twocolumn]{revtex4-1}
\usepackage[T1]{fontenc}
\usepackage{graphicx}
\usepackage{amsmath,amssymb,amsfonts,bm,dsfont,units}
\usepackage{hyperref}
\hypersetup{colorlinks=true,citecolor=blue,linkcolor=blue,urlcolor=blue,pdfstartview=FitH}
\usepackage{braket}

\newcommand{\norder}[1]{ {\mkern1mu\colon\mkern-4mu{#1}\colon\mkern-3mu} }
\newcommand{\alpRuCl}[1]{\texorpdfstring{$\alpha$-$\mathrm{RuCl}_{#1}$}{alpha-RuCl{#1}}}
\newcommand{\bigXp}[1]{ {\big\langle{#1}\big\rangle} }
\newcommand{\BigXp}[1]{ {\Big\langle{#1}\Big\rangle} }
\newcommand{\HGpFqb}[5]{\mathop{_{#1}\mkern-1.5mu F_{#2}}\mkern-4mu\big( {#3}\,; {#4} \mkern2mu\big| {#5} \big)}	% generalized hypergeometric function

\usepackage{color}
\definecolor{dkgreen}{rgb}{0,0.5,0}
\definecolor{midnightblue}{rgb}{0.39,0.58,0.93}

\newcommand{\comment}[1]{}{}
\begin{document}

\title{Electrical probes of the non-Abelian spin liquid in Kitaev materials}

\author{David Aasen}
\affiliation{Microsoft Quantum, Microsoft Station Q, University of California, Santa Barbara, California 93106-6105 USA}
\affiliation{Kavli Institute for Theoretical Physics, University of California, Santa Barbara, California 93106, USA}
\author{Roger S. K. Mong}
\affiliation{Department of Physics and Astronomy, University of Pittsburgh, Pittsburgh, PA 15260, USA}
\affiliation{Pittsburgh Quantum Institute, Pittsburgh, PA 15260, USA}
\author{Benjamin M. Hunt}
\affiliation{Department of Physics, Carnegie Mellon University, Pittsburgh, PA 15213, USA}
\affiliation{Pittsburgh Quantum Institute, Pittsburgh, PA 15260, USA}
\author{David Mandrus}
\affiliation{Department of Materials Science and Engineering,
University of Tennessee, Knoxville, TN 37996, USA}
\affiliation{Materials Science and Technology Division, Oak Ridge National Laboratory, Oak Ridge, TN 37831, USA}
\author{Jason Alicea}
\affiliation{Department of Physics and Institute for Quantum Information and Matter, California Institute of Technology, Pasadena, CA 91125, USA}
\affiliation{Walter Burke Institute for Theoretical Physics, California Institute of Technology, Pasadena, CA 91125, USA}

\date{\today}

\begin{abstract}
Recent thermal-conductivity measurements evidence a magnetic-field-induced non-Abelian spin liquid phase in the Kitaev material \alpRuCl3.
Although the platform is a good Mott insulator, we propose experiments that \emph{electrically} probe the spin liquid's hallmark chiral Majorana edge state and bulk anyons, including their exotic exchange statistics.
We specifically introduce circuits that exploit interfaces between electrically active systems and Kitaev materials to `perfectly' convert electrons from the former into emergent fermions in the latter---thereby enabling variations of transport probes invented for topological superconductors and fractional quantum Hall states. 
Along the way we resolve puzzles in the literature concerning interacting Majorana fermions, and also develop an anyon-interferometry framework that incorporates nontrivial energy-partitioning effects.  
Our results illuminate a partial pathway towards topological quantum computation with Kitaev materials.

\end{abstract}

\maketitle

%%%%%%%%%%%%%%%%%%%%%%%%%%%%%%%%%%%%%%%%%%%%%%%%
\section{Introduction}
\label{intro}

The field of topological quantum computation pursues phases of matter supporting emergent particles known as `non-Abelian anyons' to ultimately realize scalable, intrinsically fault-tolerant qubits~\cite{Kitaev:2003,TQCreview}.  
This technological promise derives from three deeply linked non-Abelian-anyon features:
First, nucleating well-separated non-Abelian anyons generates a ground-state degeneracy consisting of states that cannot be distinguished from one another by local measurements.  
Qubits encoded in this subspace enjoy built-in protection from environmental noise by virtue of local indistinguishability.  
Second, they obey non-Abelian braiding statistics.  
That is, adiabatically exchanging pairs of non-Abelian anyons effects `rigid' non-commutative rotations within the ground-state manifold---thus producing fault-tolerant qubit gates.  
And third, pairs of non-Abelian anyons brought together in space can `fuse' to at least two different types of particles; detecting the fusion outcome provides a means of qubit readout.

Fulfilling this potential requires, at an absolute minimum, synthesizing a non-Abelian host material \emph{and} developing practical means of controlling and probing the constituent anyons.  
The observed fractional quantum Hall phase at filling factor $\nu = 5/2$~\cite{Willett1987}, widely expected to realize the non-Abelian Moore-Read state or cousins thereof~\cite{MooreRead,Levin2007,Lee2007,SonDiracCFL,Banerjee}, provided the first candidate topological-quantum-computing medium.  
Non-Abelian anyons in this setting carry electric charge (e.g., $e/4$), and hence can be manipulated via gating and probed using ingenious electrical interferometry schemes~\cite{Nayak2005,Stern2006,Bonderson2006}.  
While experimental efforts in this direction continue~\cite{Willett2019},  
during the past decade intense experimental activity has focused on `engineered' two-dimensional (2D) and especially one-dimensional (1D) topological superconductors~\cite{ReadGreen,Kitaev:2001} as alternative platforms.
These exotic superconductors can be assembled from heterostructures involving ordinary, weakly correlated materials yet share similar non-Abelian properties to the Moore-Read state (for reviews see Refs.~\onlinecite{HasanKane,QiZhang,BeenakkerReview,AliceaReview,FlensbergReview,TewariReview,FranzReview,ChetanReview,SatoReview,AguadoReview,LutchynReview}).  
Specifically, the charged non-Abelian excitations in the Moore-Read state are replaced by non-Abelian defects---i.e., domain walls and superconducting vortices---that bind Majorana zero modes. 
In a topological superconductor, Majorana zero modes are equal superpositions of electrons and holes and thus carry no net charge.  They do carry a physical fermion-parity degree of freedom, however, and are thus amenable to electronic probes including tunneling spectroscopy, interferometry, Josephson measurements, etc.; see, e.g., Refs.~\onlinecite{Kitaev:2001,FuKaneInterferometer,AkhmerovInterferometer,Law,FuTeleportation,Chung2011}.
In fact, detailed blueprints exist for scalable topological quantum computation hardware based on 1D-topological-superconductor arrays, relying largely on electrical tools for operation~\cite{Karzig2017}.

Still more recently, experiments suggest the emergence of yet another variant of the Moore-Read state, but in a fundamentally different physical setting from those above: the honeycomb `Kitaev material' \alpRuCl3~\cite{Plumb2014,Kim2015}.  
As background, consider a honeycomb lattice of spin-1/2 moments governed by a Hamiltonian of the form
\begin{equation}
  H = -\sum_{\langle {\bf r r'} \rangle} K S_{\bf r}^\gamma S_{\bf r'}^\gamma - \sum_{\bf r} {\bf B} \cdot {\bf S}_{\bf r} + \cdots.
  \label{Hprimer}
\end{equation}
The first term encodes bond-dependent spin interactions, with $\gamma = x$ on the green bonds of Fig.~\ref{KitaevHoneycombFig}(a), 
$\gamma = y$ on red bonds, and $\gamma = z$ on blue bonds; 
note the strong frustration arising from these competing spin couplings, which suppresses the tendency for conventional symmetry-breaking order.  
The second term in the Hamiltonian accounts for the possible presence of a magnetic field ${\bf h}$, while the ellipsis denotes additional allowed perturbations. 

When only the $K$ term is present, the Hamiltonian reduces to Kitaev's famed exactly solvable honeycomb model~\cite{Kitaev2006}.
Here the ground state realizes a time-reversal-invariant quantum spin liquid with gapless, emergent Majorana fermions coupled to a $\mathbb{Z}_2$ gauge field.  
For this paper it is crucial to distinguish \emph{emergent} Majorana fermions from \emph{physical} Majorana fermions that appear as excitations at the boundaries of two- and three-dimensional topological superconductors.  
The latter are built from ordinary electronic degrees of freedom, whereas the former represent bona fide fractionalized quasiparticles born within a purely bosonic spin system.  
It follows that physical Majorana fermions can shuttle between the host topological superconductor and a conventional electronic medium (e.g., a lead); conversely, emergent Majorana fermions live \emph{exclusively} in the spin liquid.  

Breaking time-reversal symmetry generates even more striking physics: A non-zero magnetic field ${\bf h}$ gaps out the Majorana fermions, yielding a \emph{non-Abelian} spin liquid exhibiting a fully gapped bulk and a gapless, chiral Majorana-fermion edge state that underlies quantized thermal Hall conductance~\cite{Kitaev2006}. 
This phase supports two nontrivial quasiparticle types: massive emergent Majorana fermions and `Ising' non-Abelian anyons.
The latter can be viewed as electrically neutral counterparts of the non-Abelian anyons in the Moore-Read state.
Alternatively, they comprise deconfined cousins of non-Abelian defects in topological superconductors that bind Majorana zero modes carrying an emergent rather than physical fermion-parity degree of freedom.  

Jackeli and Khaliullin established that a class of strongly spin-orbit-coupled Mott insulators can, quite remarkably, be well-modeled by Eq.~\eqref{Hprimer} with inevitably present corrections represented by the ellipsis being `small'~\cite{Jackeli2009}.
Their pioneering result opened up the now experimentally active field of Kitaev materials whose spins interact predominantly via bond-dependent spin interactions of the type built into Kitaev's honeycomb model~\cite{Winter2017,Trebst2017,Janssen2019,Motome2019}.  
All honeycomb-lattice Kitaev materials studied to date---\alpRuCl3 included~\cite{Sears2015}---magnetically order at zero field.
Evidently perturbations beyond the $K$ term in Eq.~\eqref{Hprimer}, while nominally small, destabilize the gapless quantum spin liquid~\cite{Chaloupka2013} (various experiments nevertheless report residual fractionalization signatures at `high' energies~\cite{Sandilands2015,Nasu2016,Banerjee2016,Banerjee2017,Do2017,Kasahara2017,Wellm2018,Wang2018,Jansa2018,Widmann2019,Zhang2019}).  
In \alpRuCl3, applying a $\sim\unit[10]{T}$ in-plane magnetic field destroys the zero-field magnetic ordering~\cite{Johnson2015}.
Numerous experiments are consistent with the fascinating possibility that the system then enters the non-Abelian spin liquid phase highlighted above~\cite{Baek2017,Sears2017,Wolter2017,Leahy2017,Banerjee2018,Hentrich2018,Jansa2018,Kasahara2018,Balz2019}.  
Most strikingly, Kasahara et al.~\cite{Kasahara2018} report thermal-Hall-conductance measurements that agree well with the quantized value expected from the hallmark chiral Majorana edge mode.  
This experiment withstood some initial theoretical scrutiny~\cite{Ye2018,Vinkler2018}, and has very recently been extended in Ref.~\onlinecite{Tokoi}.

Can one plausibly exploit \alpRuCl3 (or perhaps some related Kitaev materials) for topological quantum compution?
This question is well-motivated on at least two fronts.
For one, the energy scales appear quite favorable.  
In Refs.~\onlinecite{Kasahara2018,Tokoi}, quantized thermal Hall conductance persists up to temperatures of roughly \unit[5]{K}, suggesting a spin liquid bulk gap of similar magnitude---an encouraging figure compared to the gap expected in most other candidate non-Abelian platforms 
\footnote{Any topological quantum computing platform would ideally be run at the lowest accessible temperatures.  A large gap is nevertheless desirable for suppressing errors.}.
Moreover, \alpRuCl3 affords a great deal of materials-science flexibility~\cite{Zhou2018,Zhou2019,Mashhadi2018,Mashhadi2019}: it is exfoliatable, amenable to nanofabrication, can be readily interfaced with other materials, etc.  

Manipulating and probing the anyons as required for advanced applications nevertheless poses a major outstanding challenge.  
In essence, the detailed roadmaps developed for quantum Hall and topological superconductor platforms---which again heavily invoke electrical tools---need to be largely rewritten for non-Abelian spin liquids in Kitaev materials because they are electrically inert Mott insulators.  
Two subclasses of problems naturally arise here: 
$(i)$ devising feasible techniques for creating, transporting, and fusing Ising anyons on demand in Kitaev materials and 
$(ii)$ developing schemes for unambiguously detecting individual emergent fermions and Ising anyons as well as their nontrivial statistics.
Vacancies and spin impurities appear to be promising ingredients for item $(i)$.
At least in the gapless spin liquid phase of Kitaev's honeycomb model, both have been shown to trap $\mathbb{Z}_2$-flux excitations~\cite{HoneycombVacancies,HoneycombImpurity,HoneycombImpurity2}, which evolve into Ising anyons upon entering the non-Abelian phase.  
We leave detailed investigations of this issue for future work, and instead propose a series of experiments that directly tackle item $(ii)$.  

Our primary innovation is that, counterintuitively, low-voltage electrical transport \emph{can} be profitably employed to probe the detailed structure of non-Abelian spin liquids, their Mott-insulating character notwithstanding.
We build off of seminal theory works that highlight the possibility of coherently converting physical fermions into emergent deconfined quasiparticles in Abelian spin liquids~\cite{Barkeshli2014} and non-Abelian quantum Hall phases~\cite{Barkeshli2015} to probe fractionalization \footnote{Reference~\onlinecite{Barkeshli2015} also briefly discusses applications to the non-Abelian spin liquid in Kitaev's honeycomb model, though their approach is very different from the one developed here.}.  
We pursue a complementary approach that closely relates to the physics of `fermion condensation' put on rigorous mathematical foundation in a similar setting in Ref.~\onlinecite{AasenFC}.  
Specifically, we introduce a series of circuits that interface electronically active systems---notably proximitized $\nu = 1$ integer quantum Hall states, though other choices are possible---with Kitaev materials realizing a non-Abelian spin liquid.  
Strong interactions at their interface can effectively `sew up' these very different subsystems, leading to a striking and exceedingly useful phenomenon: 
A physical electron injected at low energies on the electronically active side converts \emph{with unit probability} into an emergent fermion in the spin liquid.  

Our circuits exploit this perfect conversion process to electrically reveal (via universal conductance signatures) the spin liquid's chiral Majorana edge state, bulk emergent fermions, and bulk Ising anyons, using variations of transport techniques developed for topological superconductors and fractional quantum Hall states.  
Figures~\ref{condexp}, \ref{interference}, and \ref{psidetector} sketch the corresponding setups.
The electrical conductance of these circuits changes qualitatively upon perturbing the Kitaev material (again, an electrically inert element!), e.g., to add or remove even a single bulk emergent fermion or Ising anyon; we argue that this feature makes our predictions especially unambiguous.  
Moreover, the circuits designed to detect individual bulk quasiparticles rely on interferometric signatures that further unambiguously reveal the non-Abelian statistics of Ising anyons as well as the nontrivial mutual statistics between Ising anyons and emergent fermions.

These results collectively establish a partial roadmap towards utilizing Kitaev materials for topological quantum computation.  
En route to putting our predictions on firm footing, we introduce some nontrivial technical innovations as well.
First, we resolve an outstanding puzzle in the literature concerning interacting Majorana fermions.  
Specifically, the interaction strength required to induce an instability in a self-dual Majorana chain has been found to vary by \emph{orders of magnitude} depending on subtle variations in the microscopic interaction (for a recent review see Ref.~\onlinecite{Rahmani2019}).  
We explain this peculiar behavior as arising from interaction-dependent renormalization of kinetic energy for the Majorana chain.
Second, we analyze anyon interferometry in a new regime using a phenomenological picture combined with rigorous formalism 
 that incorporates crucial energy-partitioning effects.  
The framework that we develop here could prove valuable in a variety of other contexts.  

The remainder of the paper is organized as follows.  
We begin in Sec.~\ref{Kitaev_phenomenology} by reviewing the phenomenology of the Kitaev honeycomb model. 
Section~\ref{Interacting_Majorana} explores interacting helical Majorana fermions from several perspectives, and then Sec.~\ref{QHSL_interface} bootstraps off of those results to describe how a $\nu = 1$ quantum Hall state can be sewn (in a precise sense) to a non-Abelian spin liquid with the aid of a superconductor.
In the next three sections we introduce circuits that use this `sewing' to electrically interrogate a non-Abelian spin liquid:
Section~\ref{EdgeDetection} focuses on electrical detection of the chiral Majorana edge state, Sec.~\ref{IsingDetection} introduces a circuit that probes bulk Ising anyons, and Sec.~\ref{interferometry2} introduces an interferometer that probes both bulk Ising anyons \emph{and} emergent fermions, as well as non-Abelian statistics.  
We conclude and highlight numerous open questions in Sec.~\ref{discussion}.
Several appendices provide additional details and supplementary results on our circuits as well as interacting Majorana-fermion models.

%%%%%%%%%%%%%%%%%%%%%%%%%%%%%%%%%%%%%%%%%%%%%%%%
\section{Kitaev honeycomb model phenomenology}
\label{Kitaev_phenomenology}

To set the stage, this section reviews the phenomenology of the Kitaev honeycomb model~\cite{Kitaev2006}, focusing in particular on universal properties of the non-Abelian spin liquid phase.
We also establish various conventions here that will be employed throughout.

%%%%%%%%%%%%%%%%%%%%%%%%%%%%%%%%%%%%%%%%%%%%%%%%
\subsection{Gapless spin liquid}
\label{GaplessSec}

\begin{figure*}
		\includegraphics[width=.9\linewidth]{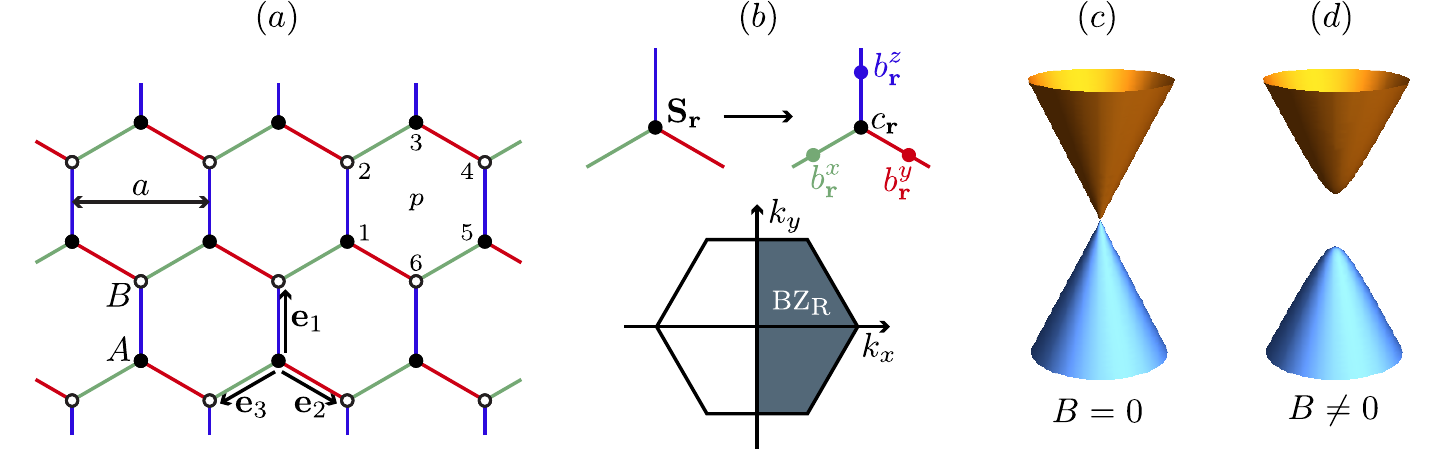}
	\caption{ (a) Lattice structure for the Kitaev honeycomb model.	
	Spins exhibit bond-dependent nearest-neighbor interactions, with the $x$, $y$, and $z$ components respectively coupling along green, red, and blue bonds.
	Vectors ${\bf e}_{1,2,3}$ point from the $A$ sublattice (solid circles) to the $B$ sublattice (open circles).
	(b) Sketch of the Majorana-fermion representation of the spin operators [Eq.~\eqref{MajoranaRep}] used for the exact solution.  The $b^{x,y,z}_{\bf r}$ operators combine to form a $\mathbb{Z}_2$ gauge field, while the $c_{\bf r}$ operators define itinerant Majorana fermions that hop between nearest-neighbor sites.  
	Hermiticity of the $c_{\bf r}$'s allows the kinetic energy to be expressed as a sum over momenta in the right half of the Brilllouin zone (${\rm BZ}_{\rm R}$, shaded region).
	(c) In the gapless spin liquid phase, the fermionic spectrum features a single massless Dirac cone.
	(d) Breaking time-reversal symmetry via an applied magnetic field opens a gap at the Dirac point, generating a non-Abelian spin liquid phase.
} 
	\label{KitaevHoneycombFig}
\end{figure*}

We start with the `pure' Kitaev honeycomb model at zero magnetic field:
\begin{align}
\label{pureKitaev}
{H}_K = -\sum_{\langle {\bf r r'} \rangle} K S_{\bf r}^\gamma S_{\bf r'}^\gamma .
\end{align}
Once again, we have $\gamma = x$, $y$, and $z$ respectively on green, red, and blue bonds of Fig.~\ref{KitaevHoneycombFig}(a).
For any hexagonal plaquette $p$, $H_K$ commutes with the operator 
\begin{align}
	W_p = S^x_1 S^y_2 S^z_3 S^x_4 S^y_5 S^z_6,
	\label{Fluxoperator}
\end{align}
where sites $1,2,\dots, 6$ around plaquette $p$ are labeled as in Fig.~\ref{KitaevHoneycombFig}(a).
The resulting extensive number of conserved quantities ultimately enables an exact solution.
To this end we re-express the spins via
\begin{align}
	S^\alpha_{\bf r} =\frac{i}{2}b^\alpha_{\bf r}c_{\bf r};
	\label{MajoranaRep}
\end{align}
on the right side $b^\alpha_{\bf r}$ and $c_{\bf r}$ denote Majorana-fermion operators that are Hermitian, square to the identity, and anticommute with one another.  
For an illustration see Fig.~\ref{KitaevHoneycombFig}(b).
Remaining faithful to the original spin-1/2 Hilbert space requires enforcing the local constraint $b^x_{\bf r} b^y_{\bf r} b^z_{\bf r} c_{\bf r} = +1$ at every site.

In the Majorana representation, the Hamiltonian becomes
\begin{align}
	{H}_K =  \frac{K}{4}\sum_{\langle {\bf r r'}\rangle} i \hat{u}_{\bf r r'} c_{\bf r}  c_{\bf r'}.
	\label{hamfermion}
\end{align}
Above we introduced link variables $\hat{u}_{\bf r r'} \equiv i b_{\bf r}^\gamma b_{\bf r'}^\gamma \in \pm 1$ that, crucially, commute with each other and with the Hamiltonian.
The link variables can thus be treated as classical parameters---thereby reducing the model to a free-fermion problem in any fixed $\hat{u}_{\bf r r'}$ configuration \footnote{Obtaining physical spin wavefunctions still requires enforcing the local constraint $D_{\bf r} \equiv b^x_{\bf r} b^y_{\bf r} b^z_{\bf r} c_{\bf r} = +1$ for all ${\bf r}$, which can be done by applying a projector $P = \prod_{\bf r} \left(\frac{1+D_{\bf r}}{2}\right)$ to many-body fermion states.  Although $[\hat{u}_{\bf r r'},D_{\bf r''}] \neq 0$, gauge-invariant quantities (e.g., the energy) can nevertheless be exactly extracted from the free-fermion limit of Eq.~\eqref{hamfermion} with fixed $\hat{u}_{\bf r r'}$ values.}.  
Physically, $\hat{u}_{\bf r r'}$ is a $\mathbb{Z}_2$ gauge field whose flux around plaquette $p$ is proportional to the conserved $W_p$ operator in Eq.~\eqref{Fluxoperator} (hence the absence of nontrivial dynamics).

%%%%%%%%%%%%%%%%%%%%%%%%%%%%%%%%%%%%%%%%%%%%%%%%
%%%%%%%%%%%%%%%%%%%%%%%%%%%%%%%%%%%%%%%%%%%%%%%%

The ground state of Eq.~\eqref{hamfermion} arises in the sector with $\mathbb{Z}_2$ gauge flux of $\pi$ through every hexagonal plaquette~\cite{LiebFlux}.  
Let us decompose the honeycomb lattice into $A$ and $B$ sublattices, and also introduce vectors ${\bf e}_{j = 1,2,3}$ that link the two sublattices; see Fig.~\ref{KitaevHoneycombFig}(a).
A convenient gauge encoding $\pi$ flux per plaquette is $\hat u_{{\bf r, r + e}_j} = +1$ for all ${\bf r}$ on sublattice $A$.  
Inserting this gauge choice into Eq.~\eqref{hamfermion} yields a Hamiltonian
\begin{equation}
  \tilde H_K = \frac{K}{4} \sum_{{\bf r} \in A} \sum_{j = 1}^3 i c_{\bf r} c_{{\bf r + e}_j}
  \label{HKtilde}
\end{equation}
that describes the ground-state flux sector.
One can view Eq.~\eqref{HKtilde} as an analogue of graphene wherein Majorana fermions hop between nearest-neighbor honeycomb sites.
To obtain the spectrum of $H_K$ we pass to momentum space, employing conventions such that
\begin{align}
  c_{{\bf r} \in A/B} &= \sqrt{\frac{2}{N_{\rm uc}}} \sum_{{\bf k} \in {\rm BZ}}e^{i {\bf k \cdot r}}c_{A/B{\bf k}}
  \nonumber \\
  &= \sqrt{\frac{2}{N_{\rm uc}}} \sum_{{\bf k} \in {\rm BZ}_{\rm R}}(e^{i {\bf k \cdot r}}c_{A/B{\bf k}} + e^{-i {\bf k \cdot r}}c_{A/B{\bf k}}^\dagger),
  \label{cFT}
\end{align}
where $N_{\rm uc}$ is the number of unit cells.
The momentum-space operators so defined satisfy $\{c_{\alpha \bf k},c_{\beta \bf k'}^\dagger\} = \delta_{\alpha\beta}\delta_{\bf k,k'}$ and 
$c_{\alpha \bf k} = c_{\alpha-{\bf k}}^\dagger$ (reflecting Hermiticity of $c_{\bf r}$).  
In the second line of Eq.~\eqref{cFT} we used the latter property to express $c_{\bf r}$ as a sum over momenta in the right half of the Brillouin zone (${\rm BZ}_{\rm R}$), i.e., ${\bf k}$ with $k_x > 0$ as shown in Fig.~\ref{KitaevHoneycombFig}(b).
Defining a two-component spinor $C_{\bf k}^\dagger = [c_{A{\bf k}}^\dagger~c_{B{\bf k}}^\dagger]$ and a function $\xi({\bf k}) = -i(K/2)\sum_j e^{-i {\bf k \cdot e}_j}$, Eq.~\eqref{HKtilde} becomes
\begin{equation}
  \tilde H_K = \sum_{{\bf k} \in {\rm BZ}_{\rm R}} C_{\bf k}^\dagger \begin{bmatrix}
0 & \xi^*({\bf k}) \\
\xi({\bf k}) & 0 
\end{bmatrix}C_{\bf k}.
\label{HKtilde2}
\end{equation}
The resulting single-particle energies are $\pm |\xi({\bf k})|^2$, and the many-particle ground state populates all negative energy levels.

This ground state realizes the gapless spin liquid phase of Kitaev's honeycomb model.  
Specifically, Eq.~\eqref{HKtilde2} describes gapless (emergent!) fermion excitations with a \emph{single} massless Dirac cone centered at momentum ${\bf Q} = \frac{4\pi}{3a}{\bf \hat{x}}$, with $a$ the lattice constant.
See Fig.~\ref{KitaevHoneycombFig}(c).
We now focus on these gapless excitations by writing ${\bf k} = {\bf Q} + {\bf q}$ and retaining only modes with `small' ${\bf q}$.
Equation~\eqref{HKtilde2} then reduces to the following effective Dirac Hamiltonian that captures low-energy fermionic excitations in the ground-state flux sector:
\begin{align}
  \mathcal{H}_{\rm eff} &= v_{\rm bulk}\int_{\bf q} \Psi_{\bf q}^\dagger(q_x \sigma^y - q_y \sigma^x)\Psi_{\bf q}
  \nonumber \\
  &= v_{\rm bulk}\int_{\bf r} \Psi^\dagger(-i \partial_x \sigma^y +i \partial_y \sigma^x)\Psi.
  \label{H_Dirac}
\end{align}
Here $v_{\rm bulk} = \sqrt{3} aK/4$, $\Psi_{\bf q} \propto C_{\bf Q + q}$, and in the last line we Fourier transformed back to real space.
Furthermore, we have employed $\hbar = 1$ units, and continue to do so throughout (for clarity however we will express the conductance quantum as $e^2/h$).
This gapless spin liquid phase also admits gapped $\mathbb{Z}_2$-flux excitations that are not captured by $\mathcal{H}_{\rm eff}$.  

Suppose that we now supplement Eq.~\eqref{pureKitaev} with generic perturbations that preserve translation symmetry and time-reversal symmetry $\mathcal{T}$, leading to a Hamiltonian of the form
\begin{equation}
  H = H_K + \cdots.
  \label{H_Kgeneric}
\end{equation}
Despite the loss of exact solvability, one can address the stability of the gapless spin liquid from the viewpoint of the effective low-energy theory.  
The original spin operators transform under $\mathcal{T}$ according to ${\bf S}_{\bf r} \rightarrow - {\bf S}_{\bf r}$.
Within the ground-state flux sector, $\mathcal{T}$ sends $c_{{\bf r} \in A} \rightarrow c_{{\bf r} \in A}$ and $c_{{\bf r} \in B} \rightarrow - c_{{\bf r} \in B}$, and in turn transforms the low-energy Dirac field via $\Psi \rightarrow \sigma^z (\Psi^\dagger)^t$.  
The only translationally invariant perturbation to Eq.~\eqref{H_Dirac} that can open an energy gap is the mass term $m \Psi^\dagger \sigma^z\Psi$---which is odd under $\mathcal{T}$ and can not appear provided time-reversal symmetry persists.
Consequently, the gapless spin liquid constitutes a stable symmetry-protected phase with some finite tolerance to the ellipsis in Eq.~\eqref{H_Kgeneric}.

%%%%%%%%%%%%%%%%%%%%%%%%%%%%%%%%%%%%%%%%%%%%%%%%
\subsection{Non-Abelian spin liquid}
\label{NonAbelianSec}

In this paper we are primarily interested in the physics resulting when time-reversal symmetry is explicitly broken by an applied magnetic field ${\bf B}$.  
The field modifies Eq.~\eqref{H_Kgeneric} to 
\begin{equation}
  H = H_K -\sum_{\bf r} {\bf B} \cdot {\bf S}_{\bf r} + \cdots.
  \label{H_Kgeneric2}
\end{equation}
On symmetry grounds~\cite{Balents2016}, the Zeeman term can be expanded in terms of low-energy degrees of freedom as ${\bf B} \cdot {\bf S}_{\bf r} \sim \beta \Psi^\dagger \sigma^z \Psi + \cdots$.
Here $\beta \propto |{\bf B}|$ is a non-universal constant that vanishes only for fine-tuned field orientations~\cite{Tokoi}, while the ellipsis denotes additional symmetry-allowed terms that are unimportant for our purposes and will henceforth be dropped.
The effective low-energy Hamiltonian accordingly now reads
\begin{align}
  \mathcal{H}_{\rm eff} &= \int_{\bf r}\Psi^\dagger[v_{\rm bulk}(-i \partial_x \sigma^y +i \partial_y \sigma^x) + m \sigma^z]\Psi
  \label{H_Dirac2}
\end{align}
with $m \propto |{\bf B}|$, and describes emergent fermions with a \emph{gapped} Dirac spectrum illustrated in Fig.~\ref{KitaevHoneycombFig}(d).
[Without the generic perturbations that we implicitly included in Eq.~\eqref{H_Kgeneric2}, the Dirac gap would scale like $B^xB^yB^z$ instead of $|{\bf B}|$.  
We stress that this fine-tuned behavior is a pathology of perturbating about the exactly solvable $H_K$ Hamiltonian as Ref.~\onlinecite{Balents2016} discusses in detail.]

The resulting field-induced phase realizes a \emph{non-Abelian} spin liquid with `Ising' topological order.
Although the bulk is fully gapped, the system's boundary hosts a single emergent chiral Majorana mode with central charge $c = 1/2$.
(One can trace the edge state's existence to the quantized half-integer thermal Hall conductance that arises from gapping out a single Dirac cone; for related problems see Refs.~\onlinecite{Haldane,Volovik,KaneFisherThermal,Cappelli}.)
Low-energy edge excitations are described by the continuum Hamiltonian
\begin{align}
\label{chiralbdry}
	\mathcal{H}_\text{edge} =  \int_x (-iv_\text{edge}\gamma \, \partial_x \gamma),
\end{align}
where $v_\text{edge}$ is a non-universal velocity%
	~\footnote{In general the edge velocity is expected to depend on details of the boundary and need not be spatially uniform, but for simplicity we ignore such complications in this paper.}, 
$x$ is a coordinate along the boundary, and $\gamma(x)$ is a Majorana-fermion field.  
(For clarity we have employed subscripts that distinguish edge and bulk velocities, though later we abandon such notation.)
Here and below we normalize continuum Majorana fields such that
\begin{align}
\{ \gamma(x), \gamma(x') \} = \frac{1}{2}\delta(x-x').
\end{align}
With this choice the energy for an edge excitation with momentum $k$ is simply $v k$.  
Note that Eq.~\eqref{chiralbdry} exhibits a global $\mathbb{Z}_2$ symmetry that sends $\gamma \rightarrow -\gamma$, which as we will see in Sec.~\ref{Sewing} has important practical consequences for the interfaces that we exploit later in this paper.  

%%%%%%%%%%%%%%%%%%%%%%%%%%%%%%%%%%%%%%%%%%%%%%%%
\begin{figure}
	\includegraphics[width=\linewidth]{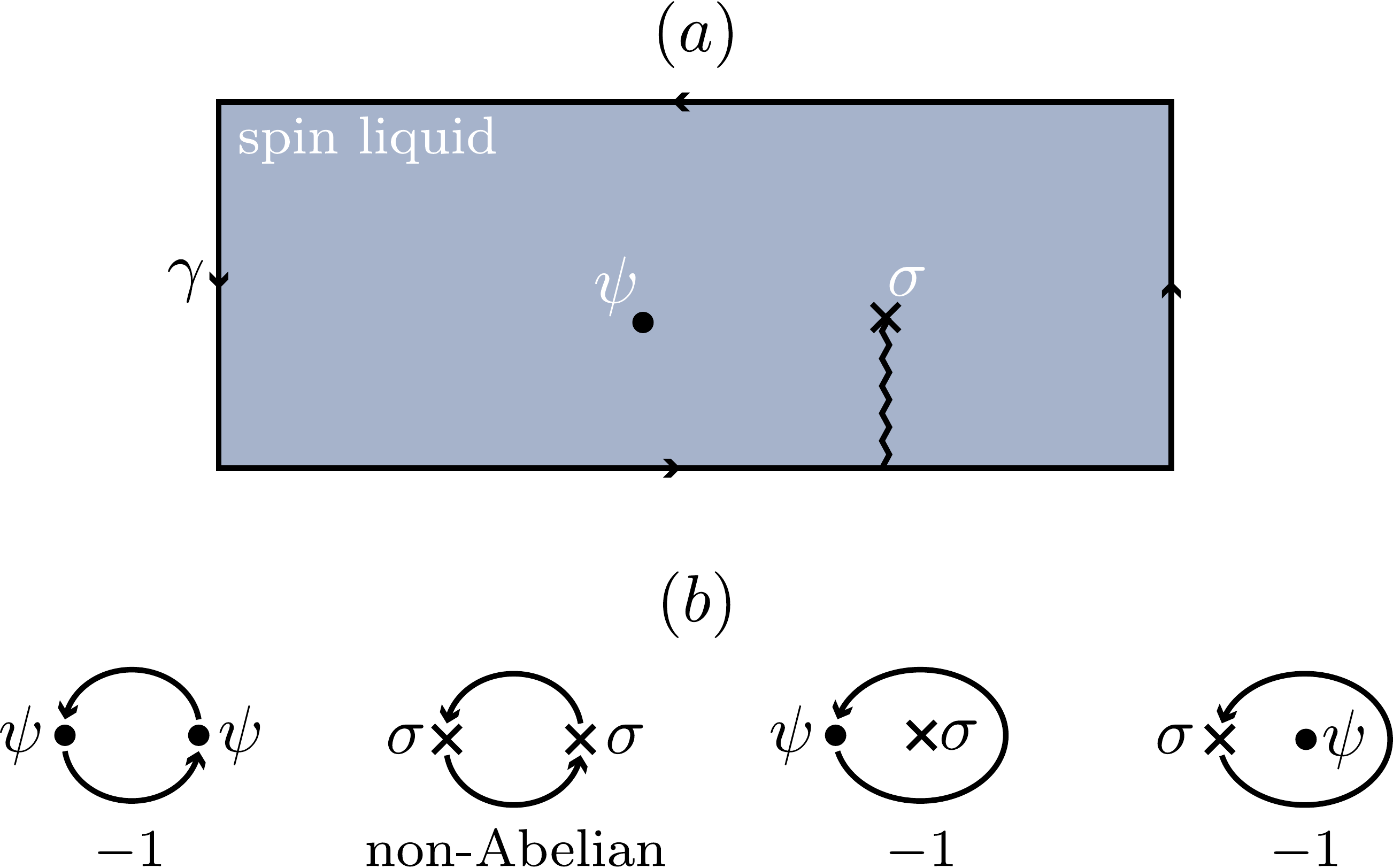}
	\caption{Non-Abelian spin liquid synopsis.
	(a) The boundary hosts a gapless chiral Majorana mode, while the bulk supports two nontrivial gapped quasiparticle types: emergent neutral fermions $\psi$ and Ising non-Abelian anyons $\sigma$. 
	Fermions acquire a minus sign on crossing the wavy line (which represents a branch cut) emanating from $\sigma$. 
	(b) Summary of quasiparticle braiding statistics.} 
	\label{NonAbelianSLsynopsis}
\end{figure}
%%%%%%%%%%%%%%%%%%%%%%%%%%%%%%%%%%%%%%%%%%%%%%%%

The bulk of the non-Abelian spin liquid supports three gapped quasiparticle types.
First, there are non-fractionalized bosonic excitations---as in any phase of matter---that we will call trivial particles labeled by $\mathds{1}$. 
Second, the system hosts more exotic gapped emergent fermions ($\psi$ particles) captured by the effective Hamiltonian in Eq.~\eqref{H_Dirac2}.
Third, and most interestingly, gapped $\mathbb{Z}_2$ flux excitations bind emergent Majorana zero modes and realize `Ising anyons' ($\sigma$ particles) with non-Abelian braiding statistics.
These quasiparticle types obey the following nontrivial `fusion rules',
\begin{align} 
	\psi \otimes \psi \cong \mathds{1}, \quad
	\sigma \otimes \sigma \cong \mathds{1} \oplus \psi,
	\quad	\psi \otimes \sigma \cong \sigma,
\end{align}
which roughly describe how they behave when brought together in space.
That is, two emergent fermions coalesce into a local boson, two Ising anyons can combine to yield \emph{either} a local boson or an emergent fermion, and Ising anyons can freely absorb emergent fermions without changing their quasiparticle type.
Viewed `in reverse', a local boson can fractionalize into a pair of emergent fermions, an individual emergent fermion can further fractionalize into a pair of Ising anyons, and pairs of Ising anyons can be pulled out of the vacuum.
Finally, $\psi$ and $\sigma$ particles exhibit not only nontrivial self-statistics, but also nontrivial mutual statistics: taking a fermion all the way around an Ising anyon, or vice versa, yields a statistical phase of $-1$.  
The above quasiparticle characteristics become essential for the circuits developed in Secs.~\ref{IsingDetection} and \ref{interferometry2}.
Figure~\ref{NonAbelianSLsynopsis} summarizes the bulk and edge content of the non-Abelian spin liquid.

%%%%%%%%%%%%%%%%%%%%%%%%%%%%%%%%%%%%%%%%%%%%%%%%
\section{Primer: Interacting helical Majorana fermions}
\label{Interacting_Majorana}

As an illuminating warm-up, next we explore gapless \emph{non-chiral} Majorana fermions propagating in 1D with strong interactions.
We proceed in two stages: first examining interfaces between non-Abelian spin liquids, and then turning to one-dimensional lattice models that harbor similar physics.  
Results obtained here carry over straightforwardly to the quantum Hall-spin liquid interfaces that we introduce in Sec.~\ref{QHSL_interface} and later exploit to electrically detect chiral Majorana edge states and bulk anyons in Kitaev materials (Secs.~\ref{EdgeDetection} through \ref{interferometry2}).

%%%%%%%%%%%%%%%%%%%%%%%%%%%%%%%%%%%%%%%%%%%%%%%%
\subsection{Sewing up non-Abelian spin liquids}
\label{Sewing}

\begin{figure}
	\includegraphics[width=\linewidth]{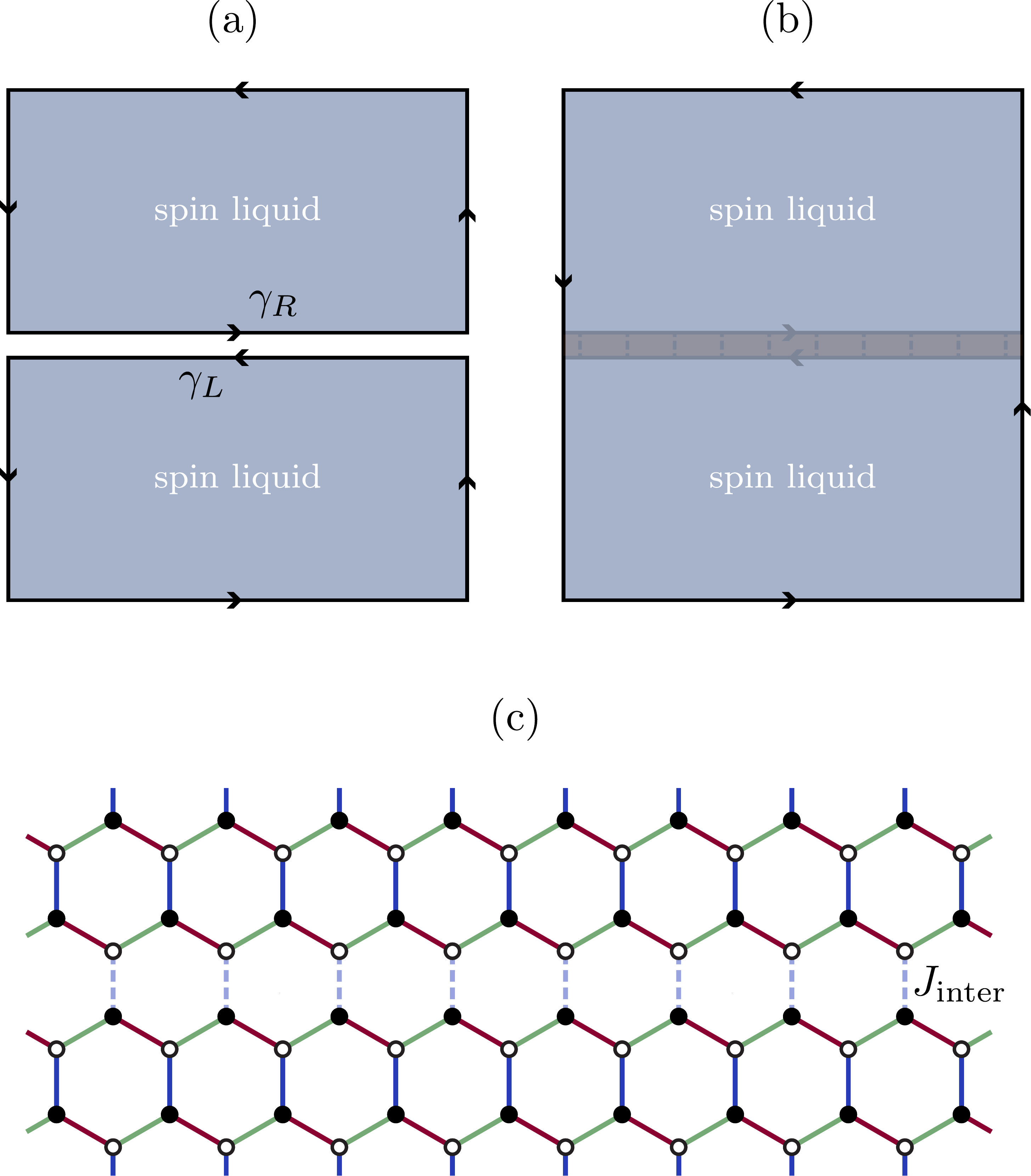}
	\caption{(a) Decoupled non-Abelian spin liquids hosting emergent chiral Majorana fermions $\gamma_R$ and $\gamma_L$ at their interface.  
	(b) Strong interactions between $\gamma_R$ and $\gamma_L$ [described by Eq.~\eqref{deltaH} with $\kappa>0$] gaps out these modes---thus sewing the two phases into a single non-Abelian spin liquid.
	(c) Microscopic view of the spin liquid interface, viewed as two Kitaev honeycomb models coupled via vertical bonds of strength $J_{\rm inter}>0$ (dashed lines); see Eq.~\eqref{Jinter}.  
	The interface is fully gapped, as in (b), when $J_{\rm inter}$ exceeds a critical value $J_c$, but otherwise hosts gapless Majorana modes, as in (a).
	} 
	\label{SL_fig}
\end{figure}

Consider the setup from Fig.~\ref{SL_fig}(a) consisting of two non-Abelian spin liquids realized in adjacent Kitaev materials. 
Physically, it is natural to anticipate that suitable hybridization between the subsystems can effectively sew them together---producing a single, uninterrupted spin liquid.  
Our goal here is to understand this sewing-up process, both from effective field theory and microscopic viewpoints.  

%%%%%%%%%%%%%%%%%%%%%%%%%%%%%%%%%%%%%%%%%%%%%%%%
%%%%%%%%%%%%%%%%%%%%%%%%%%%%%%%%%%%%%%%%%%%%%%%%

When the two layers decouple as in Fig.~\ref{SL_fig}(a), their interface hosts helical Majorana modes whose kinetic energy is described by the low-energy Hamiltonian
\begin{align}
	\mathcal{H}_0 = \int_x(-i v \gamma_R \partial_x \gamma_R + i v \gamma_L \partial_x \gamma_L).
	\label{H0}
\end{align}
Here $x$ is a coordinate along the interface, $v$ is the edge-state velocity, and $\gamma_{R}$ and $\gamma_L$ respectively denote right- and left-moving Majorana-fermion fields.  
Upon turning on interactions between the adjacent layers, the Hamiltonian becomes
$\mathcal{H} = \mathcal{H}_0 + \delta \mathcal{H}$, where $\delta \mathcal{H}$ hybridizes the helical Majorana modes.  
Crucially, the form of $\delta \mathcal{H}$ is constrained by the fact that $\gamma_R$ and $\gamma_L$ represent \emph{emergent} fermions originating from disjoint spin liquids.
In particular, only $\emph{pairs}$ of emergent fermions---which together form a boson---can tunnel across the interface.  
The simplest such interaction is given by
\begin{align}
	\delta \mathcal{H} = -\kappa \int_x (\gamma_R \partial_x \gamma_R)(\gamma_L \partial_x \gamma_L),
	\label{deltaH}
\end{align}
where the two derivatives are necessitated by Fermi statistics.  
Notice that $\mathcal{H}$ exhibits two independent $\mathbb{Z}_2$ symmetries, one corresponding to $\gamma_R \rightarrow - \gamma_R$ and the other corresponding to $\gamma_L \rightarrow -\gamma_L$.  
These symmetries can never be broken \emph{explicitly} by any physical perturbation, reflecting the fact that individual Majorana fermions $\gamma_R$ and $\gamma_L$ live only within their respective spin liquids.  

The coupling $\kappa$ is formally irrelevant at the fixed point described by the quadratic Hamiltonian $\mathcal{H}_0$.
`Weak' $\kappa$ thus has only perturbative effects, and most importantly does not gap out the helical Majorana modes.  
Evidently, sewing up the spin liquids requires strong coupling.  
At `large' $\kappa>0$, the system can lower its energy by condensing $\langle i \gamma_R \gamma_L \rangle \neq 0$---thereby \emph{spontaneously} breaking the two independent $\mathbb{Z}_2$ symmetries noted above (but preserving their product).  
For rough intuition, consider the term $-\frac{\kappa}{\delta x^2} \big[i\gamma_R(x+\delta x)\gamma_L(x+\delta x)\big] \big[i\gamma_R(x)\gamma_L(x)\big]$, which upon Taylor expanding in the microscopic length $\delta x$ generates the interaction from Eq.~\eqref{deltaH}.
The discrete form above clearly reveals that $\langle i \gamma_R \gamma_L \rangle \neq 0$ is favored provided $\kappa$ is positive.  
In the condensed regime, the interface can be modeled by an effective mean-field Hamiltonian
\begin{align}
	\mathcal{H}_{\rm MF} = \int_x(-i v \gamma_R \partial_x \gamma_R + i v \gamma_L \partial_x \gamma_L + i m \gamma_R \gamma_L),
	\label{HMF}
\end{align}
with $m \propto \langle i \gamma_R \gamma_L\rangle$ a mass whose sign, importantly, is chosen spontaneously.  
Equation~\eqref{HMF} exhibits a fully gapped spectrum, and thus describes a scenario where the two spin liquids have been sewn into one as sketched in Fig.~\ref{SL_fig}(b).  

Several consistency checks bolster the above picture.  
First, one can view the ground state $\ket{\Psi_{\rm MF}}$ of $\mathcal{H}_{\rm MF}$ as a trial wavefunction and the mass $m$ as a variational parameter.  
In Appendix~\ref{VariationalAppContinuum} we optimize $\langle \Psi_{\rm MF}|\mathcal{H}_0 + \delta \mathcal{H}|\Psi_{\rm MF}\rangle$ with respect to $m$ for varying $\kappa$.  
This analysis indeed captures a nonzero mass $m$ provided the dimensionless ratio $\kappa \Lambda^2/v$ exceeds a critical value, where $\Lambda$ is an ultraviolet momentum cutoff.
(For $\kappa <0$ the optimal mass always vanishes within this treatment.)

Second, the two nontrivial \emph{bulk} anyons of the non-Abelian spin liquid phase are encoded in the simple mean-field Hamiltonian $\mathcal{H}_{\rm MF}$ describing the gapped interface~\cite{TeoKane}.  
Neutral fermions are clearly present as gapped excitations.  
Ising non-Abelian anyons form at domain walls in which the spontaneously chosen mass $m$ changes sign; see Fig.~\ref{masspic}(b) for an illustration.  
Unpaired Majorana zero modes bind to such domain walls, leading to the hallmark degeneracy associated with Ising anyons.
Furthermore, since the sign of the mass is arbitrary, separating the domain walls by arbitrary distances costs only finite energy---i.e., the Ising anyons are bona fide deconfined quasiparticles.  
Additional insights into the domain-wall structure can be gleaned from the lattice model discussed in the Sec.~\ref{microscopics}.  

Third, the low-energy perspective presented above seamlessly connects to microscopics.
Let us add a spin-spin interaction 
\begin{align}
	\delta H = - J_{\rm inter} \sum_{({\bf r},{\bf r'})} S^z_{\bf r} S^z_{\bf r'} 
	\label{Jinter}
\end{align}
that couples spins across bonds $({\bf r},{\bf r'})$ [dashed lines in Fig.~\ref{SL_fig}(c)] that bridge the adjacent spin liquids.
At $J_{\rm inter} = 0$, one recovers decoupled spin liquids, and the interface hosts gapless Majorana modes that are stable to weak perturbations.  
On symmetry grounds, the boundary spin operators relate to continuum Majorana fields via $S^z_{\bf r} \sim \braket{S^z_{\bf r}} +  \alpha \norder{i \gamma_R \partial_x \gamma_R}$ on the lower edge of the interface and $S^z_{\bf r} \sim \braket{ S^z_{\bf r} } -\alpha \norder{i\gamma_L \partial_x \gamma_L}$ on the upper edge.  
Here angle brackets indicate ground-state expectation values, $\alpha$ is a non-universal constant, and $\norder{}$ denotes normal ordering;
the relative minus sign in the $\alpha$ pieces above reflects the opposite chirality for the two modes. 
Using this continuum expansion, the microscopic interaction $\delta H$ indeed generates the effective-Hamiltonian term in Eq.~\eqref{deltaH} with $\kappa \propto J_{\rm inter}$.
At $J_{\rm inter} = J$---corresponding to the strong-coupling limit---the system forms a single, translationally invariant non-Abelian spin liquid; here all gapless modes at the interface have clearly been vanquished.  
It follows that the spin liquids are sewn up provided $J_{\rm inter}$ exceeds a critical value $J_c$ that satisfies $0 < J_c < J$, in qualitative agreement with our continuum analysis.

%%%%%%%%%%%%%%%%%%%%%%%%%%%%%%%%%%%%%%%%%%%%%%%%
\subsection{Insights from microscopic models}
\label{microscopics}

Complementary insights can be gleaned by examining a strictly one-dimensional (1D) toy lattice model that also realizes interacting helical Majorana fermions.  
Consider an infinite chain of physical (rather than emergent) Majorana fermions $\gamma_a$ living on lattice sites $a$ and governed by the microscopic Hamiltonian
\begin{align}
	H = \sum_a(i t \gamma_a \gamma_{a+1} +i t' \gamma_a \gamma_{a+2}- U \gamma_{a-2} \gamma_{a-1}\gamma_{a+1}\gamma_{a+2});
	\label{Hpaul}
\end{align}
we will take $t>0$ for concreteness through.
References~\onlinecite{Fendley2018,Sannomiya2019} recently studied this model motivated in part by interesting connections to supersymmetry; see also Ref.~\onlinecite{Fendley2019}.  
Importantly, $H$ preserves an (anomalous~\footnote{In the purely 1D setting under consideration, $T$ symmetry is anomalous because it changes the sign of the total fermion parity operator $P = \prod_a (i \gamma_{2a}\gamma_{2a+1})$.}) translation symmetry $T$ that transforms $\gamma_a \rightarrow \gamma_{a+1}$.
At $t' = 0$ the Hamiltonian further preserves an antiunitary chiral symmetry $\mathcal{C}$ that sends $\gamma_a \rightarrow (-1)^a \gamma_a$ and $i\rightarrow -i$ \footnote{At $t' = 0$ the chain also preserves a unitary reflection symmetry that sends $\gamma_a \to (-1)^a\gamma_{-a}$, but this symmetry will not play a role in our discussion.}.

We first specialize to $t' = 0$.  In the $U = 0$ limit the chain is gapless.  
Here one can capture the low-energy physics by writing 
\begin{equation}
  \gamma_a \sim \gamma_L + (-1)^a \gamma_R,
  \label{gamma_expansion}
\end{equation}
where $\gamma_R$ and $\gamma_L$ again denote right- and left-moving Majorana fields, leading precisely to Eq.~\eqref{H0} with $v \propto t$.  
Translation symmetry $T$ sends $\gamma_R \rightarrow -\gamma_R$ (just as for one of the $\mathbb{Z}_2$ symmetries present for the spin liquid interface examined above) while $\mathcal{C}$ swaps $\gamma_R \leftrightarrow \gamma_L$.  
A mass term $i m \gamma_R\gamma_L$ is odd under $T$ and thus can never be generated explicitly by any $T$-preserving perturbation, similar to the scenario encountered in Sec.~\ref{Sewing}.  

\begin{figure}
	{Dimerization order parameter $\braket{ i\gamma_{a-1}\gamma_{a}-i\gamma_{a}\gamma_{a+1} }$}\\[-2ex]
	\includegraphics[width=0.99\linewidth]{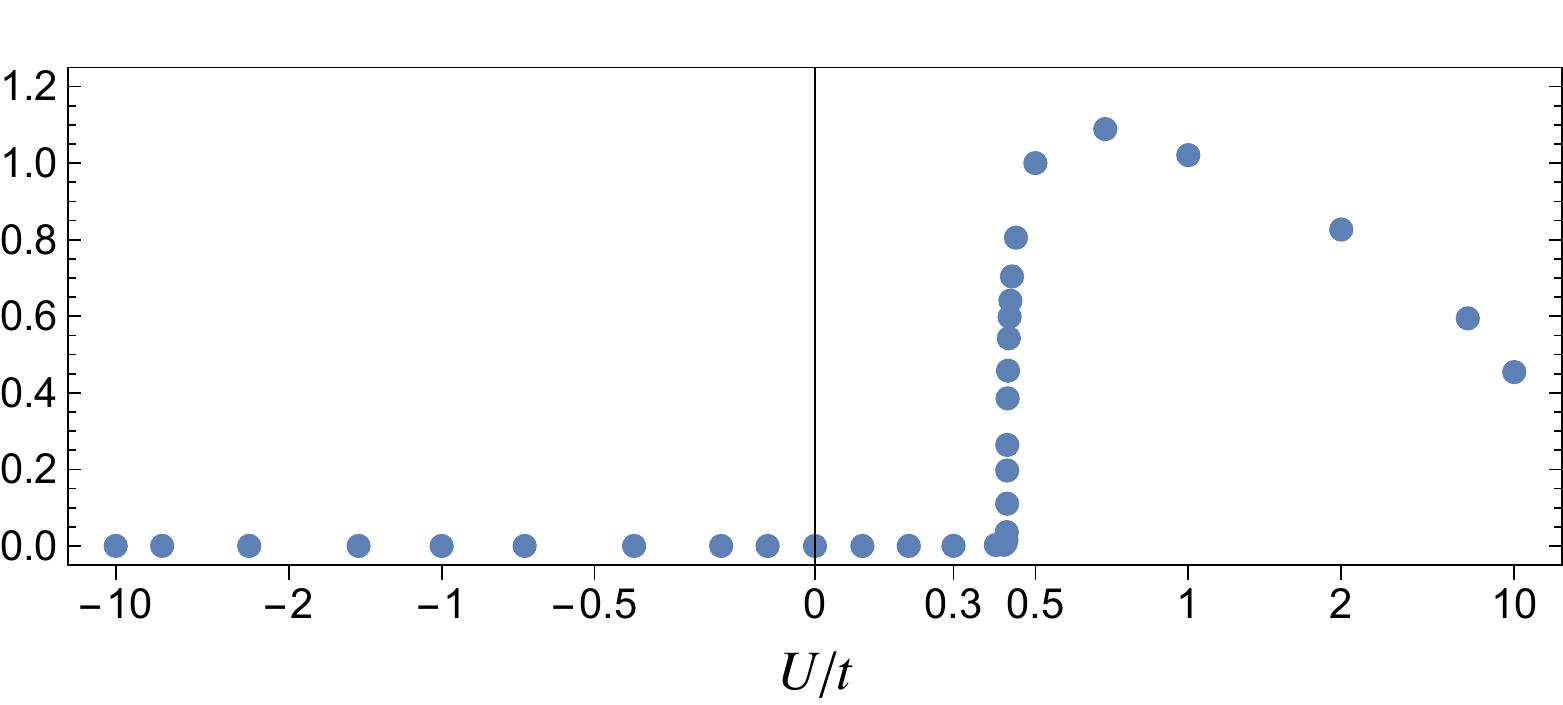}
	\caption{Dimerization order parameter $\braket{ i\gamma_{a-1}\gamma_{a}-i\gamma_{a}\gamma_{a+1} }$ versus $U/t$ obtained from DMRG simulations of the model in Eq.~\eqref{Hpaul} with $t' = 0$.
	This order parameter corresponds to the fermion mass in the continuum limit.
	For $U/t \gtrsim 0.42799(2)$, our simulations capture a gapped, spontaneously dimerized phase for the chain, in agreement with Ref.~\onlinecite{Fendley2018}.
	At smaller $U/t$ the chain instead realizes a critical state with central charge $c = 1/2$.  
	We additionally verified that this critical state extends to large negative $U/t$ values.
	}
	\label{fig:DMRG}
\end{figure}

Upon restoring non-zero $U$, the leading term that couples right- and left-movers corresponds to Eq.~\eqref{deltaH} with $\kappa \propto U$~\cite{Fendley2018}.  
Previous density-matrix renormalization-group (DMRG) simulations of Eq.~\eqref{Hpaul} (which we reproduce and extend to include $U<0$ in Fig.~\ref{fig:DMRG}) indicate that for $U\gtrsim 0.428 t$, criticality is destroyed in favor of a gapped, dimerized phase that spontaneously breaks $T$ symmetry~\cite{Fendley2018}.
In continuum language, here $\langle i \gamma_R \gamma_L\rangle \neq 0$ condenses and the helical Majorana fermions are gapped via generation of a mass $m\propto \langle i \gamma_R \gamma_L\rangle$ with arbitrary sign.  
Appendix~\ref{VariationalAppLattice} analyzes Eq.~\eqref{Hpaul} at $t' = 0$ using a variational approach that predicts spontaneous dimerization for $U \gtrsim 0.295 t$, in rough agreement with DMRG.

%%%%%%%%%%%%%%%%%%%%%%%%%%%%%%%%%%%%%%%%%%%%%%%%
%%%%%%%%%%%%%%%%%%%%%%%%%%%%%%%%%%%%%%%%%%%%%%%%

The gapped ground states at $U = t/2$ are known exactly~\cite{Fendley2018} and can be recovered by postulating `perfect' dimerization with $\mathcal{O} \equiv \langle i \gamma_{2a}\gamma_{2a+1}\rangle = -1$ and $\langle i \gamma_{2a-1} \gamma_{2a}\rangle = 0$.  
Decoupling the $U$ term using this ansatz generates a mean-field Hamiltonian $H_{\rm MF} = t\sum_a i \gamma_{2a} \gamma_{2a+1}$ for which $\mathcal{O} = -1$ in the ground state---indicating self-consistency.  
One can similarly show that the shifted dimerization with $\langle i\gamma_{2a-1} \gamma_{2a} \rangle = -1$ and $\langle i\gamma_{2a} \gamma_{2a+1} \rangle = 0$ yields a degenerate self-consistent solution.
Using Eq.~\eqref{gamma_expansion}, we have
\begin{equation}
	\braket{ i \gamma_a \gamma_{a+1} } \sim {\rm const}+(-1)^a \braket{i \gamma_R\gamma_L};
\end{equation}
it follows that these two dimerizations correspond to opposite-sign masses in the continuum formulation.

\begin{figure}
	\includegraphics[width=\linewidth]{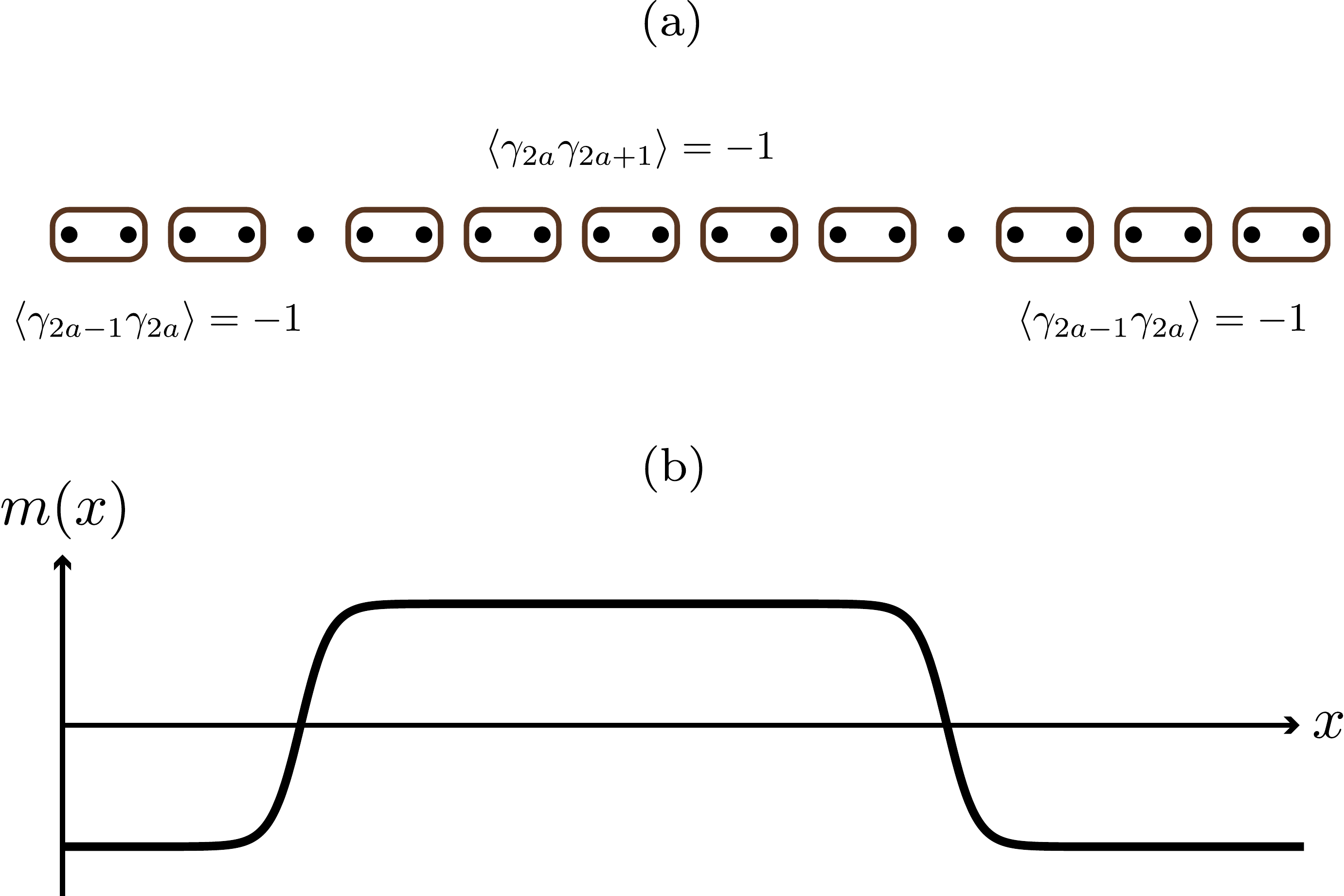}
	\caption{(a) Snapshot of domain-wall excitations at the exactly solvable, spontaneously dimerized limit of Eq.~\eqref{Hpaul} occurring for $t' = 0$ and $U = t/2$.  Neighboring sites $a,a+1$ enclosed by curved rectangles dimerize such that $\langle \gamma_a \gamma_{a+1}\rangle = -1$.  
	Domain walls at which the dimerization pattern shifts by one site trap unpaired Majorana zero modes [non-enclosed dots in (a)].
	In the low-energy limit, the dimerized chain is described at the mean-field level by Eq.~\eqref{HMF}, which includes a spontaneously generated mass $m$.  
	Microscopic domain-wall excitations from (a) correspond in continuum language to excitations at which the mass $m$ changes sign, as depicted in (b).  
	Such domain walls constitute 1D analogues of Ising anyons that arise at the spontaneously gapped spin liquid interface of Fig.~\ref{SL_fig}(b).} 
	\label{masspic}
\end{figure}

In Sec.~\ref{Sewing} we observed that the interface between two sewn-up spin liquids [Fig.~\ref{SL_fig}(b)] supports gapped emergent fermionic excitations, and that domain walls at which the mass $m$ changes sign correspond to Ising anyons hosting unpaired Majorana zero modes.  
The above mean-field ansatz at $U = t/2$, though operative in a physical-fermion system, provides an intuitive cartoon picture for these fractionalized quasiparticles%
	~\footnote{We caution, however, that physical fermions governed by the 1D lattice model do not realize Ising non-Abelian anyons in the same sense as the spin liquid.  In particular, explicitly breaking the anomalous translation symmetry $T$ in the 1D model generically confines the domain walls, whereas in the spin liquid interface no symmetry is required for their deconfinement.}.
The mean-field construction suggests that low-energy states can be labeled by domains exhibiting fixed dimerization---either $\braket{i\gamma_{2a-1}\gamma_{2a}}=-1$ or $\braket{i\gamma_{2a} \gamma_{2a+1}} = -1$---along with fermionic excitations within a given domain.
Fermionic excitations arise from flipping the sign of the dimerization expectation value at a particular bond, e.g., replacing $\langle \gamma_{2b-1}\gamma_{2b}\rangle\rightarrow +1$ for some $b$.
Figure~\ref{masspic}(a) illustrates an excited configuration with domain walls separating the two dimerization patterns, while Fig.~\ref{masspic}(b) shows the corresponding sign-changing mass profile in the continuum description.  
The domain walls clearly harbor unpaired Majorana zero modes as a consequence of the dimerization shift.  
Pairs of domain walls share a pair of Majorana zero modes, and therefore nonlocally host a single complex fermionic mode.
The occupancy of this complex fermion dictates whether the pair of domain walls `fuse' to the local vacuum or a fermion.
Comparing these quantum states to those at the interface of Fig.~\ref{SL_fig}(b), we can identify domain walls in the 1D model with Ising anyons in the spin liquid, and the excitations to which they fuse as the local vacuum or the emergent fermion.

%%%%%%%%%%%%%%%%%%%%%%%%%%%%%%%%%%%%%%%%%%%%%%%%
%%%%%%%%%%%%%%%%%%%%%%%%%%%%%%%%%%%%%%%%%%%%%%%%

More technically, we can also use this cartoon picture to relate the ground-state degeneracy of Eq.~\eqref{Hpaul} with open boundary conditions to the ground-state degeneracy of the spin liquid on a cylinder.  
For this purpose we identify the 1D chain governed by Eq.~\eqref{Hpaul} with the degrees of freedom along a path that connects the upper and lower cylinder ends. 
We can then view $U$ as a tuning potential that slices open or re-sews the cylinder as $U$ passes below or above the critical interaction strength at which spontaneous mass generation occurs. 
It is known that a topologically ordered phase on a cylinder exhibits a ground-state degeneracy given by the number of bulk anyons: three in the non-Abelian spin liquid of interest here (corresponding to anyons $\mathds{1}$, $\sigma$, and $\psi$).  
Hence we expect the gapped phase of Eq.~\eqref{Hpaul} to also admit three ground states, which is indeed the case as can be seen readily at $U = t/2$. 
In this limit, two of the ground states arise from the dimerization pattern that yields an unpaired Majorana at each end of the open chain; these states, which we label $\ket{\mathds{1}}$ and $\ket{\psi}$, can be identified with $\mathds{1}$ and $\psi$ anyons. 
The third arises from the shifted dimerization wherein the chain is fully gapped, including at the ends; this state, denoted $\ket{\sigma}$, can be identified with $\sigma$.  
As a sanity check, we can pass between the two dimerization patterns by nucleating a pair of domain walls in the bulk of the chain and then bringing one to each boundary.
If the chain begins in $\ket{\mathds{1}}$ or $\ket{\psi}$, the boundary Majorana zero modes pair up with those carried by the domain walls and create $\ket{\sigma}$ (or a locally related excited state). 
Conversely, if the chain begins in $\ket{\sigma}$, the domain walls shuttle unpaired Majorana zero modes to the boundary and thus yield $\ket{\mathds{1}}$ or $\ket{\psi}$.

\begin{figure}
	\includegraphics[width=0.8\linewidth]{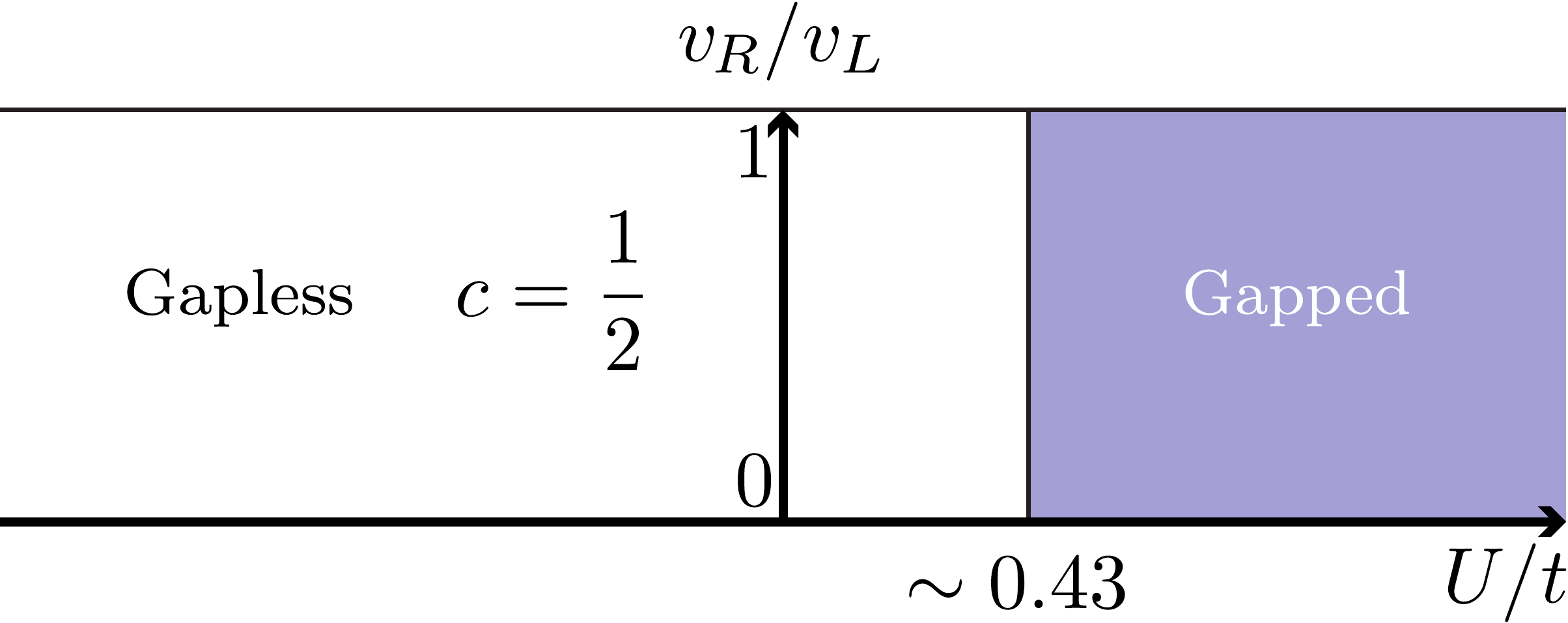}
	\caption{Schematic phase diagram of Eq.~\eqref{Hpaul} as a function of $U/t$ and velocity anisotropy induced by $t'$.  
	Here $v_{R}$ and $v_L$ respectively denote the velocities for right- and left-moving Majorana fermions at $U = 0$; Eq.~\eqref{v_anisotropy} specifies their ratio.  
	We assume $0 < v_R/v_L \leq 1$ for simplicity, though clearly the same physics arises when $v_R \leftrightarrow v_L$.
	Crucially, the phase boundary separating the critical $c = 1/2$ state and the spontaneously dimerized gapped phase depends very weakly on $v_R/v_L$---as found in our DMRG calculations.
	This observation suggests that unequal Majorana-fermion velocities, which would be expected for right- and left-movers of distinct physical origin [see, e.g., Fig.~\ref{SLQH_fig}], do not obstruct the formation of a gapped phase.  
	} 	
	\label{Ubytfig}
\end{figure}

Next we restore $t' \neq 0$.
At $U = 0$ the chain remains gapless, though the velocities $v_L$ and $v_R$ for left- and right-movers now differ due to the loss of $\mathcal{C}$ symmetry.
Explicitly, we have
\begin{align}
	\frac{v_R}{v_L} = \frac{t-2t'}{t+2t'},
	\label{v_anisotropy}
\end{align}
which vanishes as $t' \rightarrow t/2$.  
(For $t'>t/2$ additional low-energy modes appear; we will only consider $0\leq t' < t/2$ here.)  
Reference~\onlinecite{Fendley2018} found that velocity anisotropy very weakly influences the critical interaction $U$ above which the chain spontaneously dimerizes.
Our DMRG simulations confirm this result: the critical $U$ shifts by less than $1\%$ all the way down to $v_R/v_L \approx 0.07$.
For instance, we find $U_c/t \approx 0.4297(3)$ at $t'/t=0.44$, compared to $U_c/t \approx 0.42799(2)$ at $t'=0$.
Figure~\ref{Ubytfig} summarizes the phase diagram extracted from DMRG.

%%%%%%%%%%%%%%%%%%%%%%%%%%%%%%%%%%%%%%%%%%%%%%%%
%%%%%%%%%%%%%%%%%%%%%%%%%%%%%%%%%%%%%%%%%%%%%%%%

A spontaneously dimerized gapped phase can also arise in the modified model obtained by replacing the four-fermion interaction in Eq.~\eqref{Hpaul} with $U_R \gamma_{a-1} \gamma_a \gamma_{a+1} \gamma_{a+2}$.  
This seemingly innocuous microscopic modification, however, boosts the required interaction strength by three orders of magnitude: $U_R\gtrsim 250 t$~\cite{Rahmani1,Rahmani2}.
In Appendix~\ref{RahmaniApp} we explain this curious observation (among other aspects of this model's phase diagram) as arising from kinetic-energy renormalization by the $U_R$ interaction.  
In continuum language, increasing $U_R$ both generates the interaction in Eq.~\eqref{deltaH} \emph{and} increases the velocity in Eq.~\eqref{H0}, thereby sharply suppressing the onset of the strong-coupling limit where interactions dominate over kinetic energy.

%%%%%%%%%%%%%%%%%%%%%%%%%%%%%%%%%%%%%%%%%%%%%%%%
\section{Sewing a non-Abelian spin liquid to an electronic quantum Hall phase}
\label{QHSL_interface}

We have now seen two examples wherein strong interactions catalyze a condensation transition with $\langle i \gamma_R \gamma_L\rangle \neq 0$.  
At the interface between two non-Abelian spin liquids examined in Sec.~\ref{Sewing}, $\gamma_R$ and $\gamma_L$ both represent emergent fermions residing in initially separate fractionalized bosonic systems that were stitched together by the condensation.
In the strictly 1D model from Sec.~\ref{microscopics}, by contrast, both fields represent physical fermions.  
Next we we will explore a system in which a very similar condensation arises, but instead from the combination of a physical \emph{and} emergent Majorana fermion---providing a means of coherently converting one into the other.  

\begin{figure}
	\includegraphics[width=\linewidth]{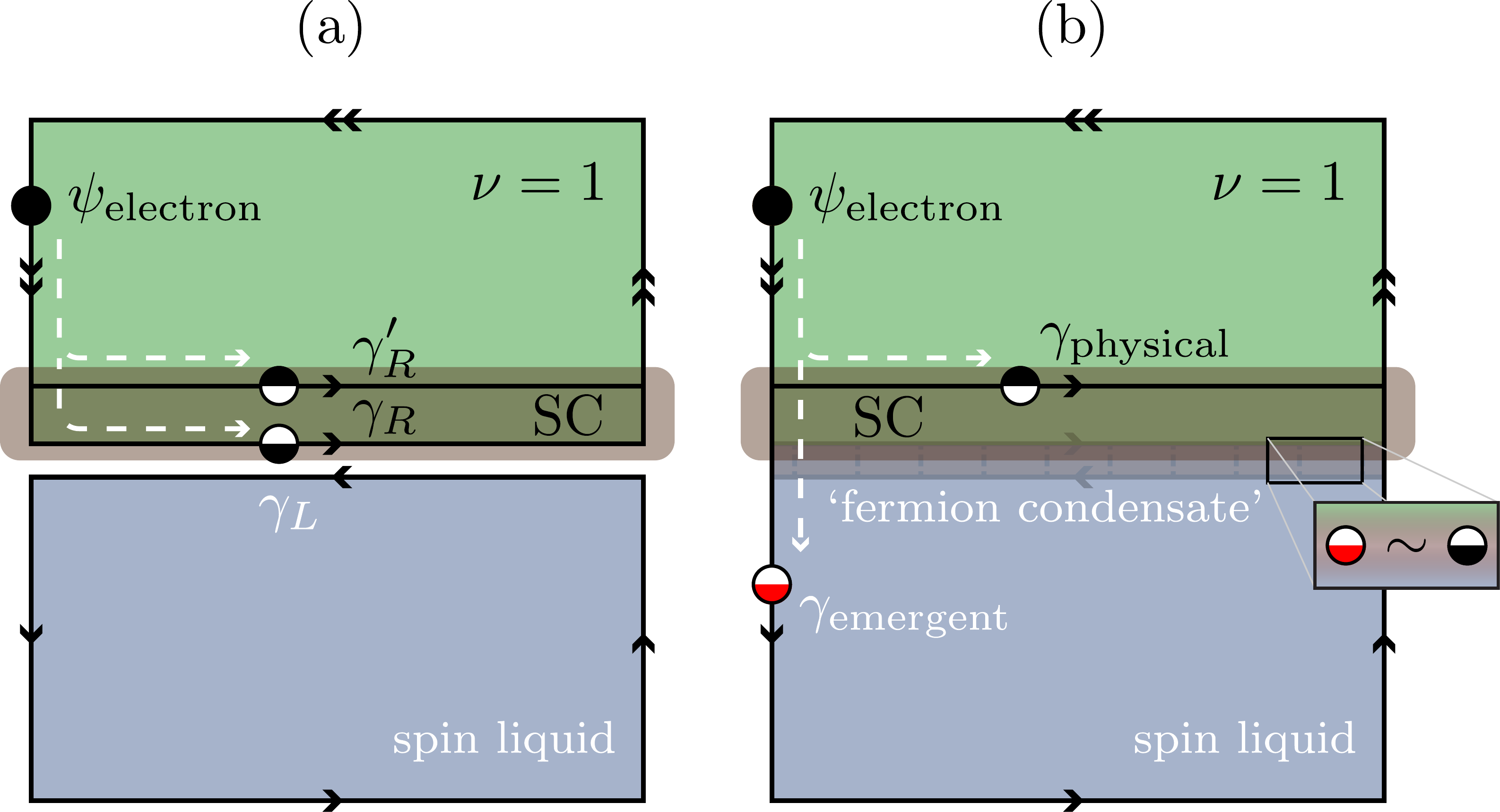}
	\caption{(a) Interface between a non-Abelian spin liquid and a proximitized $\nu = 1$ integer quantum Hall system.  
	The bare $\nu = 1$ edge state---indicated by double arrows---generically separates into two co-propagating Majorana fermions $\gamma_R$ and $\gamma_R'$ beneath the superconductor (SC) in the central region.
	(b) Strong interactions can gap out $\gamma_R$ and the spin liquid's edge Majorana fermion $\gamma_L$ via formation of a `fermion condensate' that partially sews the two subsystems together.  
	Physical Majorana fermions (black half-circles) and emergent Majorana fermions (red half-circles) are then identified at the interface; see zoom-in for an illustration.  
	A striking consequence follows: an electron injected at low energies into the $\nu = 1$ edge splinters into a pair of Majorana fermions, one of which \emph{unavoidably} enters the spin liquid.}
	\label{SLQH_fig}
\end{figure}

Figure~\ref{SLQH_fig}(a) illustrates the setup, consisting of an electronic integer quantum Hall system at filling factor $\nu = 1$ adjacent to a non-Abelian spin liquid.  
Additionally, the quantum Hall edge couples to a conventional superconductor; here and in similar setups studied in later sections, we assume fully gapped superconductivity, though we briefly discuss the role of low-lying excitations deriving from vortices and/or disorder in Sec.~\ref{discussion}.
We first present a qualitative picture for the physics that arises from interactions between these subsystems.

%%%%%%%%%%%%%%%%%%%%%%%%%%%%%%%%%%%%%%%%%%%%%%%%
%%%%%%%%%%%%%%%%%%%%%%%%%%%%%%%%%%%%%%%%%%%%%%%%

The quantum Hall edge hosts a chiral mode that can be viewed as a pair of copropagating chiral Majorana fermions (hence double arrows employed in our illustrations).  
Beneath the superconductor, the loss of charge conservation generically allows those copropagating Majorana fermions to displace from one another as shown in Fig.~\ref{SLQH_fig}(a).  
Microscopic interactions can only backscatter electrically neutral bosons---e.g., energy---between the quantum Hall and spin liquid edge states because the latter reside in an electrical insulator that hosts only emergent fermions.  
More precisely, a charge-$2ne$ excitation ($n$ is an integer) from the quantum Hall edge can neutralize by shedding its charge into the superconducting condensate%
	~\footnote{Spin-orbit coupling facilitates such processes.  For instance, spin-orbit interactions in the superconductor generically yield a triplet component at the interface, enabling injection and removal of Cooper pairs from the quantum Hall system even if the quantum Hall edge state exhibits perfect spin polarization.},
	and then backscatter into the spin liquid edge mode.  
Sufficiently strong backscattering events of this nature \emph{partially} sew up the quantum Hall state and non-Abelian spin liquid.  
That is, the emergent chiral Majorana fermion from the latter gaps out with `half' of the $\nu = 1$ edge mode, leaving a single physical chiral Majorana edge mode behind.
As sketched in Fig.~\ref{SLQH_fig}(b) an electron injected at low energies into the `naked' part of the quantum Hall edge then splinters into a pair of Majorana fermions, one of which \emph{unavoidably} enters the non-Abelian spin liquid as an emergent fermion.

For a more formal analysis, we write the effective Hamiltonian for the interface as 
\begin{equation}
  \mathcal{H} = \mathcal{H}_{\rm SL} + \mathcal{H}_{\rm QH} + \delta \mathcal{H}.
  \label{H_SL_QH}
\end{equation}
The first term,
\begin{equation}
  \mathcal{H}_{\rm SL} = \int_x(i v_L \gamma_L\partial_x\gamma_L),
\end{equation}
describes the spin liquid's emergent chiral-Majorana edge state with velocity $v_L$.
The second governs the proximitized $\nu = 1$ edge and takes the form
\begin{equation}
  \mathcal{H}_{\rm QH} = \int_x [-i u_R \psi_R^\dagger \partial_x \psi_R + \Delta(i\psi_R \partial_x \psi_R + H.c.)],
  \label{H_QH}
\end{equation}
where $\psi_R$ is a complex fermion operator that removes electrons from the edge state, $u_R$ is the associated velocity, and $\Delta$ is the proximity-induced pairing amplitude.  
Passing to a Majorana representation via $\psi_R = \gamma_R + i \gamma_R'$, one can equivalently write~\cite{Gamayun2017}
\begin{equation}
  \mathcal{H}_{\rm QH} = \int_x (-i v_R \gamma_R \partial_x \gamma_R - i v_R' \gamma_R' \partial_x \gamma_R');
  \label{H_QH2}
\end{equation} 
the velocities for the constituent co-propagating Majorana fermions $\gamma_R$ and $\gamma_R'$ are $v_{R} = u_R - 2\Delta$ and $v_R' = u_R + 2\Delta$, respectively.  

The final term in Eq.~\eqref{H_SL_QH} encodes interactions between the spin liquid and quantum Hall edge states.  
Suppose that $\gamma_R$ resides closest to $\gamma_L$ as in Fig.~\ref{SLQH_fig}.  
It is then reasonable to assume that interactions predominantly couple these fermions, so we take $\delta H$ to be given precisely by Eq.~\eqref{deltaH}.  
Useful insight follows from re-expressing $\delta H$ in terms of complex fermions $\psi_R$: $\delta H = -\kappa/4 \int_x(\gamma_L \partial_x\gamma_L)(\psi_R^\dagger \partial_x \psi_R + \psi_R \partial_x \psi_R + H.c.)$.
Here we see that interactions transfer electrically neutral dipoles as well as Cooper pairs from the quantum Hall edge to the spin liquid, consistent with our preceding physical picture.

The full Hamiltonian in Eq.~\eqref{H_SL_QH} reduces to the model studied in the previous subsections, supplemented by a decoupled sector for the $\gamma_R'$ chiral Majorana fermion.  
We immediately conclude that strong interactions can condense $\langle i \gamma_R \gamma_L \rangle \neq 0$ and partially gap out the interface, again in line with the above physical picture.
This conclusion holds even if the velocities $v_L$ and $v_R$---which bear no relation---differ significantly; recall Sec.~\ref{microscopics}.

In the present context, the transition to a state with $\langle i\gamma_R \gamma_L\rangle \neq 0$ is sometimes referred to as `fermion condensation' since the condensed object involves only \emph{one} physical fermion (specifically $\gamma_R$).  
A precise mathematical formulation of this striking phenomenon can be found in Ref.~\onlinecite{AasenFC}; see also Refs.~\onlinecite{Barkeshli2014,Barkeshli2015,Wan2017} for related earlier applications.  
The essential role played by proximity-induced superconductivity also becomes clear from this vantage point.
The condensate $\langle i \gamma_R \gamma_L \rangle \sim \langle i(\psi_R + \psi_R^\dagger)\gamma_L \rangle \neq 0$ clearly does not preserve $\mathrm{U}(1)$ charge conservation for the physical fermions.  
Without externally imposed superconductivity, interactions would thus need to \emph{spontaneously} break $\mathrm{U}(1)$ in order to partially gap the interface, which can not transpire due to the quasi-1D nature of the interface.  
Finally, we note that essentially the same fermion condensation transition can arise from interfacing a non-Abelian spin liquid with other electronic platforms, including conventional spinful 1D wires.  
We discuss alternative setups further in Sec.~\ref{discussion}.

%%%%%%%%%%%%%%%%%%%%%%%%%%%%%%%%%%%%%%%%%%%%%%%%
\section{Electrical detection of chiral Majorana edge modes in non-Abelian spin liquids}
\label{EdgeDetection}

\begin{figure}
	\includegraphics[width=\linewidth]{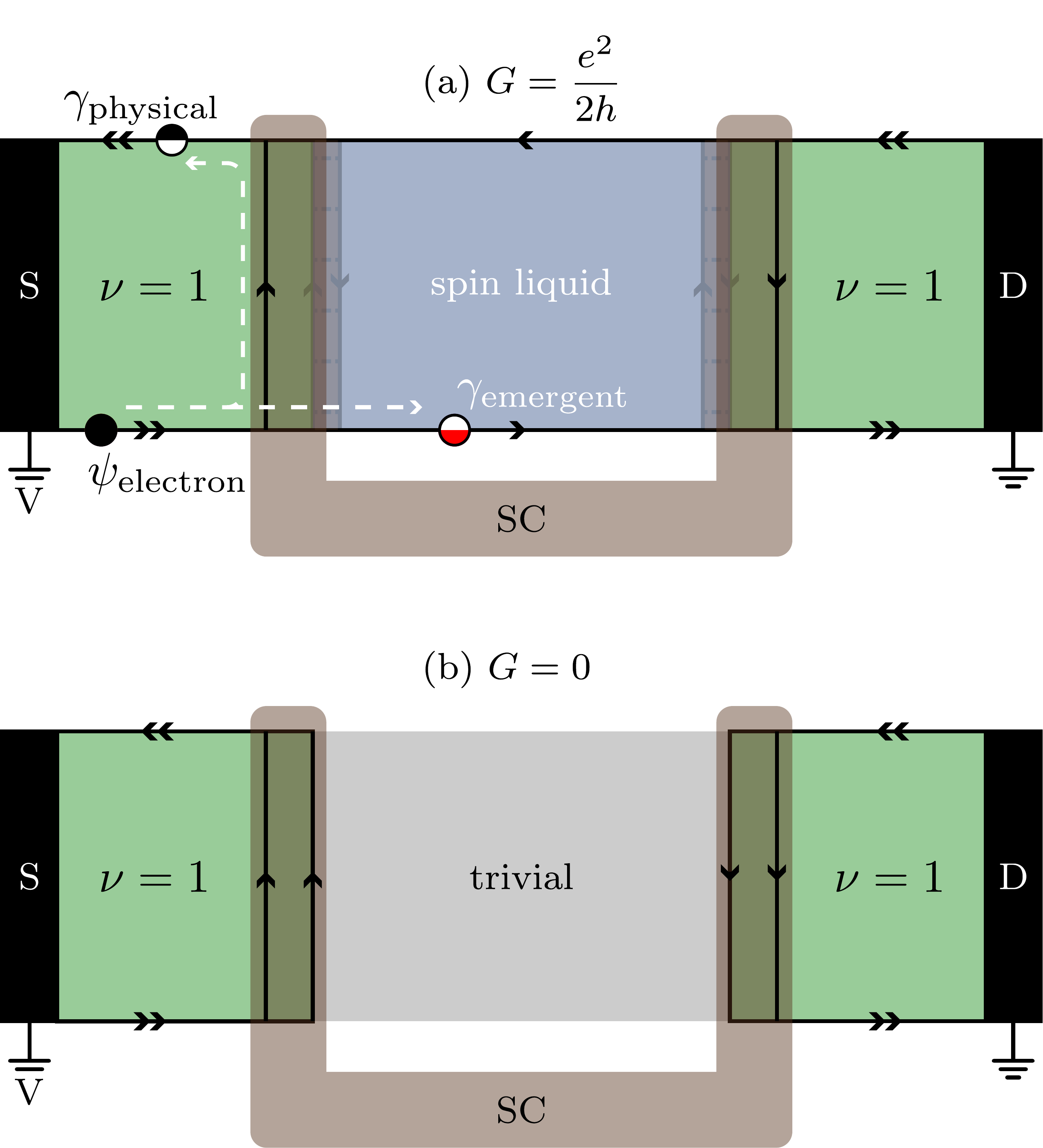}
	\caption{(a) Circuit that \emph{electrically} detects the spin liquid's emergent chiral Majorana edge state. 
	Any electrical current that flows between the source (S) and drain (D) necessarily passes through the central superconductor (and not the spin liquid). 
	Nevertheless, the perfect conversion between physical and emergent fermions at the quantum Hall--spin liquid interfaces dictates the conductance---which at zero bias is quantized at $G = \frac{e^2}{2h}$.
	(b) Control circuit in which the spin liquid is replaced with a trivial phase (e.g., a magnetically ordered state).  
	Elimination of the physical fermion $\leftrightarrow$ emergent fermion conversion processes that underlie transport in (a) leads to vanishing  zero-bias conductance $G = 0$ in this case.} 
	\label{condexp}
\end{figure}

As a first application of the phenomena developed in Sec.~\ref{QHSL_interface}, we introduce a scheme for \emph{electrically} detecting the spin liquid's emergent chiral Majorana edge state.  
Figure~\ref{condexp}(a) sketches the relevant circuit.  Here a pair of $\nu = 1$ quantum Hall systems flank a non-Abelian spin liquid.  
The interfaces are partially gapped by fermion condensation, facilitated by superconductors that are floating but exhibit negligible charging energy; note that the superconductors connect on the bottom end.  
A source on the far left is biased with voltage $V$, while a drain on the far right is grounded.
We stress that no electrical current flows through spin liquid---which still realizes a good Mott insulator.  
Any current instead passes between the quantum Hall systems via the intervening floating superconductor.  
Nevertheless, the spin liquid is by no means a spectator: electrons from the source propagate chirally along the $\nu = 1$ edge, then partially convert into emergent chiral Majorana fermions in the spin liquid, and finally re-enter the $\nu = 1$ edge as physical fermions on the other end.  
This inevitable conversion between physical and emergent fermions ultimately dictates the circuit's electrical transport characteristics as we will see.

%%%%%%%%%%%%%%%%%%%%%%%%%%%%%%%%%%%%%%%%%%%%%%%%
%%%%%%%%%%%%%%%%%%%%%%%%%%%%%%%%%%%%%%%%%%%%%%%%

We focus on vanishingly small temperature $T$ and bias voltage $V$, where the conductance attains a universal quantized value of $G = \frac{e^2}{2h}$.  
To understand this value, first observe that the left and right halves constitute identical resistors in series; hence $G = g/2$ with $g$ the conductance for (say) the left half.  
One can deduce $g$ as follows.  
An incident electron from the source splits up into an emergent Majorana fermion in the spin liquid and a physical Majorana fermion that reflects back to the source [see Fig.~\ref{condexp}(a)].  
The physical Majorana fermion is, by definition, equal part electron and equal part hole, implying probability $1/2$ for Andreev reflection.  
Thus $g = \frac{1}{2} \times \frac{2e^2}{h}$, where the factor of $2$ in the numerator arises because each Andreev process injects a Cooper pair into the superconductor.  
The overall conductance is then $G = \frac{e^2}{2h}$ as advertised.  
Although we are focusing on $V\rightarrow 0$ here, the conductance remains $G \approx \frac{e^2}{2h}$ for voltages below the gap scale of the fermion condensate.  

What happens if the emergent chiral Majorana fermions disappear entirely from the setup?  
One can readily arrange this scenario, e.g., by changing the magnetic field such that the Kitaev material exits the non-Abelian spin liquid phase.
The resulting circuit---which furnishes an essential control experiment---appears in Fig.~\ref{condexp}(b).
Most importantly, fermion condensation is now precluded, so that both $\nu = 1$ edge states must simply `turn around' at the interface with the central trivial region.  
In this case the probability for Andreev reflection \emph{vanishes} at asymptotically low energies, yielding conductance $G = 0$.

The absence of Andreev processes can be most simply understood as a consequence of Fermi statistics: the induced pairing term for the $\nu = 1$ edge state, $\Delta(i \psi_R\partial_x \psi_R + H.c.)$, necessarily contains a derivative and thus vanishes with the incident electron's momentum.  
Alternatively, as the two co-propagating Majorana fermions traverse the superconductor, they generally acquire different phase factors, in turn `rotating' the incident electron in particle-hole space and generating Andreev processes.  
The phase difference explicitly reads $\delta \phi = (k-k')L$, where $k$ and $k'$ denote the wavevectors of the two Majorana fermions as they pass through the superconducting region of length $L$.
At finite incident energy the wavevectors differ, i.e., $k \neq k'$, due to unequal velocities for the Majorana fermions in that region; recall Eq.~\eqref{H_QH2}.
As the incident electron energy vanishes, however, $k,k'\rightarrow 0$ and hence $\delta \phi \rightarrow 0$%
	~\footnote{We stress the importance of a $\nu = 1$ edge in the arguments presented here.  For a $\nu = 2$ quantum Hall system, by contrast, Andreev processes do \emph{not} freeze out at low energies.  In this alternative setting, Fermi statistics allows a pairing term $\Delta(\psi_{R1} \psi_{R2} + H.c.)$, where $\psi_{R1}$ and $\psi_{R2}$ describe fermions in the two edge channels at the $\nu = 2$ boundary.  Such a pairing term does not vanish at zero momentum.  As a corollary, at zero incident electron energy the momenta for the modes beneath the superconductor need not vanish---allowing a finite $\delta \phi$ even at asymptotically low energies.}.
An incoming electron must then exit the superconductor as an electron, yielding zero net current across the circuit in Fig.~\ref{condexp}(b).
In this control scenario the conductance remains $G \approx 0$ up to voltages $V \sim \frac{1}{e L}\frac{v_R v_R'}{|v_R-v_R'|}$, at which $\delta \phi$ becomes appreciable~\cite{Gamayun2017}. 

The contrast between the two circuits in Fig.~\ref{condexp} is particularly striking given that they differ solely in the properties of an electrically inert element.  
Nontrivial conductance quantization for Fig.~\ref{condexp}(a) relies on emergent chiral Majorana fermions in the non-Abelian spin liquid together with fermion condensation, and thus constitutes an electrical signature of both phenomena.  

Figure~\ref{condexp}(a) closely resembles the quantum anomalous Hall--superconductor heterostructures studied theoretically in Refs.~\onlinecite{Chung2011,Wang2015,Lian2016,ChenLaw,Chen2018,LianWang} and experimentally in Ref.~\onlinecite{WangMajorana}, where precisely the same quantized conductance was proposed as a signature of \emph{physical} chiral Majorana edges states at the boundary of a two-dimensional topological superconductor.  
In that context alternative quantization mechanisms that do not invoke chiral Majorana modes have also been introduced (e.g., disorder and dephasing, or if the superconductor behaves as a normal contact~\cite{JiWen,Huang2018,Kayyalha}; see also the critical discussion in Ref.~\onlinecite{LianVaezi}).
If operative in our setups, such trivial mechanisms would---at most---depend weakly on the precise phase of matter realized in the Kitaev material, thus yielding similar transport characteristics for \emph{both} circuits in Fig.~\ref{condexp}.  
Observing the qualitatively different conductances predicted for Figs.~\ref{condexp}(a) and (b) would therefore strongly suggest against these alternative interpretations.

%%%%%%%%%%%%%%%%%%%%%%%%%%%%%%%%%%%%%%%%%%%%%%%%
\section{Electrical detection of bulk Ising non-Abelian anyons}
\label{IsingDetection}

\begin{figure}
	\includegraphics[width=\linewidth]{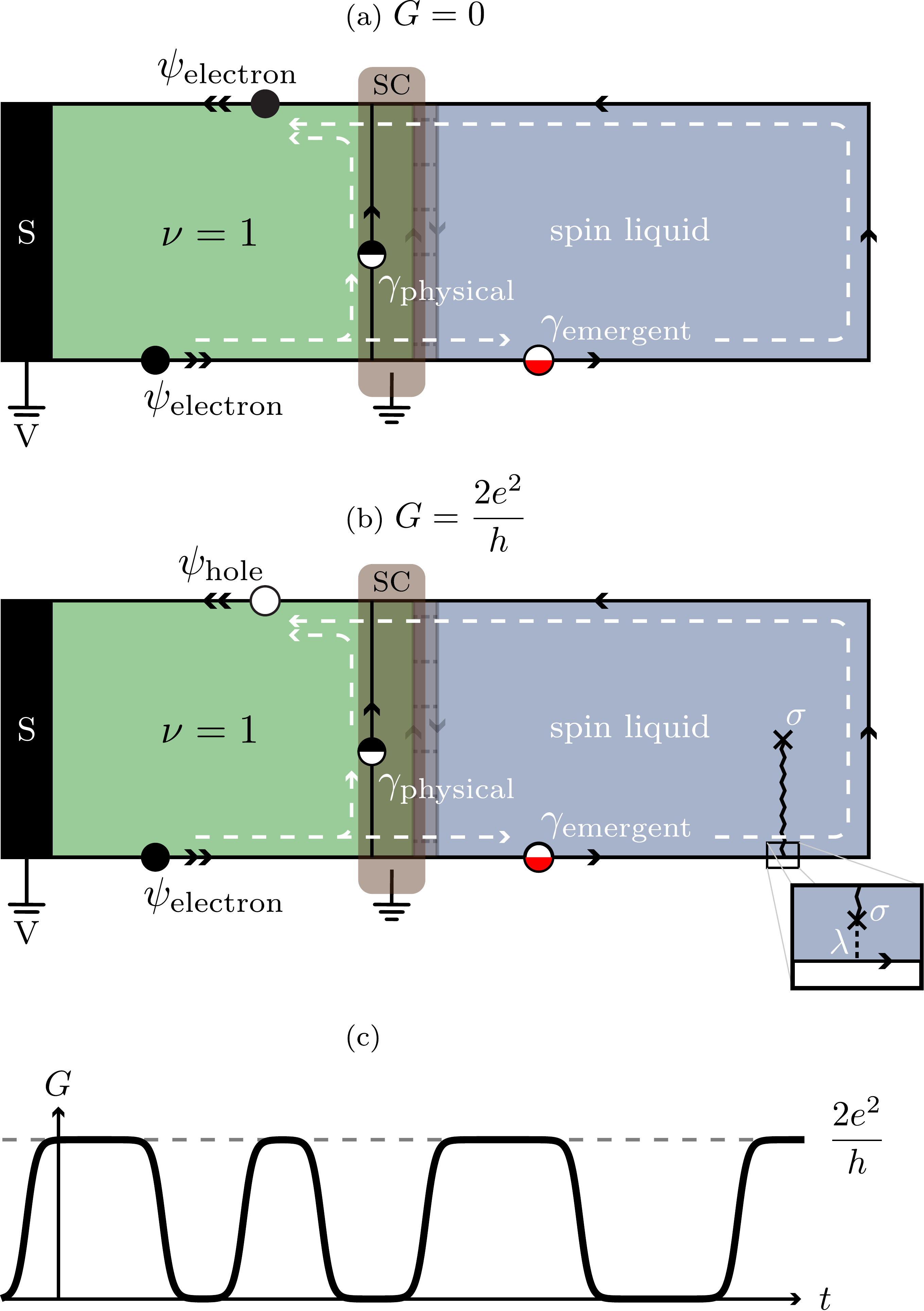}
	\caption{(a,b) Electrical detection of bulk Ising anyons via quantized zero-bias conductance $G$. 
	Current flows from the source to the grounded superconductor.
	In (a) the spin liquid contains no Ising anyons in the bulk. 
	Here electrons injected at zero energy along the lower $\nu = 1$ edge splinter into physical and emergent Majorana fermions, but simply recombine into electrons at the upper $\nu = 1$ edge.
	No current flows into the superconductor and hence $G = 0$.
	In (b) a pair of bulk Ising anyons has been pulled out of the vacuum, with one of those anyons dragged to the gapless edge.
	Nontrivial mutual statistics between emergent fermions and the single remaining bulk Ising anyon causes incident zero-energy electrons to recombine perfectly into \emph{holes} at the upper $\nu = 1$ edge.
	Each injected electron transmits a Cooper pair into the superconductor, yielding $G = \frac{2e^2}{h}$.
	Zoom-in: The Ising anyon dragged to the boundary couples to the chiral Majorana edge state with strength $\lambda$ [see Eq.~\eqref{IsingEdge}].
	(c) If random thermal processes toggle the system between configurations (a) and (b), $G$ exhibits telegraph noise as a function of time, switching stochastically between $G = 0 $ and $\frac{2e^2}{h}$.  
	 } 
	\label{interference}
\end{figure}

One can also employ electrical transport to detect individual bulk Ising anyons in a non-Abelian spin liquid---in fact using a slightly simpler circuit compared to Fig.~\ref{condexp}.  
Consider now the setups shown in Fig.~\ref{interference}.  
There, a single $\nu = 1$ quantum Hall system is partially sewn to a spin liquid via fermion condensation mediated by a grounded superconductor. 
A source at bias voltage $V$ on the left generates a current $I$ that flows through the superconductor to ground.
More precisely, electrons emanating from the source $(i)$ propagate along the lower $\nu = 1$ edge, then $(ii)$ fragment into emergent and physical Majorana fermions that acquire a phase difference $\delta\phi$ upon encircling the spin liquid, and finally $(iii)$ recombine into either electrons, holes, or superpositions thereof depending on $\delta \phi$.  
Recombination into a hole in this final step indicates absorption of a Cooper pair into the superconductor, thereby contributing to the current $I$.  
We are interested in the conductance $G = dI/dV$ in the limit $T,V \rightarrow 0$ (but see the next section for an extension to finite $V$).  
Just as we saw in Sec.~\ref{EdgeDetection}, the spin liquid---although electrically inert---plays a decisive role in electrical transport: 
One of two distinct universal quantized conductances emerges depending on the quasiparticle configuration in the spin liquid.

%%%%%%%%%%%%%%%%%%%%%%%%%%%%%%%%%%%%%%%%%%%%%%%%
%%%%%%%%%%%%%%%%%%%%%%%%%%%%%%%%%%%%%%%%%%%%%%%%

Suppose first that the spin liquid's interior is devoid of Ising non-Abelian anyons as in Fig.~\ref{interference}(a).
Here the phase difference acquired in stage $(ii)$ is simply 
\begin{equation}
  \delta \phi = k_p L_p- k_e L_e,
  \label{phase_diff1}
\end{equation} 
with $k_p$ the momentum of the physical Majorana fermion as it travels the distance $L_p$ between the lower and upper $\nu = 1$ edge, and $k_e$ and $L_e$ the analogous quantities for the emergent Majorana fermion.  
The limit $k_p,k_e \rightarrow 0$ yields $\delta \phi \rightarrow 0$, implying that at asymptotically low energies incident electrons recombine into outgoing electrons with unit probability.
To summarize, stages $(i)$ through $(iii)$ proceed according to
\begin{equation}
  \psi_{\rm electron} \rightarrow \gamma_{\rm physical} + i \gamma_{\rm emergent} \rightarrow \psi_{\rm electron}
\end{equation}
as Fig.~\ref{interference}(a) illustrates.
No current flows into the superconductor, and therefore $G = 0$.  
Note the similarity to the physics encountered for the control circuit in Fig.~\ref{condexp}(b).

Next, imagine nucleating a pair of Ising non-Abelian anyons and then dragging one of those anyons to a \emph{gapless} part of the spin liquid edge~\cite{Fendley2009}.
The resulting setup, shown in Fig.~\ref{interference}(b), contains a single Ising anyon in the bulk.  
At asymptotically low incident-electron energies, the emergent Majorana fermion in stage $(ii)$ \emph{acquires an additional minus sign} upon crossing the Ising anyon absorbed at the edge [i.e., at the termination of the wavy line in the inset of Fig.~\ref{interference}(b); see below for further details].
This all-important minus sign reflects the nontrivial mutual statistics between emergent fermions and Ising anyons in the spin liquid (recall Sec.~\ref{NonAbelianSec}).  
It follows that $\delta \phi \rightarrow \pi$ as $k_p,k_e \rightarrow 0$, implying that at low energies incident electrons recombine into outgoing \emph{holes} with unit probability.  
Stages $(i)$ through $(iii)$ can then be summarized as
\begin{eqnarray}
  \psi_{\rm electron} &\rightarrow& \gamma_{\rm physical} + i \gamma_{\rm emergent} 
  \nonumber \\
  &\rightarrow& \gamma_{\rm physical} - i \gamma_{\rm emergent}  \rightarrow \psi_{\rm hole};
\end{eqnarray}
see Fig.~\ref{interference}(b).
The perfect `Andreev conversion' of electrons into holes yields nontrivially quantized conductance $G = \frac{2e^2}{h}$.  

More generally, if the fragmented emergent and physical Majorana fermions encircle $n_\sigma$ Ising anyons in the interior of the spin liquid, the phase difference at low energies is $\delta \phi = \pi n_\sigma$, yielding zero-bias conductance
\begin{equation}
  G = {\rm mod}(n_\sigma,2) \frac{2e^2}{h}.
  \label{Ggeneral}
\end{equation}
The even-odd effect encoded in Eq.~\eqref{Ggeneral} represents a `smoking gun' electrical transport signature of bulk Ising non-Abelian anyons.  
We stress that one can, at least in principle, toggle between the two quantized conductances by \emph{locally} perturbing the spin liquid far from any electrically active circuit components.  
Trivial origins for such exotic behavior would appear to require almost divine intervention.
At present, however, it remains unclear how to feasibly manipulate Ising anyons so as to probe the even-odd effect in a systematic experiment.
A worthwhile preliminary study could instead rely on thermal fluctuations and/or noise to stochastically drag Ising anyons on and off of the gapless spin liquid edge.  
Such processes would change $n_\sigma$ as a function of time, leading to telegraph noise with the conductance switching between $G = 0$ and $\frac{2e^2}{h}$ as sketched in Fig.~\ref{interference}(c).  

The circuits in Figs.~\ref{interference}(a,b) can be viewed as cousins of `$\mathbb{Z}_2$ interferometers' designed to electrically probe physical Majorana fermions in proximitized topological-insulator surfaces~\cite{FuKaneInterferometer,AkhmerovInterferometer}.  
In that context the conductance exhibits an analogous even-odd effect, but in the number of superconducting $h/(2e)$ vortices threaded through the device.  
Similar to Sec.~\ref{EdgeDetection}, fermion condensation has allowed us to adapt such techniques developed for exotic superconductors to probe non-Abelian quasiparticles in a Mott insulator.  

As a technical aside, above we envisioned creating Fig.~\ref{interference}(b) by dragging an Ising anyon to the gapless boundary of the spin liquid.  
But if this Ising anyon resides some `small' distance from the edge [see zoom-in from Fig.~\ref{interference}(b)], how can one quantify whether the fragmented physical and emergent Majorana fermions encircle only the bulk Ising anyon (corresponding to $n_\sigma = 1$), or also the Ising anyon near the boundary (corresponding to $n_\sigma = 2$)?
Following Ref.~\onlinecite{Fendley2009}, this question can be addressed using a minimal model in which the gapless edge hybridizes with the Majorana zero mode $\gamma$ localized to the adjacent Ising anyon.  The Hamiltonian reads
\begin{equation}
 \mathcal{H} = \int_x[-i v \gamma_R \partial_x \gamma_R + i \lambda \gamma_R \gamma \delta(x)],
 \label{IsingEdge}
\end{equation}
where $\gamma_R$ describes the emergent gapless Majorana fermion and $\lambda$ is the coupling strength to $\gamma$, assumed to reside at position $x = 0$.
Note that $\lambda^2/v$ defines an energy scale for the hybridization.

Reference~\onlinecite{Fendley2009} showed that an incident Majorana fermion with energy $E$ acquires a phase shift
\begin{equation}
  e^{i\phi(E)} = \frac{2E + i \lambda^2/v}{2E-i\lambda^2/v}
  \label{gen_phase_shift}
\end{equation}
due to the $\lambda$ coupling.  
The `high' and `low' energy limits of this result can be captured intuitively as follows.  
At incident energies $E \gg \lambda^2/v$, the gapless edge and the adjacent Ising anyon essentially decouple; in this `high-energy' regime the physical and Majorana fermions should be viewed as encircling \emph{both} Ising anyons in Fig.~\ref{interference}(b).  
No additional $\pi$ phase shift arises, though finite, non-universal conductance generically emerges due to Andreev processes (which only freeze out at low energies). 
At $E \ll \lambda^2/v$, one can project onto Hamiltonian eigenstates by sending $\gamma_R(x) \rightarrow {\rm sgn}(x) \tilde \gamma_R(x)$, where $\tilde\gamma_R(x)$ is a slowly varying chiral Majorana fermion.
In terms of $\tilde \gamma_R$, the $\lambda$ term in Eq.~\eqref{IsingEdge} disappears due to the sign change introduced above, so that $\mathcal{H} \rightarrow \int_x(-iv\tilde\gamma_R\partial_x\tilde\gamma_R)$.
In this precise sense, the adjacent Ising anyon has been absorbed by the gapless edge---its only trace is the $\pi$ phase shift inherent in the definition of $\tilde \gamma_R$.  
Hence the physical and emergent Majorana fermions should now be viewed as encircling \emph{only} the bulk Ising anyon in Fig.~\ref{interference}(b).  
Our transport analysis focused on the asymptotic low-energy limit, where the latter scenario prevails.  
Both extremes captured above are consistent with the general formula in Eq.~\eqref{gen_phase_shift}.

%%%%%%%%%%%%%%%%%%%%%%%%%%%%%%%%%%%%%%%%%%%%%%%%
\section{Interferometric detection of neutral fermions, Ising anyons, and non-Abelian statistics}
\label{interferometry2}

The circuit introduced in Sec.~\ref{IsingDetection} reveals bulk Ising anyons but is oblivious to the presence of bulk neutral fermions.  
This dichotomy arises because an emergent fermion living at the boundary acquires a statistical minus sign upon encircling an Ising anyon, in turn influencing the electrical conductance, whereas encircling a neutral fermion yields a trivial statistical phase.  
Here we study an interferometer that enables emergent fermions injected from a lead (with the aid of fermion condensation) to splinter into unpaired Ising anyons---which exhibit nontrivial braiding statistics with both bulk Ising anyons \emph{and} neutral fermions, leading to conductance signatures of both quasiparticle types.

%%%%%%%%%%%%%%%%%%%%%%%%%%%%%%%%%%%%%%%%%%%%%%%%
\begin{figure}
	\includegraphics[width=\linewidth]{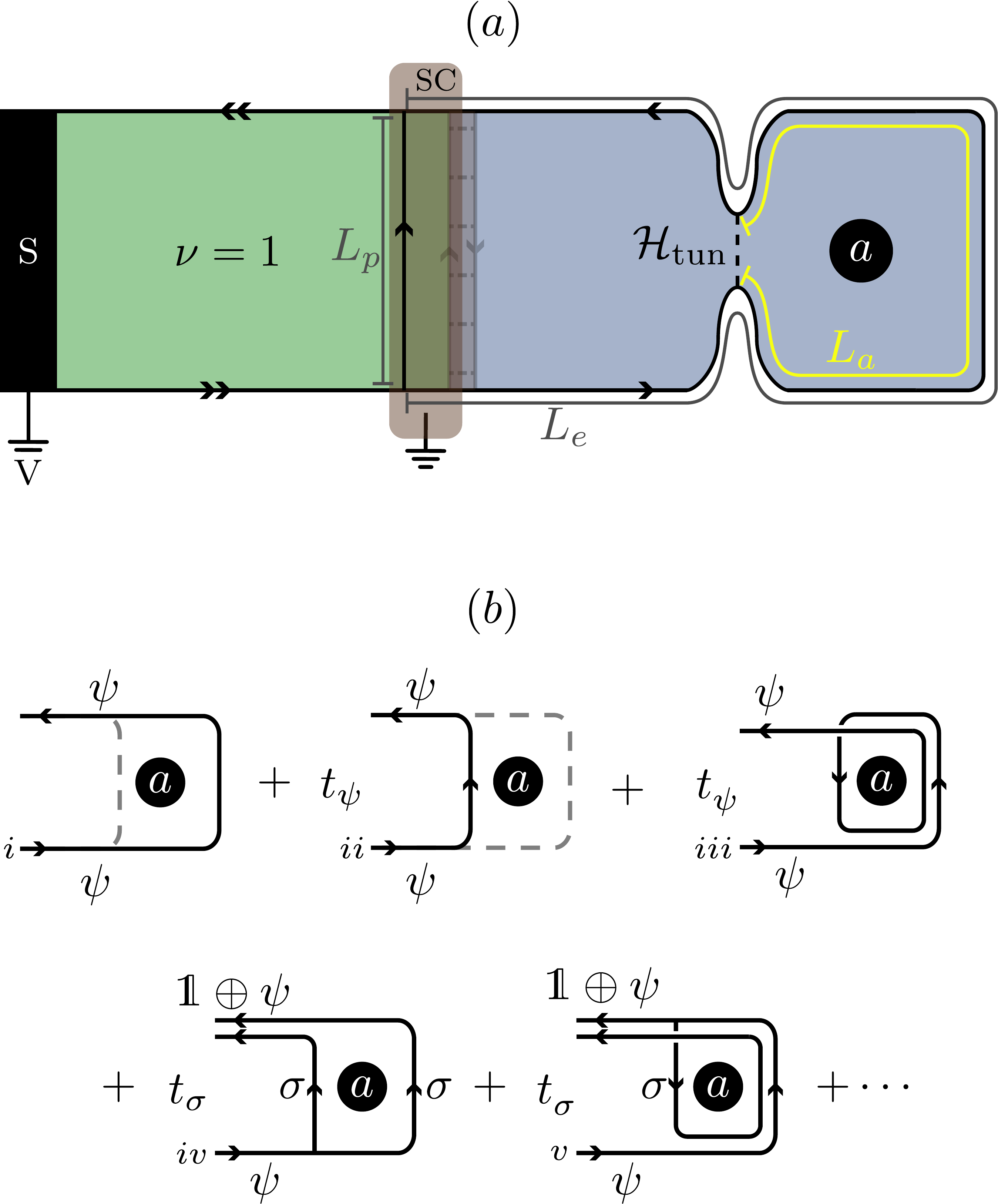}
	\caption{(a) Interferometer that electrically detects both bulk Ising anyons and bulk emergent fermions.  
	The geometry can be viewed as Fig.~\ref{interference} with a constriction in the spin liquid, governed by $\mathcal{H}_{\rm tun}$ in Eq.~\eqref{Htun}.
	At the constriction, incident emergent Majorana fermions can either tunnel across ($t_\psi$ process) or fractionalize into a pair of Ising anyons ($t_\sigma$ process)---one hopping across and the other encircling a bulk quasiparticle of type $a = I, \psi$, or $\sigma$.  
	Nontrivial braiding statistics among the anyons in the spin liquid yields $a$-dependent electrical conductances [Eqs.~\eqref{G1psi} and \eqref{Gsigma}] that enable readout of the quasiparticle type.
	Most notably, non-Abelian statistics between Ising anyons vanquishes first-order conductance corrections from $t_\sigma$ events.
	(b) Illustration of the five paths taken by incident emergent fermions up to first order in $\mathcal{H}_{\rm tun}$.  
	For details see Secs.~\ref{sec:interf_1psi} and \ref{sec:interf_sigma}.  
	The ellipsis denotes higher-order contributions not included here.} 
	\label{psidetector}
\end{figure}
%%%%%%%%%%%%%%%%%%%%%%%%%%%%%%%%%%%%%%%%%%%%%%%%

The device we consider appears in Fig.~\ref{psidetector}(a) and can be viewed as Fig.~\ref{interference}(a) with a constriction. 
At the constriction the upper and lower spin liquid edges couple via a Hamiltonian 
\begin{align}
  \mathcal{H}_{\rm tun} = e^{-i \pi h_\sigma} t_\sigma \sigma(x_2)\sigma(x_1) +  e^{-i \pi h_\psi} t_\psi \gamma(x_2)\gamma(x_1)
  \label{Htun}
\end{align}
with $\sigma(x)$ the Ising CFT field and $t_\sigma, t_\psi \geq 0$ real couplings.  
The $t_\sigma$ and $t_\psi$ terms respectively shuttle Ising anyons and emergent fermions between positions $x_1$ on the lower edge and $x_2$ on the upper edge.  
(For a detailed discussion of the $t_\sigma$ term see Ref.~\onlinecite{Fendley2007}.)
The phase factors in Eq.~\eqref{Htun} involve the topological spin $h_\sigma = 1/16$ of an Ising anyon and $h_\psi = 1/2$ of a fermion, and are required for Hermiticity.  
Note that $e^{-i \pi h_\psi} = -i$ represents the usual imaginary coefficient accompanying a Majorana-fermion bilinear in the Hamiltonian; we employ this form simply to parallel the $t_\sigma$ term.

Ising-anyon tunneling constitutes a relevant perturbation to the fixed point describing decoupled edges, and in the asymptotic low-energy limit effectively chops the spin liquid in two at the constriction~\cite{Fendley2007}.
Fermion tunneling, by contrast, is marginal.  
Throughout we work in a regime---to be quantified below---where both tunneling terms can be regarded as weak.  
Incident fermions at the lower edge then bypass the constriction and take `the long way around' with nearly unit probability, enabling a perturbative treatment of $\mathcal{H}_{\rm tun}$.

Let us then examine Fig.~\ref{psidetector}(a) at temperatures $T \rightarrow 0$ and finite bias voltages $V$ below the gap scale for the fermion condensate. 
We are interested in the conductance $G(V)$ when a quasiparticle of type $a = \mathds{1},\sigma,$ or $\psi$ resides in the right half of the interferometer.  
Figure~\ref{psidetector}(b) sketches the five contributing processes up to first order in $t_\psi$ and $t_\sigma$.  
Path $(i)$ corresponds to the dominant process whereby the incident edge emergent fermion bypasses the constriction and goes around quasiparticle $a$ (as necessarily occurs in Fig.~\ref{interference}).  
In path $(ii)$, the incident fermion short-cuts across the constriction via the fermion-tunneling term $t_\psi$.  
In path $(iii)$, the fermion travels to the upper side of the constriction before similarly tunneling via $t_\psi$.  
Paths $(iv)$ and $(v)$ invoke Ising-anyon tunneling $t_\sigma$.  
In $(iv)$ the incident fermion travels to the lower end of the constriction, then splinters into a pair of Ising anyons that recombine at the upper edge into either a trivial particle or an emergent fermion depending on $a$.
And in $(v)$, the incident fermion travels to the upper side of the constriction before similarly splintering into Ising anyons.
In what follows we examine these processes within a phenomenological treatment that we eventually connect to more formal analyses given in Appendices~\ref{fermion_tunneling} and \ref{app:Ising_tunneling} (see also the analyses of related interferometers, e.g., in Refs.~\onlinecite{Bishara2008,Bonderson2010,Nilsson2010}). 
We start with the case $a = \mathds{1}$ or $\psi$ and then consider $a = \sigma$.

%%%%%%%%%%%%%%%%%%%%%%%%%%%%%%%%%%%%%%%%%%%%%%%%
\subsection{Interferometer with \texorpdfstring{$a = \mathds{1}$ or $\psi$}{a = 1 or a = psi}}
\label{sec:interf_1psi}

When $a = \mathds{1}$ or $\psi$, the splintered Ising anyons from paths $(iv)$ and $(v)$ of Fig.~\ref{psidetector}(b) necessarily recombine into an outgoing emergent fermion at the upper edge (braiding $\sigma$ around either $\mathds{1}$ or $\psi$ preserves the Ising anyons' fusion channel).  
Thus, in all five paths, emergent Majorana fermions incident from below necessarily exit the interferometer as fermions.
The conductance simply follows from the phase accumulated en route.  
We now separately examine each path from Fig.~\ref{psidetector}(b).

{\bf Path $(i)$.}~Consider an emergent Majorana fermion with momentum $k_e$ that travels a distance $L_e$ the long way around the interferometer.  The associated quantum amplitude reads 
\begin{equation}
  A_i = e^{i k_e L_e},
  \label{phi_i}
\end{equation}
which is simply the phase acquired by the fermion.  

For the remaining cases, it will be useful to express the amplitude for path $(p)$ as $A_p = w_p e^{i \phi_p}$; here $w_p$ encodes local physics at the constriction and $e^{i \phi_p}$ captures the phase accumulated due to propagation along the boundary.  Note that $w_{ii}$ and $w_{iii}$ are proportional to $t_\psi$ while $w_{iv}$ and $w_{v}$ are proportional to $t_\sigma$.  

{\bf Path $(ii)$.}~If the fermion propagates to the lower end of the constriction and then tunnels across, the amplitude is
\begin{equation}
  A_{ii} = w_{ii} e^{i k_e(L_e-L_a)},
  \label{phi_ii}
\end{equation}
where $L_a$ is the perimeter of the region enclosing $a$ as sketched in Fig.~\ref{psidetector}(a).

{\bf Path $(iii)$.}~Let $L_c$ be the distance between the constriction and either end of the fermion condensate [see Fig.~\ref{psidetector}(a)].  The amplitude for path $(iii)$ can then be written
\begin{align}
  A_{iii} = w_{iii}e^{i k_e(L_c+L_a)} e^{i k_e(L_a + L_c)}.
   \label{phi_iii}
\end{align}
The first exponential represents the phase acquired upon traveling to the upper part of the constriction, and the second is the phase acquired by the fermion after tunneling across the constriction and returning to the fermion condensate. 
Noting that $L_e = 2L_c + L_a$, we can simplify Eq.~\eqref{phi_iii} to
\begin{equation}
  A_{iii} = w_{iii} e^{i k_e(L_e+L_a)}.
  \label{phi_iii2}
\end{equation}

{\bf Path $(iv)$.}~When the incident Majorana fermion splinters in path $(iv)$, its momentum $k_e$ can partition among the resulting pair of Ising anyons in various ways that are compatible with energy conservation \footnote{For a discussion of energy partitioning in the Luttinger-liquid context, see Ref.~\onlinecite{LLpartitioning}.}.  
Suppose for now that the Ising anyon tunneling across the constriction carries momentum $k_1$, while the Ising anyon that takes the long way around carries $k_2 = k_e-k_1$.  
The amplitude for this event is
\begin{align}
  A_{iv}(k_1)
  &= w_{iv}(k_e,k_1)e^{i k_e L_c} e^{i k_1 L_c} e^{i k_2(L_a+L_c)} (-1)^{n_\psi}
  \nonumber \\
  &= w_{iv}(k_e,k_1) e^{i k_e L_e} e^{-i k_1 L_a}  (-1)^{n_\psi}.
  \label{path2b}
\end{align}
Here the factor $w_{iv}$ generically depends on both $k_e$ and $k_1$ as a consequence of the relevance of the Ising-anyon tunneling term.  
The first exponential on the top line  is the phase acquired by the fermion as it travels to the constriction; 
the second is the phase acquired by the Ising anyon that tunnels across the constriction and travels to the upper end of the fermion condensate;
and the third is the phase acquired by the Ising anyon that travels the long way around.
In the last factor, $n_\psi$ is the number of bulk neutral fermions (mod 2) enclosed in the right half of the interferometer, i.e., $n_\psi = 0$ if $a = \mathds{1}$ while $n_\psi = 1$ if $a = \psi$.  
The all-important additional $\pi$ phase that arises when $n_\psi = 1$ reflects the nontrivial mutual statistics between Ising anyons and neutral fermions, and ultimately allows the interferometer to detect the latter bulk quasiparticle type.  

Events corresponding to distinct, physically permissible $k_1$ values must be integrated over since the wavefunction will consist of a weighted sum over all such energy partitionings.  
In particular, the pair of splintered Ising anyons must both carry positive momentum and energy, yielding the inequality $0\leq k_1 \leq k_e$ for path $(iv)$. 

{\bf Path $(v)$.}~Suppose now that an incident emergent Majorana fermion in path $(v)$ similarly splinters into one Ising anyon carrying momentum $k_1$ across the constriction and another that carries momentum $k_2 = k_e-k_1$ past the constriction.
The corresponding amplitude is
\begin{align}
  A_{v}(k_1)
  &= w_{v}(k_e,k_1)e^{i k_e (L_c + L_a)}e^{i k_1(L_a + L_c)} e^{i k_2 L_c} (-1)^{n_\psi} 
  \nonumber \\
  &= w_{v}(k_e,k_1) e^{i k_e L_e} e^{i k_1L_a} (-1)^{n_\psi}.
\end{align}
The first three exponentials in the top line respectively denote the phase acquired by the fermion prior to splintering, the Ising anyon that tunnels across the constriction, and the Ising anyon that bypasses the constriction.
The $(-1)^{n_\psi}$ factor once again reflects the braiding statistics between Ising anyons and neutral fermions.

Physically permissible $k_1$ values must be integrated over, as in path $(iv)$, but now the allowed range differs.
Indeed here the Ising anyon that tunnels across the constriction can carry arbitrary positive momentum since the pair of Ising anyons that combines on the upper end of the interferometer always carries total momentum $k_e$ regardless of the magnitude of $k_1$.
For path $(v)$ we thus have the inequality $0\leq k_1 < \infty$ (neglecting an ultraviolet momentum cutoff for simplicity).  

Upon summing over the five paths, the amplitude describing the arrival of the emergent Majorana fermion at the upper end of the fermion condensate is
\begin{align} \begin{split}
	A &= A_i + A_{ii} + A_{iii}
\\	&\quad + \int_0^{k_e}\!\!\! dk_1\, A_{iv}(k_1) + \int_0^\infty\!\!\! dk_1\, A_v(k_1) .
\end{split} \end{align}
Inserting the above expressions for $A_i$ through $A_v$ yields
\begin{align} \begin{split}
	A &= e^{i k_e L_e} \Bigg\{ 1 + \Big[ w_{ii}e^{-i k_e L_a} + w_{iii} e^{i k_e L_a} \Big]
\\	&\quad+  (-1)^{n_\psi}\bigg[\int_0^{k_e} dk_1 w_{iv}(k_e,k_1) e^{-i k_1 L_a} 
\\	&\qquad\qquad+ \int_0^{\infty} dk_1 w_{v}(k_e,k_1) e^{i k_1 L_a} \bigg]\Bigg\}.
\label{Acomplete}
\end{split} \end{align}
We can further constrain the form of the amplitude using dimensional analysis.  
Since the scaling dimension of the Majorana fermion $\gamma(x)$ is $1/2$, $t_\psi$ has units of ${\rm energy} \times {\rm length}$; similarly, the Ising field $\sigma(x)$ scaling dimension is $1/16$, and so $t_\sigma$ has units of ${\rm energy} \times ({\rm length})^{1/8}$.  
Thus we can write
\begin{align}
  w_{ii,iii} &= \left[\frac{t_\psi}{v_e}\right]\alpha_{ii,iii}, 
  \label{wii_iii}
  \\
  w_{iv,v}(k_e,k_1) &= \left[\frac{t_\sigma}{v_ek_e^{7/8}}\right]\frac{f_{iv,v}(k_1/k_e)}{k_e}
  \label{wiv_v}
\end{align}
with $v_e$ the emergent-fermion edge-state velocity, $\alpha_{ii}$ and $\alpha_{iii}$ numerical factors, and $f_{iv}$ and $f_v$ dimensionless scaling functions.
Notice that the bracketed factors above are dimensionless.

\begin{figure}
	\includegraphics[width=0.98\linewidth]{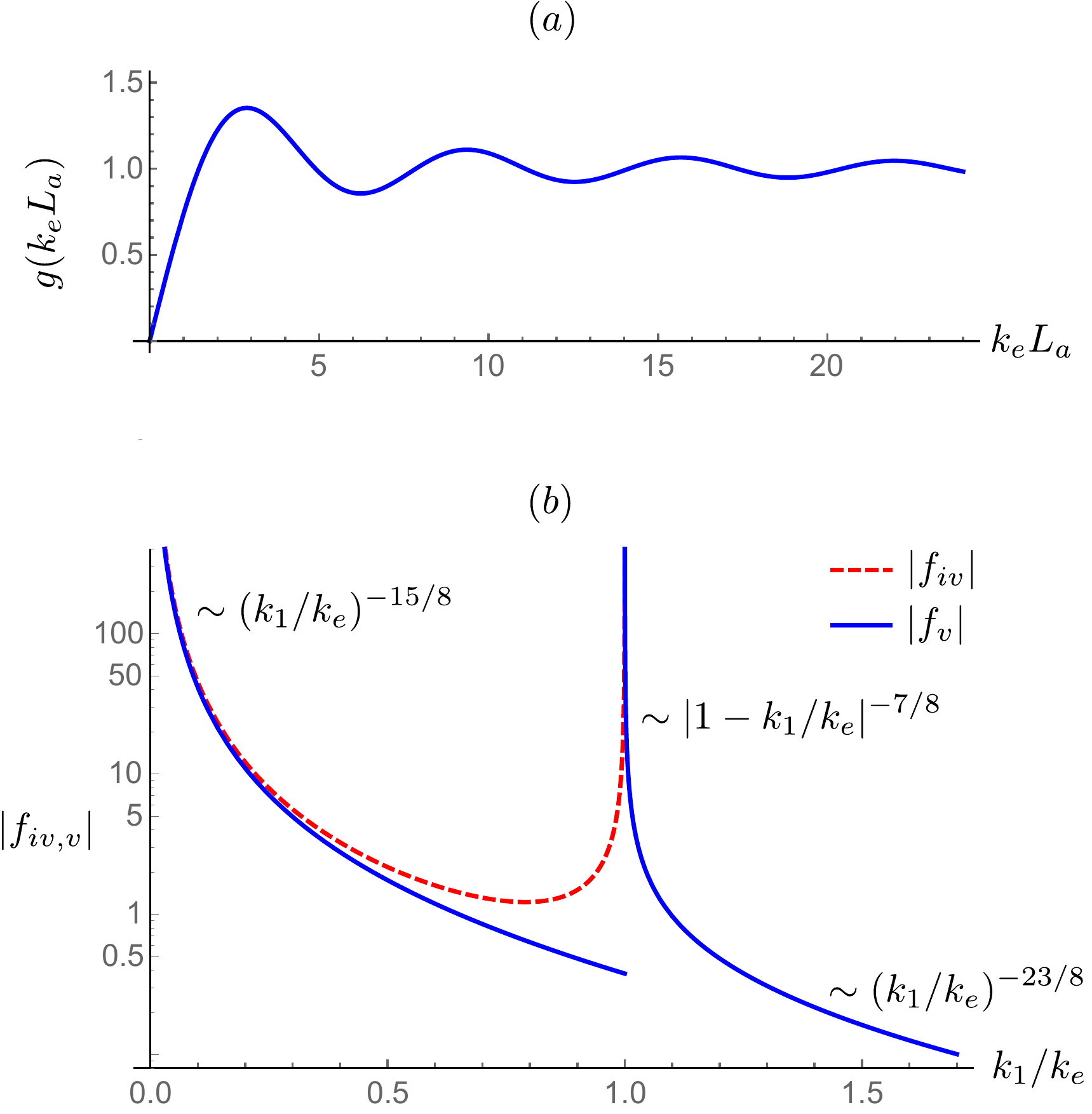}
	\caption{(a) Plot of $g(k_e L_a)$ [Eq.~\eqref{gHGF}] versus $k_e L_a$, where $k_e$ denotes the incoming emergent-fermion momentum and $L_a$ is the length defined in Fig.~\ref{psidetector}(a).   	
	This function determines the $t_\sigma$ correction to the fermion transmission amplitude [Eq.~\eqref{Ag}], and thus also the conductance [Eq.~\eqref{G1psi}], for the interferometer in Fig.~\ref{psidetector}(a) when $a = I$ or $\psi$.	
	(b) Magnitude of the scaling functions $f_{iv}$ (red dashed line) and $f_v$ (blue line) obtained by inverting Eq.~\eqref{gu} as carried out in Appendix~\ref{app:Ising_tunneling_inversion}.  	
	These scaling functions govern energy partitioning associated with Ising-anyon tunneling in paths $(iv)$ and $(v)$ summarized in Fig.~\ref{psidetector}(b) and discussed in Sec.~\ref{sec:interf_1psi}; recall Eq.~\eqref{wiv_v}. 
	In the horizontal axis $k_1$ is the momentum carried by an Ising anyon that tunnels across the constriction.
	The limiting scaling behavior displayed above implies that Ising anyons tunnel primarily carrying momentum $\ll k_e$ and secondarily carrying momentum very near $k_e$.
	}
	\label{gf_fig}
\end{figure}

Determining the remaining unspecified quantities in $A$ requires explicit calculations.  
For the fermion-tunneling paths, Appendix~\ref{fermion_tunneling} presents a standard Heisenberg-picture analysis that yields
\begin{equation}
  \alpha_{ii} = -\alpha_{iii} = -\frac{1}{2}.
  \label{alpha_def}
\end{equation}
Equations~\eqref{wii_iii} through \eqref{alpha_def} then allow us to express the amplitude as
\begin{align}
  A = e^{i k_e L_e} \bigg{[} 1 &+ i \frac{t_\psi}{v_e} \sin(k_e L_a) 
  \nonumber \\
  &+ i (-1)^{n_\psi} \frac{2t_\sigma L_a^{7/8}}{v_e} g(k_e L_a)\bigg{]},
  \label{Ag}
\end{align}
where we defined
\begin{align}
  g(u) &= \frac{1}{2iu^{7/8}}\bigg{[} \int_0^1\! dy\, e^{-i u y} f_{iv}(y)
+ \int_0^\infty\!\!\! dy\, e^{i u y} f_{v}(y)\bigg{]}.
  \label{gu}
\end{align}
Treating the Ising-anyon tunneling paths demands a more sophisticated conformal field theory analysis carried out in Appendix~\ref{app:Ising_tunneling}.  
There we show that
\begin{align}
  g(u) = \frac{\pi u}{4} \HGpFqb{1}{2}{\tfrac{1}{2}}{1,2}{-\tfrac{1}{4}u^2}
  \label{gHGF}
\end{align}  
with $\mathop{_1\mkern-1.5mu F_2}$ a generalized hypergeometric function.
Figure~\ref{gf_fig}(a) plots $g(k_e L_a)$ versus $k_e L_a$.  
The perturbative regime stipulated earlier requires%
~\footnote{One might have naively guessed that Ising-anyon tunneling instead admits a perturbative treatment provided the incoming fermion momentum $k_e$ is sufficiently large that $t_\sigma/(v_e k_e^{7/8}) \ll 1$.  
		When this inequality holds, it would appear that one is probing the system at high energies for which $t_\sigma$ has not yet flowed to strong coupling.  
		The fallacy in this argument stems from energy partitioning.  
		Regardless of the magnitude of $k_e$, at the constriction the incident fermion splinters into Ising anyons that share the incident energy in all permissible ways.  
		In particular, the allowed partitionings include cases where an Ising anyon tunneling across the constriction carries arbitrarily small momentum, and for such events $t_\sigma$ can not be regarded as weak.  
		Consequently, finite length $L_a$ is required to define a perturbative regime, corresponding to the quoted inequality $t_\sigma L_a^{7/8}/v_e \ll 1$.} 
\begin{equation}
  \frac{t_\psi}{v_e} \ll 1~~{\text{and}}~~ \frac{t_\sigma L_a^{7/8}}{v_e} \ll 1
\end{equation}
so that fermion and Ising-anyon tunneling across the constriction contribute only small corrections to the amplitude in Eq.~\eqref{Ag}.

%%%%%%%%%%%%%%%%%%%%%%%%%%%%%%%%%%%%%%%%%%%%%%%%
%%%%%%%%%%%%%%%%%%%%%%%%%%%%%%%%%%%%%%%%%%%%%%%%

To make contact with our phenomenological picture, we can invert Eq.~\eqref{gu} to extract the $f_{iv}(k_1/k_e)$ and $f_v(k_1/k_e)$ scaling functions that quantify the energy partitioning.  
Appendix~\ref{app:Ising_tunneling_inversion} pursues this (nontrivial) exercise; for a summary see Fig.~\ref{gf_fig}(b).  
Both scaling functions exhibit a leading divergence at $k_1 \rightarrow 0$ and a subleading divergence at $k_1 \rightarrow k_e$.
It follows that Ising anyons tunnel across the constriction primarily carrying `small' momentum and secondarily carrying momentum near $k_e$. 
More explicitly, $f_{iv,v}(k_1/k_e) \sim (k_1/k_e)^{-15/8}$ at small $k_1$; 
the exponent implies that the $k_e$ dependence of the weights defined in Eq.~\eqref{wiv_v} drops out at $k_1 \rightarrow 0$, i.e., in this regime the Ising-anyon tunneling probability becomes \emph{independent of the incident fermion momentum}.  
Furthermore, $f_{iv}(k_1/k_e) \sim (1-k_1/k_e)^{-7/8}$ as $k_1$ approaches $k_e$ from below and $f_{v}(k_1/k_e) \sim (k_1/k_e-1)^{-7/8}$ as $k_1$ approaches $k_e$ from above.  
Notice that $f_v(k_1/k_e)$ does not diverge as $k_1$ approaches $k_e$ from below---hence for path $(v)$ Ising anyons tunnel far more efficiently with momentum slightly \emph{larger} than $k_e$ compared to momentum just below $k_e$.

It is instructive to examine some limits of the function $g(k_e L_a)$.  
At small arguments one finds
\begin{equation}
  g(k_e L_a\ll 1) \approx \frac{\pi}{4} k_e L_a,
  \label{gsmall}
\end{equation}
i.e., the amplitude correction from Ising-anyon tunneling vanishes linearly with the fermion momentum.  
In this limit of our perturbative analysis one can use an operator product expansion (OPE) to fuse the Ising CFT fields in Eq.~\eqref{Htun}, yielding
\begin{equation}
  e^{-i \pi h_\sigma} t_\sigma \sigma(x_2)\sigma(x_1) \rightarrow {\rm const} \times t_\sigma L_a^{15/8} \gamma \partial_x \gamma.
  \label{OPE}
\end{equation}
The term on the right side (which is a descendent of the identity) is irrelevant, which explains the linear vanishing of the amplitude correction as $k_e L_a \rightarrow 0$.  

At $k_e L_a \gg 1$ one instead finds
\begin{equation}
  g(k_e L_a\gg 1) \approx 1-\frac{\cos(k_e L_a)}{k_e L_a}.
  \label{glarge}
\end{equation}
Contrary to the purely oscillatory amplitude correction from fermion tunneling, the amplitude correction from Ising-anyon tunneling thus tends to an $L_a$-dependent constant as $k_e L_a \rightarrow \infty$, with subdominant oscillations that decay with a $1/(k_e L_a)$ prefactor.  
The former feature reflects the propensity of Ising anyons to tunnel across the constriction with vanishingly small momentum.  
Oscillations in the latter piece come from Ising anyons that tunnel with momentum near $k_e$, while the decay arises because of the finite spread in the allowed momentum carried.

The tunneling Hamiltonian invoked in Eq.~\eqref{Htun} could of course be amended in various ways, e.g., by allowing fermions and Ising anyons to tunnel over a finite range of positions between the upper and lower sides of the constriction (as opposed to discrete points $x_{1,2}$) or by adding derivatives that effectively make the tunnel couplings momentum dependent.
It is thus important to address which features of the amplitude correction in Eq.~\eqref{Ag} are generic.  
The linear vanishing of both the $t_\psi$ and $t_\sigma$ corrections with $k_e$ [cf.~Eq.~\eqref{gsmall}] is certainly universal (though the prefactor of course is not). 
At $k_e \rightarrow 0$ one can exploit an OPE similar to Eq.~\eqref{OPE} to reduce arbitrary fermion tunneling and Ising-anyon tunneling terms to irrelevant descendants of the identity.
The leading irrelevant term is $\gamma \partial_x \gamma$, and the derivative ensures amplitude corrections $\propto k_e$ as $k_e \rightarrow 0$.

We further expect that the exponents governing the power-law divergences in the $f_{iv,v}$ scaling functions shown in Fig.~\ref{gf_fig}(b) are universal.
At $k_e L_a \gg 1$, these divergences determine the leading and sub-leading $k_e L_a$ dependence of the $t_\sigma$ correction specified by Eq.~\eqref{glarge}---which would then also be universal.  
Note, however, that the prefactors of the two terms in Eq.~\eqref{glarge} will depend on details of the tunneling Hamiltonian.
The generic saturation of the $t_\sigma$ correction to an $L_a$-dependent constant at $k_e L_a \rightarrow \infty$ admits an intuitive explanation: 
Energy partitioning invariably suppresses oscillations as $k_e L_a$ increases, 
whereas processes for which Ising anyons tunnel across the constriction carrying vanishingly small momentum naturally leave a $k_e$-independent correction.
[Again, the Ising-anyon tunneling weights in Eq.~\eqref{wiv_v} do not depend on the incident fermion momentum in the $k_1 \rightarrow 0$ limit.]

We are now ready to extract the conductance.  
The net phase acquired by an incident emergent Majorana fermion is given by
\begin{equation}
   e^{i \phi_{\rm emergent}}  = \frac{A}{|A|}.
   \label{phi_emergent_def}
 \end{equation}
Moreover, the phase acquired by a physical Majorana fermion with momentum $k_p$ that travels the length $L_{p}$ of the fermion condensate is 
\begin{equation}
  e^{i \phi_{\rm physical}} = e^{i k_p L_{p}},
\end{equation}
yielding a phase difference
\begin{align}
  &\delta \phi_{a = \mathds{1},\psi}(k_e,k_p) = k_p L_p - k_e L_e 
  \nonumber \\
  &~~~~- \frac{t_\psi}{v_e} \sin(k_e L_a) - (-1)^{n_\psi}\frac{2t_\sigma L_a^{7/8}}{v_e} g(k_eL_a)
\end{align}
to first order in $t_\psi, t_\sigma$; cf.~Eq.~\eqref{phase_diff1}.  
Suppose next that an incident electron injected from the lead into the lower $\nu= 1$ edge of Fig.~\ref{psidetector}(a) carries energy $E$.
The Majorana-fermion momenta are then $k_e = E/v_e$ and $k_p = E/v_p$, where $v_p$ is the physical Majorana fermion's edge velocity. 
As reviewed in Appendix \ref{ConductanceApp} the conductance at bias voltage $V$  is
\begin{align}
  G_{a = \mathds{1},\psi}(V) &= \frac{e^2}{h}\bigg{\{}1- \cos \left[\delta \phi_{a = \mathds{1},\psi}\left(\frac{eV}{v_e},\frac{eV}{v_p}\right)\right]\bigg{\}}.
\end{align}
Finally, expanding in $t_\psi$ and $t_\sigma$ yields
\begin{align}
  &G_{a = \mathds{1},\psi}(V) \approx \frac{e^2}{h}\bigg{\{}1-\cos\left[eV\left(\frac{L_p}{v_p}-\frac{L_e}{v_e}\right)\right]
  \nonumber \\
  &-\frac{t_\psi}{2v_e} \cos\left[eV\left(\frac{L_p}{v_p}-\frac{L_e+L_a}{v_e}\right)\right]
  \nonumber \\
  &+\frac{t_\psi}{2 v_e} \cos\left[eV\left(\frac{L_p}{v_p}-\frac{L_e-L_a}{v_e}\right)\right]
  \nonumber \\
  &-(-1)^{n_\psi}\frac{2t_\sigma L_a^{7/8}}{v_e} g\left(eV\frac{L_a}{v_e}\right) \sin\left[eV\left(\frac{L_p}{v_p}-\frac{L_e}{v_e}\right)\right]\bigg{\}}.
  \label{G1psi}
\end{align}
In the first line the oscillatory voltage dependence reflects the periodic revival and destruction of Andreev processes as the phase difference accumulated by the physical and emergent Majorana fermions varies in path $(i)$.
The next two lines encode corrections from fermion tunneling across the constriction, hence the dependence on the shifted path lengths $L_e \pm L_a$.  
And by far most importantly, the correction from Ising-anyon tunneling in the last line reveals the presence of individual emergent fermions by virtue of the $n_\psi$ dependence.

%%%%%%%%%%%%%%%%%%%%%%%%%%%%%%%%%%%%%%%%%%%%%%%%
\subsection{Interferometer with \texorpdfstring{$a = \sigma$}{a = sigma}}
\label{sec:interf_sigma}

Suppose now that $a = \sigma$ (which turns out to be far simpler to analyze compared to the $a = \mathds{1}$ or $\psi$ cases).
We assume that the bulk Ising anyon's `partner' has been dragged to an adjacent gapless part of the edge in the right half of the interferometer.
Path $(i)$ acquires an additional $\pi$ phase relative to Eq.~\eqref{phi_i} due to the nontrivial mutual statistics between fermions and Ising anyons.  
By contrast, the phases from paths $(ii)$ and $(iii)$---wherein the edge fermion encircles the bulk Ising anyon an even number of times---conform exactly to Eqs.~\eqref{phi_ii} and \eqref{phi_iii2}, respectively.
In paths $(iv)$ and $(v)$ the incident emergent fermion splinters into two Ising anyons at the constriction, one of which now encircles a bulk Ising anyon.  
This braiding process changes the fusion channel for the splintered edge Ising anyons from $\psi$ to $\mathds{1}$.  
More physically, in paths $(iv)$ and $(v)$ the incident emergent fermion exits the interferometer as a trivial boson, and hence these paths no longer contribute to the electrical conductance.  

The amplitude from Eq.~\eqref{Ag} accordingly becomes
\begin{align}
  A = e^{i k_e L_e} \bigg{[} -1 + i \frac{t_\psi}{v_e} \sin(k_e L_a)\bigg{]}.
  \label{Ag2}
\end{align}
Following precisely the same steps outlined in the preceding section, one obtains a conductance
\begin{align}
  &G_{a = \sigma}(V) \approx \frac{e^2}{h}\bigg{\{}1+\cos\left[eV\left(\frac{L_p}{v_p}-\frac{L_e}{v_e}\right)\right]
  \nonumber \\
  &-\frac{t_\psi}{2v_e} \cos\left[eV\left(\frac{L_p}{v_p}-\frac{L_e+L_a}{v_e}\right)\right]
  \nonumber \\
  &+\frac{t_\psi}{2v_e} \cos\left[eV\left(\frac{L_p}{v_p}-\frac{L_e-L_a}{v_e}\right)\right]\bigg{\}}.
  \label{Gsigma}
\end{align}
In the limit $t_\sigma = t_\psi = 0$ the interferometer effectively reduces to the setup in Figs.~\ref{interference}(a) and (b) that allowed electrical detection of Ising anyons.  
Indeed Eqs.~\eqref{G1psi} and \eqref{Gsigma} respectively yield $G = 0$ and $G = \frac{2e^2}{h}$ at $V \rightarrow 0$, in agreement with the analysis from Sec.~\ref{IsingDetection}.
Allowing quasiparticle tunneling across the constriction additionally reveals the non-Abelian statistics of Ising anyons as manifested by the disappearance of the oscillatory $t_\sigma$ correction in Eq.~\eqref{G1psi} when a bulk Ising anyon sits in the interferometer loop.  
Qualitatively similar physics appears in quantum Hall interferometers introduced in Refs.~\onlinecite{Nayak2005,Stern2006,Bonderson2006}.

%%%%%%%%%%%%%%%%%%%%%%%%%%%%%%%%%%%%%%%%%%%%%%%%
\section{Discussion}
\label{discussion}

A growing body of evidence supports the realization of a non-Abelian spin liquid phase in the Kitaev material \alpRuCl3~\cite{Baek2017,Sears2017,Wolter2017,Leahy2017,Banerjee2018,Hentrich2018,Jansa2018,Kasahara2018,Balz2019,Tokoi}.
This remarkable development strongly motivates proposals for probing \emph{individual} fractionalized excitations in honeycomb Kitaev materials, as required for eventual topological quantum computing applications.
The Mott-insulating nature of the host platform renders the problem both interesting and nontrivial.  
We introduced a strategy based, counterintuitively, on universal low-voltage \emph{electrical} transport in novel circuits designed to perfectly convert electrons into emergent Majorana fermions born in the spin liquid, and vice versa.
Similar techniques can be adapted to any bosonic topologically ordered phase hosting gapless emergent-fermion edge states.

\begin{figure}
	\includegraphics[width=\linewidth]{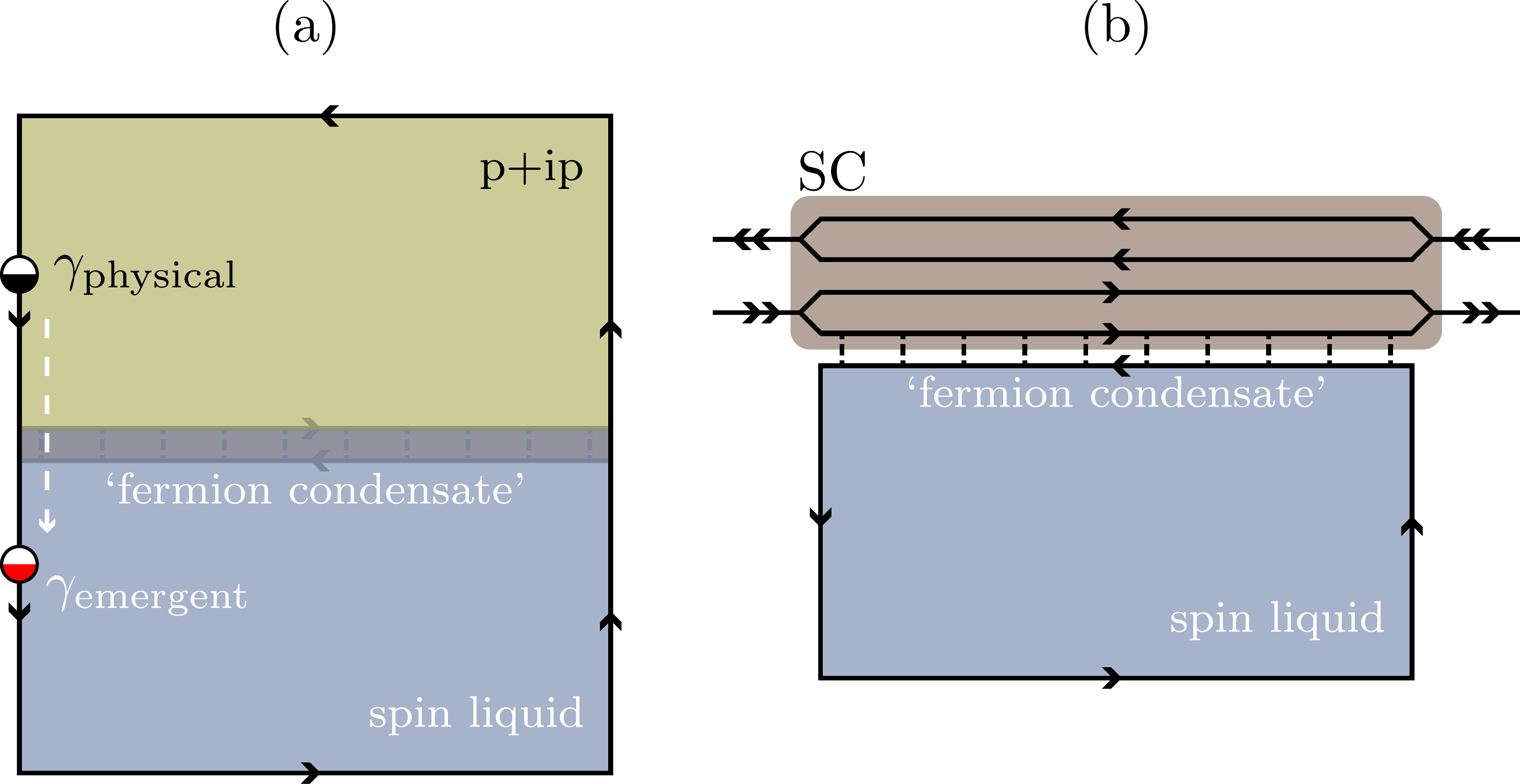}
	\caption{Variations on Fig.~\ref{SLQH_fig} in which the $\nu = 1$ quantum Hall system is replaced by (a) a 2D spinless $p+ip$ superconductor and (b) a proximitized 1D wire.  
	Both alternatives also enable physical fermion $\leftrightarrow$ emergent fermion conversion via fermion condensation.
	In (a), the physical chiral Majorana edge state of a spinless 2D $p+ip$ superconductor gaps out with the spin liquid's emergent chiral Majorana edge mode---yielding a fully gapped interface between the two subsystems.  
	Physical Majorana fermions from the former thus invariably enter the spin liquid as emergent Majorana fermions.
	In (b), right- and left-moving electron channels (double arrows) separate into Majorana modes beneath the proximitizing superconductor.
	Fermion condensation arises when one of those Majorana modes gaps out with the spin liquid's edge state.
	(Among the three remaining modes beneath the superconductor, only one must be gapless.)
	} 
	\label{DiscussionFig}
\end{figure}

Perfect physical fermion $\leftrightarrow$ emergent fermion conversion transpires when the non-Abelian spin liquid's chiral Majorana edge state gaps out with a counterpropagating 1D electronic channel---forming a `fermion condensate'.  
Throughout this paper we considered proximitized $\nu = 1$ integer quantum Hall systems as the source of 1D electrons participating in fermion condensation.
We adopted this choice due to the wide availability of $\nu = 1$ states and for convenience in defining electrical transport quantities.  
As a possible variation, one could replace the quantum Hall system with a spinless 2D $p+ip$ superconductor, which supports `half' of a $\nu = 1$ edge state and thus admits a fully gapped interface with the spin liquid.  
Alternatively, one could employ proximitized 1D wires instead of 2D topological phases at the expense of introducing additional electronic channels.
For illustrations see Fig.~\ref{DiscussionFig}.  
Exploring transport characteristics of circuits employing such variations would certainly be worthwhile.
Moreover, developing a detailed microscopic understanding of the interaction mediating fermion condensation [recall Eq.~\eqref{deltaH}] remains an important open problem even for our quantum Hall-based setups.

%%%%%%%%%%%%%%%%%%%%%%%%%%%%%%%%%%%%%%%%%%%%%%%%
%%%%%%%%%%%%%%%%%%%%%%%%%%%%%%%%%%%%%%%%%%%%%%%%

In Sec.~\ref{EdgeDetection} we saw that quantum Hall--Kitaev material--quantum Hall circuits (Fig.~\ref{condexp}) reveal the spin liquid's emergent chiral Majorana edge state via quantized zero-bias \emph{charge} conductance.
This purely electrical fingerprint complements the well-known quantized thermal Hall signature of a Majorana edge mode~\cite{Kitaev2006,Kasahara2018,Tokoi}, and relies critically on the fermion condensation that underlies our anyon-detection schemes.
Section~\ref{IsingDetection} explored a somewhat simpler quantum Hall--Kitaev material device (Fig.~\ref{interference}) designed to electrically detect bulk Ising non-Abelian anyons.
In particular, the circuit admits zero-bias conductance quantized at either $G = 0$ or $G = 2e^2/h$ depending on whether an even or odd number of Ising anyons resides in the bulk.  
This striking even-odd effect reflects the nontrivial mutual braiding statistics between Ising anyons and emergent fermions.
Taking the same circuit and adding a constriction within the spin liquid leads to the interferometer explored in Sec.~\ref{interferometry2} (Fig.~\ref{psidetector}).
At the constriction an emergent fermion can splinter into a pair of Ising anyons---one taking a shortcut across the pinch and the other continuing along the spin liquid edge.  
The electrical conductance is then sensitive to the presence of bulk Ising anyons \emph{and} bulk emergent fermions since Ising anyons exhibit nontrivial braiding statistics upon encircling either quasiparticle type.  
Notably, \emph{non-Abelian} braiding statistics is manifested as a vanishing of certain conductance corrections when an odd number of bulk Ising anyons sits in the interferometer [cf.~Eqs.~\eqref{G1psi} and \eqref{Gsigma}]---similar to the physics encountered in fractional quantum Hall architectures~\cite{Nayak2005,Stern2006,Bonderson2006}.  

Analyzing such circuits with realistic imperfections poses another worthwhile direction for future investigation.  
For instance, the superconductors in practice will likely contain low-energy degrees of freedom due to disorder and/or vortices.  
Electrons from the $\nu = 1$ edge could directly hop onto these low-lying modes without encountering the fermion condensate or Andreev reflecting, thereby providing a parallel conduction channel that modifies the conductances predicted in this paper.  
Thermally excited quasiparticles in the spin liquid can also produce unwanted errors, e.g., due to processes wherein an edge emergent fermion splinters into Ising anyons that enclose a thermally excited bulk quasiparticle residing near the boundary.
For our purposes, these corrections must be sufficiently small that $(i)$ a sharp contrast remains between the nontrivial and control circuits in Fig.~\ref{condexp} and $(ii)$ the quasiparticle-dependent conductances predicted for Figs.~\ref{interference} and \ref{psidetector} remain discernible.  

For initial anyon-detection experiments, one could rely on nature to thermally cycle between various bulk quasiparticle configurations---leading to telegraph noise wherein the conductance randomly cycles among the predicted values as a function of time [see, e.g., Fig.~\ref{interference}(c)].  
Deterministic, real-time anyon control is nevertheless clearly desirable.  
To this end it is essential to develop practical means of trapping Ising anyons and emergent fermions in the spin liquid.
We anticipate that this subtle energetics issue can be profitably addressed by studying the Kitaev honeycomb model supplemented by generic symmetry-allowed perturbations.  
Finally, developing complementary methods of detecting individual bulk anyons that mitigate the experimental requirements of our scheme poses a key challenge.  
We hope that this work helps stimulate such efforts with the ultimate aim of crafting a realistic roadmap towards topological quantum computation with Kitaev materials.

%%%%%%%%%%%%%%%%%%%%%%%%%%%%%%%%%%%%%%%%%%%%%%%%
\begin{acknowledgments}
It is a pleasure to thank Chao-Ming Jian, Stevan Nadj-Perge, Achim Rosch, Ady Stern, and especially Paul Fendley for stimulating conversations.
This work was supported by
	a postdoctoral fellowship from the Gordon and Betty Moore Foundation, under the EPiQS initiative, Grant GBMF4304;
	the Army Research Office under Grant Award W911NF-17-1-0323; 
	the National Science Foundation through grants DMR-1723367 (JA) and DMR-1848336 (RM); 
	the Caltech Institute for Quantum Information and Matter, an NSF Physics Frontiers Center with support of the Gordon and Betty Moore Foundation through Grant GBMF1250; 
	the Walter Burke Institute for Theoretical Physics at Caltech; 
	and the Gordon and Betty Moore Foundation's EPiQS Initiative, Grant GBMF8682 to JA.
BMH acknowledges support from the Department of Energy under the Early Career award program (\#DE-SC0018115).	
DM acknowledges support from the Gordon and Betty Moore Foundation’s EPiQS Initiative, Grant GBMF9069.
Finally, we acknowledge the 2018 Erice conference on \emph{Majorana Fermions and Topological Materials Science}, where this work was initiated.  
\end{acknowledgments}

%%%%%%%%%%%%%%%%%%%%%%%%%%%%%%%%%%%%%%%%%%%%%%%%
%%%%%%%%%%%%%%%%%%%%%%%%%%%%%%%%%%%%%%%%%%%%%%%%
%%%%%%%%%%%%%%%%%%%%%%%%%%%%%%%%%%%%%%%%%%%%%%%%
\clearpage
\appendix

%%%%%%%%%%%%%%%%%%%%%%%%%%%%%%%%%%%%%%%%%%%%%%%%
\section{Variational analysis of interacting Majorana fermions}

%%%%%%%%%%%%%%%%%%%%%%%%%%%%%%%%%%%%%%%%%%%%%%%%
\subsection{Continuum model}
\label{VariationalAppContinuum}
Here we employ a variational approach to study the interacting continuum Hamiltonian $\mathcal{H}_0 + \delta \mathcal{H}$ defined in Eqs.~\eqref{H0} and \eqref{deltaH}.  
Specifically, we view the ground state of the free-fermion Hamiltonian $\mathcal{H}_{\rm MF}$ from Eq.~\eqref{HMF} as a trial wavefunction for the interacting problem, treating the mass $m$ as a variational parameter.
Note that sending $\gamma_R \rightarrow -\gamma_R$ leaves the interacting Hamiltonian invariant but sends $m \rightarrow -m$. 
Without loss of generality we therefore consider trial wavefunctions with $m\geq 0$ below.  

The analysis is most conveniently carried out in terms of momentum-space Majorana fermions defined with Fourier-transforms conventions
\begin{align}
\gamma_A(x) &= \frac{1}{\sqrt{2}} \int \frac{dk}{2\pi} e^{ikx} \gamma_{A,k} \\
\gamma_{A,k} & = \sqrt{2} \int dx e^{-ikx} \gamma_A(x)
\end{align} 
for $A = L$ or $R$.
The corresponding commutation relations are given by 
\begin{align} 
\{ \gamma_A(x), \gamma_B(y) \} &= \frac{1}{2} \delta_{AB}\delta(x-y)\\
\{ \gamma_{A,k}, \gamma_{B,q} \} &=2 \pi \delta_{AB}\delta(k+q).
\end{align}
Upon passing to momentum space, the free-fermion Hamiltonian $\mathcal{H}_{\rm MF}$ can be readily diagonalized, yielding
single-particle excitation energies $E(k) = \sqrt{(vk)^2 + (m/2)^2}$ and correlation functions
\begin{align}
\langle \gamma_{R,k} \gamma_{R,q} \rangle &= 2\pi \delta(k+q) \frac{1}{2} \bigg[ 1+\frac{vk}{E(k)} \bigg]
\label{corr1} \\
\langle \gamma_{L,k} \gamma_{L,q} \rangle &= 2\pi \delta(k+q) \frac{1}{2} \bigg[ 1-\frac{vk}{E(k)} \bigg]\\
\langle \gamma_{R,k} \gamma_{L,q} \rangle &= 2\pi \delta(k+q) \frac{i}{4}  \frac{m}{E(k)}.
\label{corr3}
\end{align}
Here and in the remainder of this Appendix expectation values are taken with respect to the ground state of $\mathcal{H}_{\rm MF}$.

Equations~\eqref{corr1} through \eqref{corr3} allow one to efficiently evaluate the trial energy density $\mathcal{E}_{\rm trial} \equiv \langle\mathcal{H}_0 + \delta \mathcal{H}\rangle/L$ ($L$ is the length of the interface).  
The results are conveniently expressed in terms of integrals
\begin{align}
I_{\alpha} = \int_{-2 v \Lambda/m}^{2v\Lambda/m} du \frac{u^\alpha}{\sqrt{1+u^2}},
\end{align} 
where $\Lambda$ is a momentum cutoff for the interacting Majorana fermions.
The kinetic-energy piece reads
\begin{equation}
  \frac{\langle \mathcal{H}_0 \rangle}{L} = -\frac{m^2}{16\pi v}I_2.
\end{equation}
For the interactions, we first write  
\begin{align}
\frac{\langle \delta \mathcal{H}\rangle}{L} = \frac{\kappa}{4}\int_{q_1,q_2,q_3,q_4}q_2 q_4 \langle \gamma_{R,q_1}\gamma_{R,q_2} \gamma_{L,q_3} \gamma_{L,q_4}\rangle
\end{align}
and then evaluate the expectation value in the integrand using Wick's theorem:
\begin{align} \begin{aligned}
	\braket{ \gamma_{R,q_1}\gamma_{R,q_2} \gamma_{L,q_3} \gamma_{L,q_4} } &=
	\langle \gamma_{R,q_1}\gamma_{R,q_2}\rangle \langle \gamma_{L,q_3} \gamma_{L,q_4}\rangle
	\\	&\quad - \langle \gamma_{R,q_1} \gamma_{L,q_3}\rangle \langle \gamma_{R,q_2} \gamma_{L,q_4}\rangle
	\\	&\quad + \langle \gamma_{R,q_1} \gamma_{L,q_4}\rangle \langle \gamma_{R,q_2}\gamma_{L,q_3} \rangle.
\end{aligned} \end{align}
Some algebra yields
\begin{equation}
    \frac{\langle \delta \mathcal{H}\rangle}{L} = -\frac{\kappa m^4}{1024 \pi^2 v^4} (I_2^2 + I_0 I_2).
\end{equation}
Summing the contributions above gives our trial energy density,
\begin{align}
\label{MFNRG}
\mathcal{E}_{\rm trial}  = - \frac{m^2}{16\pi v}I_2 - \frac{\kappa m^4}{1024 \pi^2 v^4} (I_2^2 + I_0 I_2). 
\end{align}

\begin{figure}[htbp]
\begin{center} 
\includegraphics[width=0.45\textwidth]{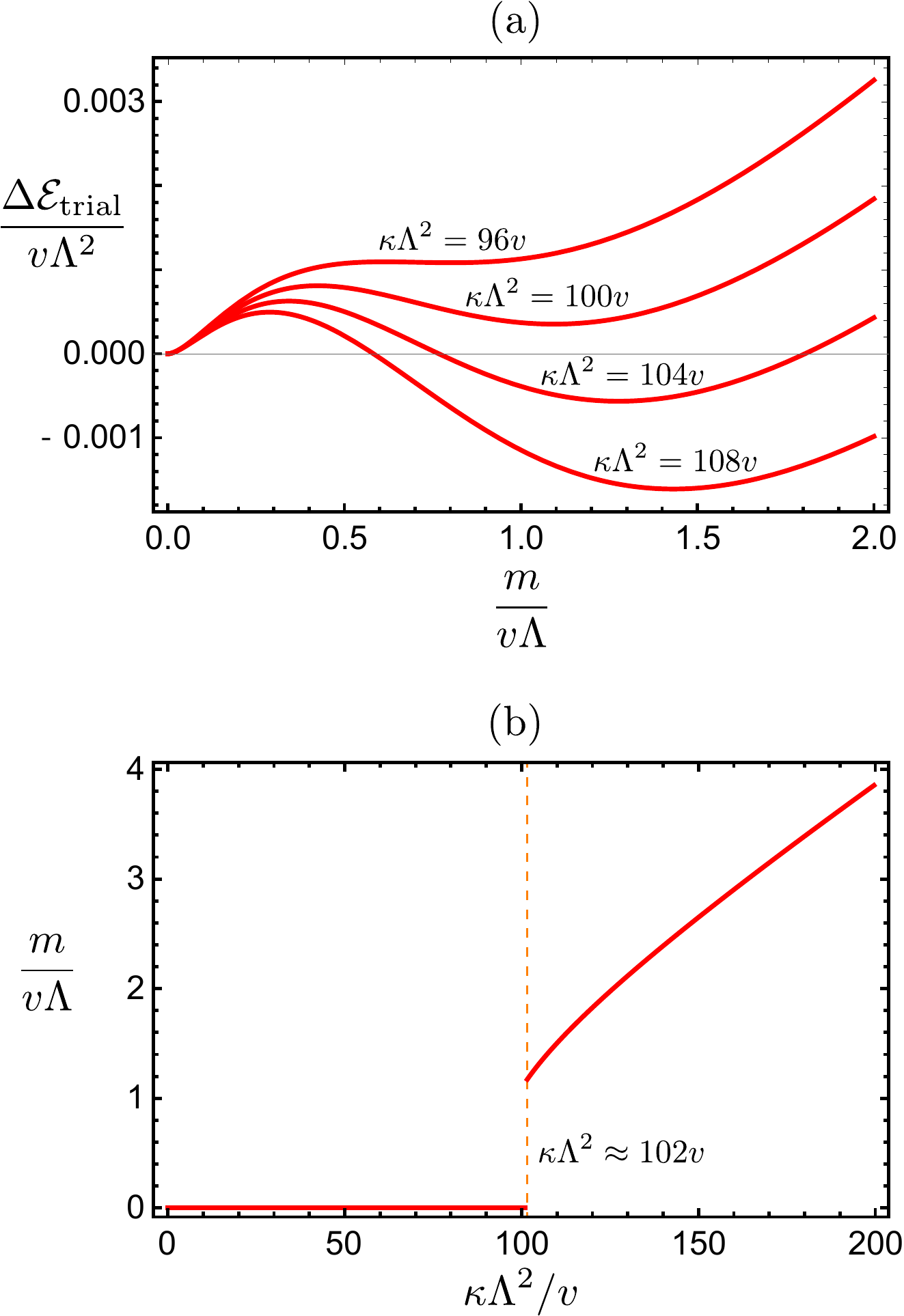}
\caption{
(a) Variational energy difference $\Delta \mathcal{E}_{\rm trial}$ [Eq.~\eqref{DEtrial}] versus mean-field mass $m$ for varying interaction strengths $\kappa$ in the continuum model given by Eqs.~\eqref{H0} and \eqref{deltaH}.
For $\kappa \Lambda^2 \gtrsim 102v$, the variational energy is minimized with nonzero $m$, signaling the spontaneous opening of a gap.  
(b) Optimal mass as a function of interaction strength. 
In the normalizations adopted above, $v$ is the velocity and $\Lambda$ is the cutoff.
}
\label{jumpdisc}
\end{center}
\end{figure}

We now minimize \eqref{MFNRG} with respect to $m$. 
For $\kappa < 0$ the minimization always yields $m = 0$. 
With $\kappa >0$, however, a nontrivial solution does arise beyond a critical interaction strength.  
It is convenient to examine 
\begin{equation}
  \Delta \mathcal{E}_{\rm trial} \equiv \mathcal{E}_{\rm trial} - \mathcal{E}_{\rm trial}|_{m = 0},
  \label{DEtrial}
\end{equation}
which quantifies the change in energy density due to a nonzero mass $m$.  
Figure~\ref{jumpdisc}(a) plots $\Delta \mathcal{E}_{\rm trial}/(v\Lambda^2)$ versus $m/(v \Lambda)$ for varying dimensionless interaction strengths $\kappa \Lambda^2/v$, while Fig.~\ref{jumpdisc}(b) displays the optimized mass as a function of $\kappa \Lambda^2/v$. 
A first-order jump in the optimal mass appears at a critical interaction strength 
\begin{equation}
  (\kappa \Lambda^2/v)_c \approx 102.
  \label{kappa_c}
\end{equation}
The large numerical factor on the right-hand side naively suggests that spontaneous mass generation requires implausibly large interactions. 
We stress that such a conclusion is generally incorrect.  
In the next subsection we will see that lattice models with microscopic interaction strength $U_{\rm microscopic}$ yield $\kappa = c U_{\rm microscopic}$, where $c$ is a number of order $\sim10^2\mbox{--}10^3$.
Thus even modest microscopic interactions translate into `large' $\kappa$ values that can exceed the threshold in Eq.~\eqref{kappa_c}, consistent with conclusions from more rigorous microscopic analyses reviewed in Secs.~\ref{Sewing} and \ref{microscopics}.

%%%%%%%%%%%%%%%%%%%%%%%%%%%%%%%%%%%%%%%%%%%%%%%%
%%%%%%%%%%%%%%%%%%%%%%%%%%%%%%%%%%%%%%%%%%%%%%%%

%%%%%%%%%%%%%%%%%%%%%%%%%%%%%%%%%%%%%%%%%%%%%%%%
\subsection{One-dimensional lattice model}
\label{VariationalAppLattice}

Next we will similarly study interacting Majorana fermions on an $N$-site chain with periodic boundary conditions, governed by a Hamiltonian 
\begin{align}
  H &= H_0 + \Delta H
  \label{latt_H}
  \\
  H_0 &= i t\sum_a \gamma_a\gamma_{a+1}
  \\
  \delta H &= -g\sum_a \gamma_a\gamma_{a+m} \gamma_{a+n}\gamma_{a+p}
  \label{latt_deltaH}
\end{align}
with $t>0$ and $0<m<n<p$.  
We are specifically interested in quantifying the regime of interaction strength over which the chain spontaneously dimerizes.  
To this end we employ a trial wavefunction given by the ground state of the free-fermion Hamiltonian
\begin{align}
\label{MeanFieldH}
H_{\text{MF}} = i\sum_a[t+(-1)^a \bar m] \gamma_a \gamma_{a+1},
\end{align}
where $\bar m$ is a variational parameter that, when non-zero, indicates spontaneous dimerization.  
Note that the translation $\gamma_a \rightarrow \gamma_{a+1}$ leaves $H$ invariant but flips the sign of $\bar{m}$ in $H_{\rm MF}$. 
The energies for configurations with $\bar m$ and $-\bar m$ thus necessarily match, so that we need only consider variational wavefunctions with $\bar m\geq 0$ in what follows.

The spectrum and eigenstates of Eq.~\eqref{MeanFieldH} can be readily constructed by going to momentum space.
For our purposes here it suffices to report correlation functions of Majorana bilinears $\langle \gamma_a \gamma_b\rangle$, from which all other correlations can be deduced using Wick's theorem.  
(All expectation values presented in this Appendix are taken with respect to the ground state of $H_{\rm MF}$.)
In particular, we find
\begin{align} \begin{aligned}
	\langle \gamma_a\gamma_b \rangle &= \delta_{ab} + \left[(-1)^{a+b}-1\right] 4it f_{s,a-b}
	\\ &\quad + \left[(-1)^a-(-1)^b\right] 4i \bar m f_{c,a-b}.
\end{aligned} \end{align}
Above we introduced functions
\begin{align}
  f_{s,x} &= \int_{-\pi/2}^{\pi/2} \frac{dk}{2\pi}\frac{\sin k\sin(kx)}{E(k)}
  \\
  f_{c,x} &= \int_{-\pi/2}^{\pi/2} \frac{dk}{2\pi}\frac{\cos k\cos(kx)}{E(k)},
\end{align}
where $E(k) = \sqrt{(4t \sin k)^2 + (4\bar m\cos k)^2}$ are the single-particle excitation energies for $H_{\rm MF}$.  

We can now straightforwardly obtain our trial energy density $\mathcal{E}_{\rm trial} = \langle H_0 + \delta H\rangle/N$.  
The hopping contribution is simply
\begin{equation}
  \frac{\langle H_0\rangle}{N} = -8t^2 f_{s,1}.
\end{equation}
The interaction term can be treated using the decomposition
\begin{align}
\label{PaulInteraction}
\langle \gamma_{a} \gamma_{a+m} \gamma_{a+n} \gamma_{a+p}   \rangle &= 
\langle \gamma_{a} \gamma_{a+m}\rangle \langle \gamma_{a+n} \gamma_{a+p}   \rangle \nonumber\\
&-
\langle \gamma_{a} \gamma_{a+n} \rangle \langle \gamma_{a+m} \gamma_{a+p}   \rangle  \nonumber \\
&+\langle \gamma_{a} \gamma_{a+p} \rangle \langle \gamma_{a+m} \gamma_{a+n}   \rangle,
\end{align}
yielding
\begin{equation}
  \frac{\langle \delta H\rangle}{N} = -16g(F_{mnp}-F_{nmp}+F_{pmn}),
\end{equation}
where
\begin{align}
  F_{xyz} &= [(-1)^x-1]\{[(-1)^{y+z}-1]t^2 f_{s,x}f_{s,y-z}
  \nonumber \\
  &+[(-1)^y-(-1)^z]\bar m^2 f_{c,x} f_{c,y-z}\}.
\end{align}
Our total trial energy density is then
\begin{equation}
  \mathcal{E}_{\rm trial} = -8t^2 f_{s,1}-16g(F_{mnp}-F_{nmp}+F_{pmn}).
\end{equation}
Below it will also be useful to consider the difference 
\begin{equation}
  \Delta \mathcal{E}_{\rm trial} \equiv \mathcal{E}_{\rm trial} - \mathcal{E}_{\rm trial}|_{\bar m = 0}.
  \label{DEtrial2}
\end{equation}

%%%%%%%%%%%%%%%%%%%%%%%%%%%%%%%%%%%%%%%%%%%%%%%%
\begin{figure}[htbp]
\begin{center}
\includegraphics[width=0.45\textwidth]{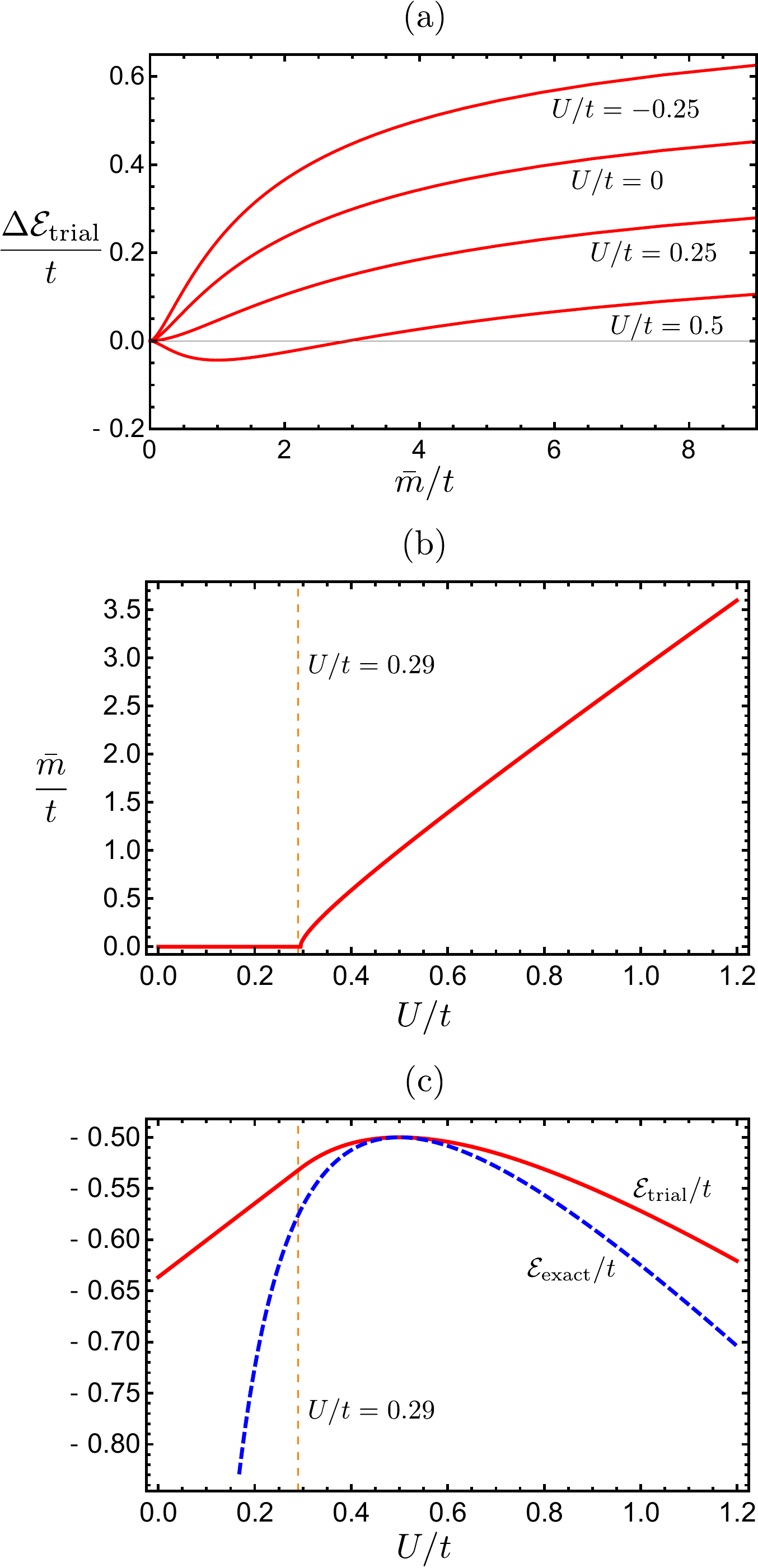}
\caption{Variational results for the lattice model given by Eqs.~\eqref{latt_H} through \eqref{latt_deltaH} with $m = 1, n = 3, p = 4$ and $g = U$.
(a) Variational energy difference $\Delta \mathcal{E}_{\rm trial}$ [Eq.~\eqref{DEtrial2}] versus mean-field dimerization order parameter $\bar m$ for varying interaction strengths $U$.
At $U\gtrsim 0.29 t$ the variational energy is minimized by nonzero $\bar m$, indicating spontaneous dimerization (as captured by DMRG, but at slightly larger interaction strengths $U \gtrsim 0.428t$~\cite{Fendley2018}).
(b) Optimal variational $\bar m$ as a function of interaction strength.
(c) Variational ground-state energy $\mathcal{E}_{\rm trial}$ (solid red line) and exact ground-state energy $\mathcal{E}_{\rm exact}$ (dashed blue line)~\cite{Fendley2018} versus $U/t$.  
The two converge at $U = t/2$, at which point the variational approach becomes exact.
}
\label{LatticeData}
\end{center}
\end{figure}
%%%%%%%%%%%%%%%%%%%%%%%%%%%%%%%%%%%%%%%%%%%%%%%%

Let us specialize to the lattice model in Eq.~\eqref{Hpaul} with $t' = 0$, for which $g = U$ and $m = 1, n = 3, p = 4$.  
In this case the trial energy density explicitly reads
\begin{align}
  \mathcal{E}_{\rm trial} = -8t^2 f_{s,1} -64U[t^2(f_{s,3}^2 - f_{s,1}^2)
  + \bar m^2(f_{c,1}^2 - f_{c,3}^2)].
  \label{EtrialPaul}
\end{align}
At this point it is instructive to relate Eq.~\eqref{EtrialPaul} to the energy density in Eq.~\eqref{MFNRG} for the continuum Hamiltonian. 
This exercise proceeds by $(i)$ introducing a momentum cutoff in the $f_{s,x}$ and $f_{c,x}$ integrals, $(ii)$ expanding the numerator in the integrands to order $k^2$, $(iii)$ replacing $\sin k\rightarrow k$ and $\cos k \rightarrow 1$ in $E(k)$, and $(iv)$ dropping a term proportional to $\bar m^6 I_2^2$ that can only arise upon including higher-momentum terms in the continuum model.  
Some algebra yields the relations
\begin{equation}
  v = 4t, ~~m = 8\bar m, ~~ \kappa = 512U.
\end{equation}
Note especially the large numerical prefactor in front of $U$ in connection with the discussion at the end of Appendix~\ref{VariationalAppContinuum}.

We now minimize Eq.~\eqref{EtrialPaul} with respect to $\bar m$.
Figure~\ref{LatticeData}(a) plots $\Delta \mathcal{E}_{\rm trial}/t$ versus $\bar m/t$ for several $U/t$ values.  
The optimized $\bar m$ as a function of $U/t$ appears in Fig.~\ref{LatticeData}(b).  
Our variational analysis predicts spontaneous dimerization---now via a continuous transition in contrast to the continuum model---for $U \gtrsim 0.29t$.
This prediction agrees reasonably well with DMRG, which yields spontaneous dimerization for $U \gtrsim 0.428t$~\cite{Fendley2018}.  
Interestingly, for $U = t/2$ our variational ansatz actually becomes exact since in this limit the interaction admits an exact self-consistent mean-field decoupling; recall the discussion in Sec.~\ref{microscopics}.  
Figure~\ref{LatticeData}(c) shows the optimized variational energy density together with the exact energy density extracted from Ref.~\onlinecite{Fendley2018}, which indeed agree at $U = t/2$.

%%%%%%%%%%%%%%%%%%%%%%%%%%%%%%%%%%%%%%%%%%%%%%%%
\begin{figure}[htbp]
\begin{center}
\includegraphics[width=0.45\textwidth]{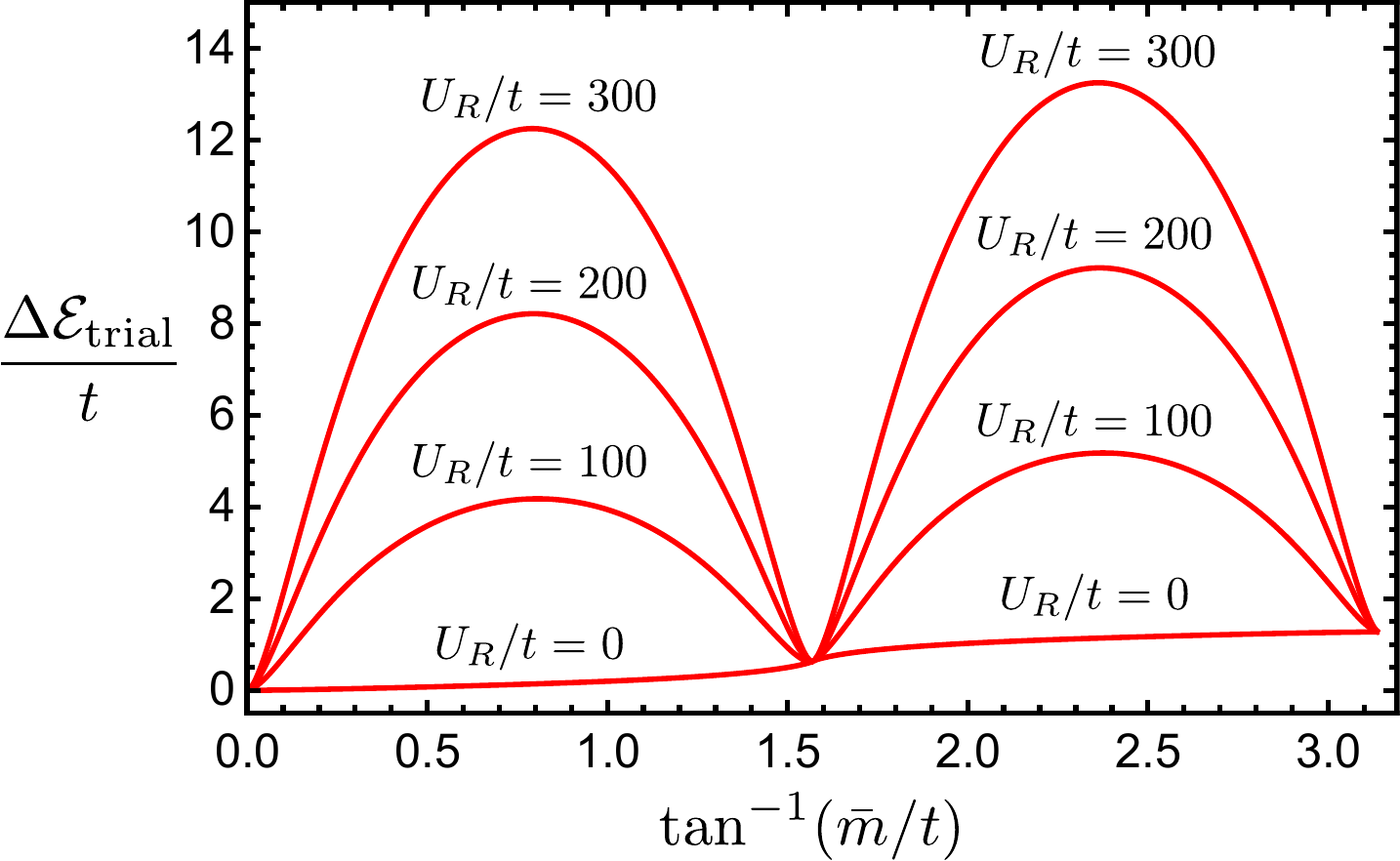}
\caption{Variational energetics for the lattice model given by Eqs.~\eqref{latt_H} through \eqref{latt_deltaH} with $m = 1, n = 2, p = 3$ and $g = -U_R$.  
The plot shows the variational energy difference $\Delta \mathcal{E}_{\rm trial}$ [Eq.~\eqref{DEtrial2}] versus $\tan^{-1}(\bar m/t)$, with $\bar m$ the mean-field dimerization order parameter, at various interaction strengths.
For \emph{any} $U_R$ the variational energy is minimized at $\bar m = 0$---precluding a transition at the mean-field level.
DMRG calculations do capture a dimerization instability, but only at extremely strong interactions strengths $U_R/t \gtrsim 250$~\cite{Rahmani1,Rahmani2}.
Appendix~\ref{RahmaniApp} explains the striking difference between the behavior of the two models examined here and in Fig.~\ref{LatticeData}.
}
\label{MeanfieldEnergyprime}
\end{center}
\end{figure}
%%%%%%%%%%%%%%%%%%%%%%%%%%%%%%%%%%%%%%%%%%%%%%%%

Finally, suppose that we instead take $g = -U_R$ and $m = 1, n = 2, p = 3$.  
This choice corresponds to a different interaction in which four adjacent Majorana fermions interact with strength $U_R$.  
We will only consider $U_R > 0$ in our variational analysis since negative $U_R$ values generate additional gapless modes beyond those in Eqs.~\eqref{H0} and \eqref{deltaH} (see Appendix~\ref{RahmaniApp} for a synopsis); our trial wavefunctions are thus not expected to adequately capture the physics at $U_R < 0$.  
The trial energy density for this model becomes
\begin{align}
  \mathcal{E}_{\rm trial} & = -8t^2 f_{s,1} -64U_R[t^2(f_{s,1}^2 + f_{s,1}f_{s,3})
  \nonumber\\
  &+ \bar m^2(f_{c,1}^2 - f_{c,1}f_{c,3})].
  \label{EtrialRahmani}
\end{align}
Upon connecting to the continuum energy density as outlined above, one finds the relations 
\begin{equation}
  v = 4t, ~~m = 8\bar m, ~~\kappa = 256U_R.
\end{equation}
Once again a large prefactor appears in the expression for $\kappa$ in agreement with the alternative derivation from Ref.~\onlinecite{Rahmani1}.
Based on our continuum analysis, one might therefore expect that a spontaneous dimerization transition sets in once $U_R/t>0$ becomes of order unity, as arose for the alternative interaction $U$.  
Curiously, however, minimizing the lattice energy density in Eq.~\eqref{EtrialRahmani} yields an optimal $\bar m$ that vanishes for \emph{any} $U_R>0$.  
Figure~\ref{MeanfieldEnergyprime} illustrates this conclusion by plotting $\Delta \mathcal{E}_{\rm trial}/t$ versus $\tan^{-1}(\bar m/t)$ for varying $U_R/t$ values.  
More rigorous DMRG simulations do capture a dimerization transition but only at extremely large $U_R/t$.
The following Appendix resolves the apparent discrepancy between our continuum and lattice analyses and explains the curious suppression of the instability for the $U_R$ interaction.

%%%%%%%%%%%%%%%%%%%%%%%%%%%%%%%%%%%%%%%%%%%%%%%%
\section{Interacting Majorana fermions in an alternative microscopic model}
\label{RahmaniApp}

Consider the microscopic model
\begin{eqnarray}
  H_R &=& H_0 + \delta H
  \label{Hrahmani} \\
  H_0 &=& i t\sum_a \gamma_a \gamma_{a+1}
  \\
  \delta H &=& U_R \sum_a \gamma_{a-1} \gamma_{a}\gamma_{a+1}\gamma_{a+2}
\end{eqnarray}
defined with periodic boundary conditions.  
This model preserves both translation symmetry $T$ and chiral symmetry $\mathcal{C}$ defined in Sec.~\ref{microscopics}.  
References~\onlinecite{Rahmani1,Rahmani2} extensively studied the phase diagram; here we note the following features:
$(i)$ A single pair of gapless, counterpropagating Majorana fermions [Eq.~\eqref{H0}] captures the low-energy physics over the broad interval $-0.28 \lesssim U_R/t \lesssim 250$.  Throughout the central charge is $c = 1/2$.  
$(ii)$ The chain spontaneously dimerizes, thus gapping the Majorana fermions, for $U_R/t\gtrsim 250$---which again reflects a vastly stronger interaction strength compared to that required for dimerization in Eq.~\eqref{Hpaul}.  
$(iii)$ For $-2.86 \lesssim U_R/t \lesssim -0.285$ the chain remains gapless, but the low-energy physics is described by \emph{three} pairs of counterpropagating Majorana fermions.  The central charge accordingly becomes $c = 3/2$.  
Below we will explain, within a unified framework, both the extreme robustness of the $c = 1/2$ phase to interactions as well as the onset of the $c = 3/2$ phase.  

Let us start from the non-interacting limit, $U_R = 0$.  Here the chain is diagonalized by passing to momentum space via
\begin{equation}
  \gamma_a = \frac{\sqrt{2}}{\sqrt{N}} \sum_{k} e^{i k a} \gamma_k,
\end{equation}
with $N$ the number of Majorana sites in the chain.  In our conventions $\{\gamma_k,\gamma_{k'}\} = \delta_{k,-k'}$.  Note also that self-Hermiticity of $\gamma_a$ implies that $\gamma_k = \gamma_{-k}^\dagger$; this relation allows a complete description of the chain using operators acting in half of the Brillouin zone.  In particular, one finds
\begin{equation}
  H_0 = \sum_{-\pi <k<0} \epsilon_0(k) \gamma_k^\dagger \gamma_k,
\end{equation}
where 
\begin{equation}
  \epsilon_0(k) = 4 t |\sin k|
\end{equation}
is the Majorana-fermion kinetic energy.  The ground state follows by taking $\gamma_k^\dagger \gamma_k = 0$ for all momenta in the left half of the Brillouin zone.  Single-particle excitations are obtained by taking $\gamma_p^\dagger \gamma_p = 1$ for some momentum $p$, incurring an energy cost of $\epsilon_0(p)$.  In the low-energy theory, left- and right-moving Majorana fermions correspond to excitations near momentum $0$ and $-\pi$, respectively.  
  
Interactions can generate additional symmetry-allowed hopping processes, as already observed in Ref.~\onlinecite{Rahmani2}, which in turn renormalize the kinetic energy.  
Reference~\onlinecite{Rahmani2} explored these effects within self-consistent mean field theory.
We will instead capture kinetic-energy renormalization via an exact rewriting of interactions analogous to normal ordering.  
Specifically, we will organize four-fermion terms so that all matrix elements vanish identically in the subspace consisting of states with either zero or one single-particle excitation.  
To obtain this form one can first express $\delta H$ in terms of Majorana operators with momenta in the left Brillouin zone half, and then use anticommutation relations to move all $\gamma_k^\dagger$ operators to the left of $\gamma_k$.  
The final expression takes the form 
\begin{equation}
  \delta H = \delta H_{\rm int} + \delta H_{\rm KE}, 
  \label{decomposition}
\end{equation}
where $\delta H_{\rm int}$ vanishes within the zero- and one-excitation subspace as desired and $\delta H_{\rm KE}$ contains the kinetic-energy renormalization.  
After some algebra we explicitly find
\begin{eqnarray}
  \delta H_{\rm KE} &=& \sum_{-\pi<k<0}\delta \epsilon(k)  \gamma_k^\dagger \gamma_k
  \label{deltaHKE}
  \\
  \delta \epsilon(k) &=& U_R \left[\frac{56}{3\pi} |\sin k | - \frac{8}{\pi} \sin(3k)\right].
  \label{deltaepsilon}
\end{eqnarray}
The first term in Eq.~\eqref{deltaepsilon} reflects renormalization of the nearest-neighbor hopping amplitude $t$, while the second represents an interaction-induced third-neighbor hopping.  

The decomposition of interactions employed above is very useful, at least for sufficiently small $|U_R|/t$, since the expectation value of the energy for single-excitation states can be read off immediately.  
Specifically, upon including the bare component, the total kinetic energy for an excitation with momentum $k$ becomes $\epsilon(k) \equiv \epsilon_0(k) + \delta \epsilon(k)$.  
The dispersion for right- and left-moving Majorana fermions in the continuum limit follows from expanding $\epsilon(k)$ near $k = -\pi$ and $0$; one finds $\epsilon \sim \pm v k$ with
\begin{equation}
  v = 4t + \frac{128}{3\pi}U_R
\end{equation}
the renormalized velocity.  

For $U_R<0$, interactions reduce $v$, and at a critical value $U_R^* = -\frac{3\pi}{32}t \approx -0.295 t$ the velocity \emph{vanishes}.  
Below this value the renormalized kinetic energy $\epsilon(k)$ supports additional pairs of gapless Majorana fermions---changing the central charge from $c = 1/2$ to $3/2$.  
Remarkably, the critical interaction strength $U_R^*$ extracted from our treatment agrees \emph{quantitatively} with DMRG predictions.  
We note that the mean-field treatment performed in Ref.~\onlinecite{Rahmani2} recovers similar quantitative agreement.

For $U_R>0$, interactions instead enhance $v$.  
Recall from Sec.~\ref{Sewing} that in the low-energy description, the critical phase with central charge $c = 1/2$ becomes unstable to spontaneous mass generation when $\kappa \Lambda^2/v$ becomes of order unity, where $\Lambda$ is a momentum cutoff and $\kappa$ is the coupling from Eq.~\eqref{deltaH}.  
In the present context we have $\kappa \propto U_R$.  
Upward renormalization of $v$ clearly boosts the robustness of the $c = 1/2$ phase against interactions, though we are unable to quantitatively obtain the critical value of $U_R \sim 250$ at which DMRG finds an instability.  

\begin{figure}
	{Dimerization order parameter $\braket{ i\gamma_{a-1}\gamma_{a}-i\gamma_{a}\gamma_{a+1} }$}\\[-2ex]
	\includegraphics[width=\linewidth]{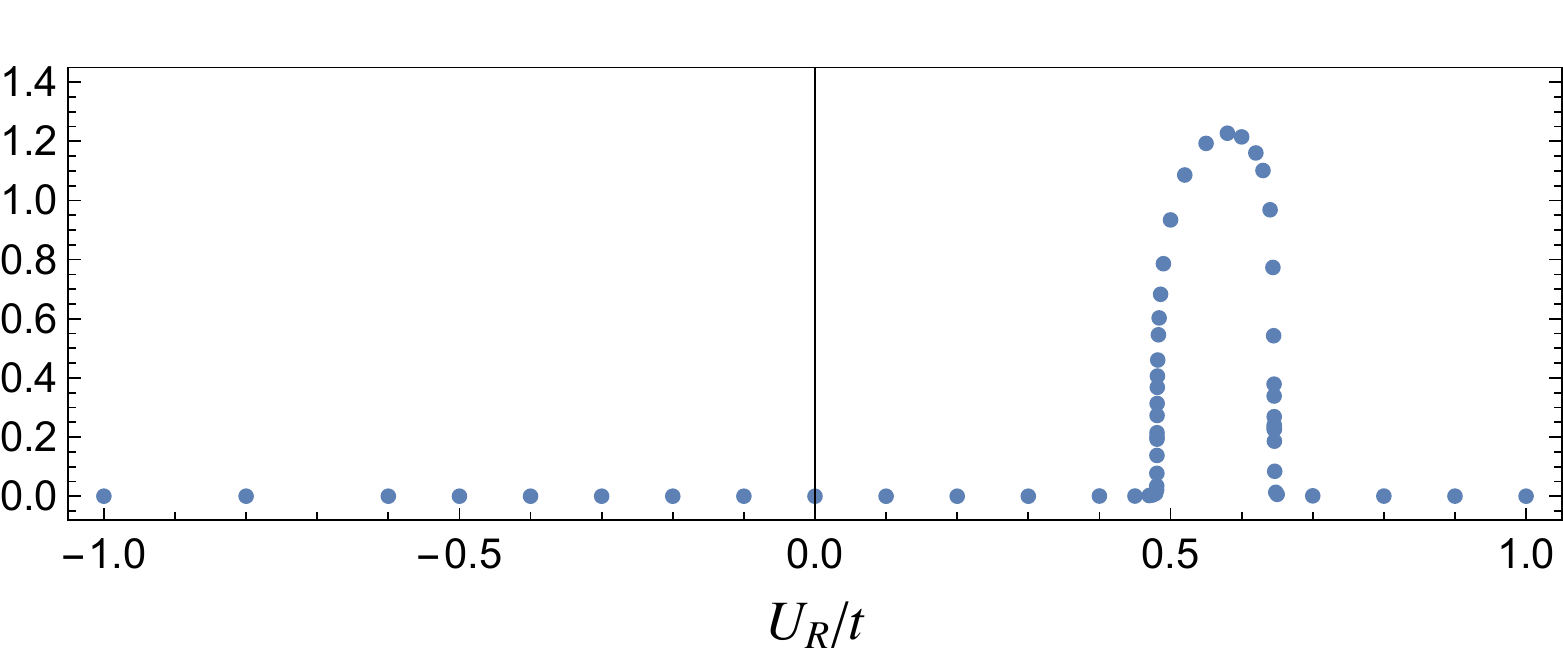}
	\caption{
		Dimerization order parameter $\braket{ i\gamma_{a-1}\gamma_{a}-i\gamma_{a}\gamma_{a+1} }$ versus $U_R/t$ obtained from DMRG simulations of the Hamiltonian in Eq.~\eqref{H_Rprime}.  
		This Hamiltonian is the same as Eq.~\eqref{Hrahmani} but with the interaction-induced kinetic-energy renormalization subtracted off.  
		The subtraction reduces the critical interaction strength required for spontaneous dimerization by \emph{three orders of magnitude}---from $U_R/t \approx 250$ down to $U_R/t \approx 0.48$.		
		Outside of the dome-shaped dimerized region, the system realizes a gapless state with central charge $c = 1/2$ (at least over the $U_R/t$ window shown).
		Re-entrance of the $c = 1/2$ critical state at $U_R/t \gtrsim 0.65$ can also be understood from the subtraction, as explained in Appendix~\ref{RahmaniApp}.}
	\label{fig:mod_rzfa_orderpar}	
\end{figure}

We can, nevertheless, stringently test the scenario above.  
If kinetic-energy renormalization indeed pushes the spontaneous-dimerization transition to extremely large $U_R$ values, then removing this renormalization should reduce the critical $U_R$ by three orders of magnitude.  
Remarkably, we indeed find that such a dramatic reduction.  
Consider the modified Hamiltonian
\begin{equation}
 H_R' = H_R -\delta H_{\rm KE},
 \label{H_Rprime}
\end{equation}
which is identical to Eq.~\eqref{Hrahmani} except that the kinetic-energy renormalization is subtracted off, thus yielding a $U_R$-independent velocity.  
DMRG simulations find a transition to a gapped phase in this model at $U_R/t \approx 0.48098(1)$, comparable to the critical interaction strength obtained from the alternative model in Eq.~\eqref{Hpaul} at $t' = 0$.
See Fig.~\ref{fig:mod_rzfa_orderpar}.
Furthermore, at least over the range of $U_R$ shown in the figure, the transition to a $c = 3/2$ phase at $U_R < 0$ has also been removed (as expected upon removal of the kinetic-energy renormalization).

Conspicuously, Fig.~\ref{fig:mod_rzfa_orderpar} also reveals a re-entrant $c = 1/2$ critical phase for $U_R/t = 0.64585(3)$.
We can explain this feature as well by examining Eq.~\eqref{H_Rprime}, which can equivalently be written as
\begin{align}
  H_R' &= i t\sum_a \gamma_a \gamma_{a+1} + U_R \sum_a\bigg{[}\gamma_{a-1} \gamma_a \gamma_{a+1}\gamma_{a+2} 
  \nonumber \\
  &- \frac{2i}{\pi}\left(\frac{7}{3}\gamma_a \gamma_{a+1} + \gamma_a \gamma_{a+3}\right)\bigg{]}.
  \label{HRprime1}
\end{align}
%The second line is just $\delta H_{\rm KE}$ expressed in real space.
Up to moderate values of $U_R/t$, it is natural to interpret the second line---which is just $\delta H_{\rm KE}$ expressed in real space---as a correction to the bare kinetic energy on the first line.
However, at sufficiently large $U_R/t$ this `correction' overwhelms the bare piece, suggesting the following alternative viewpoint.  
Let us trivially rewrite $H_R'$ as
\begin{align}
  H_R' &= \sum_a \left[it \gamma_a \gamma_{a+1} - \frac{4i}{\pi} U_R \left(\frac{7}{3}\gamma_a \gamma_{a+1} + \gamma_a \gamma_{a+3}\right)\right]
  \nonumber \\
  &+ U_R \sum_a\bigg{[}\gamma_{a-1} \gamma_a \gamma_{a+1}\gamma_{a+2} 
  \nonumber \\
  &~~~~~~~~~~~~+ \frac{2i}{\pi}\left(\frac{7}{3}\gamma_a \gamma_{a+1} + \gamma_a \gamma_{a+3}\right)\bigg{]}
  \label{HRprime2}
\end{align}
and view the top line as our new `bare' kinetic term.  
For $U_R/t > 3\pi/16 \approx 0.59$, all of the corresponding kinetic-energy eigenvalues flip sign compared to the kinetic energy from the $t$ term alone---thus completely changing the character of the associated non-interacting ground state.
The second and third lines represent an interaction that has no nontrivial matrix elements in the subspace with zero or one single-particle excitations about that modified ground state.
[Note the relative sign between the final terms in Eqs.~\eqref{HRprime1} and \eqref{HRprime2}.]
In this $U_R/t$ regime the top line yields a velocity for right- and left-movers of $v = \frac{256}{3\pi}U_R-4t$.
Once again we end up with a kinetic-energy scale that that grows with $U_R$, so that the dimerization instability is naturally suppressed beyond a critical value of $U_R/t$ as observed in DMRG.

%%%%%%%%%%%%%%%%%%%%%%%%%%%%%%%%%%%%%%%%%%%%%%%%
%%%%%%%%%%%%%%%%%%%%%%%%%%%%%%%%%%%%%%%%%%%%%%%%

As an additional sanity check, one can extract kinetic-energy renormalization arising from the four-fermion interaction in Eq.~\eqref{Hpaul} via exactly the same procedure leading to Eq.~\eqref{decomposition} above.  The result takes the form in Eq.~\eqref{deltaHKE} where now
\begin{equation}
  \delta \epsilon(k) = -\frac{64}{3\pi}U |\sin^3k|.
\end{equation}
Near $k = 0$ and $-\pi$, $\delta \epsilon(k) \propto k^3$, indicating that velocity renormalization \emph{vanishes} in this model.  
Thus no such suppression of the dimerization instability is expected in our scenario~\footnote{Actually, in the full microscopic model $\delta \epsilon(k)$ reduces the overall bandwidth even though the velocity remains fixed.  This effect likely yields a slightly smaller critical $U$ compared to what would occur if the bandwidth was also fixed.}, and indeed a transition occurs at the modest value $U \approx 0.428t$~\cite{Fendley2018}.
Reference~\onlinecite{Fendley2018} further studied Eq.~\eqref{Hrahmani} with $\delta H$ replaced by yet another interaction,
\begin{equation}
  \delta H' = U_y \sum_a(\gamma_{a-2}\gamma_a\gamma_{a+1}\gamma_{a+2}-\gamma_{a-2}\gamma_{a-1}\gamma_a\gamma_{a+2}).
\end{equation}
In this case we find that $\delta \epsilon(k) = 0$---i.e., $U_y$ produces no kinetic-energy renormalization at all.  
DMRG simulations find a transition at a similarly modest value $U_y \approx 0.45t$~\cite{Fendley2018}.
Together the results above strongly support our explanation for the anomalously strong interaction strength required for dimerization in Eq.~\eqref{Hrahmani}, which has heretofore remained enigmatic.

Zooming out, we see from this discussion that microscopic details matter when dealing with instabilities arising from `strong' interactions that are irrelevant at weak coupling.  
The insights obtained here can potentially be exploited to concoct new models that, depending on the desired outcome, either enhance or suppress the effects of such strong irrelevant interactions.

%%%%%%%%%%%%%%%%%%%%%%%%%%%%%%%%%%%%%%%%%%%%%%%%
\section{Analysis of fermion tunneling across a constriction}
\label{fermion_tunneling}
\begin{figure}[htbp]
\begin{center}
\includegraphics[width=0.8\linewidth]{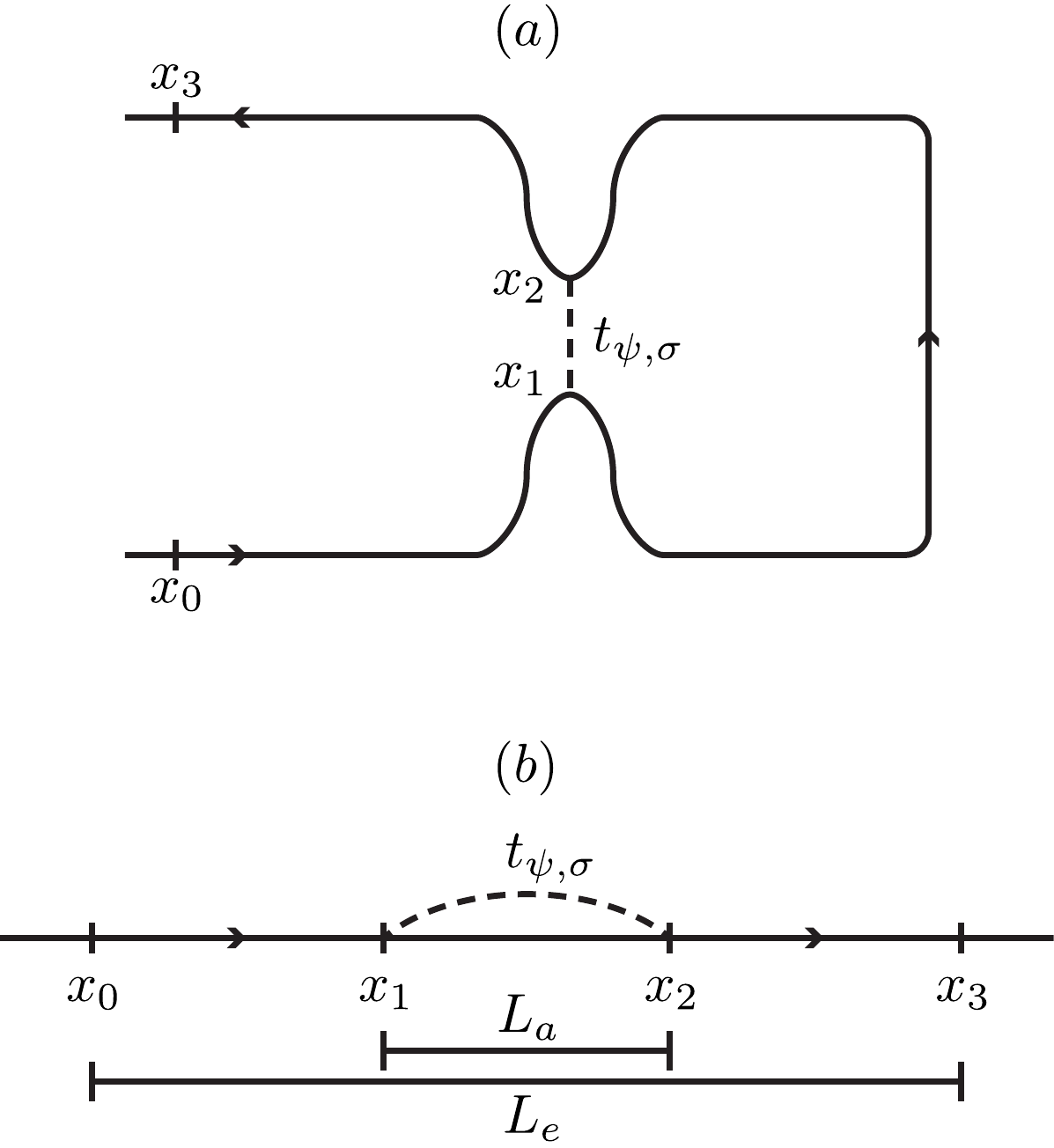}
\caption{(a) Geometry and coordinates used to explicitly analyze the emergent-fermion transmission amplitude in the interferometer from Fig.~\ref{psidetector}. 
Appendices~\ref{fermion_tunneling} and \ref{app:Ising_tunneling} respectively treat the cases where only fermions tunnel (with coupling $t_\psi$) and only Ising anyons tunnel (with coupling $t_\sigma$) across the constriction.  
(b) `Unfolded' version of (a), not to scale.
In terms of lengths shown in Fig.~\ref{psidetector}(a), we have $x_2-x_1 = L_a$ and $x_3-x_0 = L_e$.
}
\label{coords_fig}
\end{center}
\end{figure}

Here we study the interferometer in Fig.~\ref{psidetector} in the limit where only fermions are allowed to tunnel across the constriction.
That is, we take the fermion-tunneling amplitude $t_\psi \neq 0$ but set $t_\sigma = 0$ in Eq.~\eqref{Htun}---in which case we arrive at a free-fermion scattering problem that admits an exact solution.
Our goal is to deduce the phase $e^{i \phi_{\rm emergent}}$ acquired by an emergent fermion that travels from position $x_0$ before the constriction to position $x_3$ after the constriction.
 Figure~\ref{coords_fig}(a) illustrates the interferometer geometry, while Fig.~\ref{coords_fig}(b) shows an `unfolded' version.
For simplicity we consider the case where no nontrivial quasiparticles reside in the bulk of the interferometer, i.e., we assume $a = I$ in Fig.~\ref{psidetector}.

%%%%%%%%%%%%%%%%%%%%%%%%%%%%%%%%%%%%%%%%%%%%%%%%
%%%%%%%%%%%%%%%%%%%%%%%%%%%%%%%%%%%%%%%%%%%%%%%%

Evaluating the Heisenberg equation of motion $\partial_t \gamma = i [ \mathcal{H}_0 + \mathcal{H}_{\rm tun}, \gamma ]$, with $\mathcal{H}_0$ the chiral Majorana kinetic energy, one finds
\begin{align}
\partial_t \gamma(x,t) = -v_e \partial_x \gamma(x,t) &+ 
\frac{i}{2} t_\psi e^{- i \pi h_\psi}\big{[}\delta(x-x_1)\gamma(x_2,t) 
\nonumber \\
&-\delta(x-x_2) \gamma(x_1,t)\big{]}.
\label{gamma_EOM}
\end{align}
As before, $v_e$ is the emergent-fermion edge velocity, while $x_1$ and $x_2$ respectively denote positions on the lower and upper sides of the constriction (see again Fig.~\ref{coords_fig}).
By solving the equation of motion one can relate $\gamma(x_3,t)$ to $\gamma(x_0,0)$.  
The phase of interest follows from the equal-time relation $\gamma(x_3,0) = e^{i \phi_{\rm emergent}} \gamma(x_0,0)$, so hereafter we focus on the solution at $t = 0$.  

Away from $x = x_{1,2}$ Eq.~\eqref{gamma_EOM} reduces to a standard chiral wave equation.  Suppose that $x_j^+$ denotes a coordinate slightly larger than $x_j$ while $x_j^-$ denotes a coordinate slightly smaller than $x_j$, and let $k_e$ be the incident emergent-fermion momentum. 
One immediately finds
\begin{align}
  \gamma(x_1^-,0) &= e^{i k_e(x_1- x_0)}\gamma(x_0,0)
  \label{solA}
  \\
  \gamma(x_2^-,0) &= e^{i k_e(x_2 - x_1)}\gamma(x_1^+,0) 
  \label{solB}
  \\
  \gamma(x_3,0) &= e^{i k_e(x_3 - x_2)}\gamma(x_2^+,0).
  \label{solC}
\end{align}
(In the exponentials above we replaced $x_j^\pm \rightarrow x_j$ since the difference is inconsequential.)
Next, integrating Eq.~\eqref{gamma_EOM} over a small region enclosing $x_{1}$ and similarly for $x_2$ yields the linear relations
\begin{align}
\label{x1eqns}
0 &= \gamma(x_1^+,0)-\gamma(x_1^-,0) + \tilde t_\psi [\gamma(x_2^+,0)+\gamma(x_2^-,0) ]\\
\label{x2eqns}
0 &=\gamma(x_2^+,0)-\gamma(x_2^-,0) - \tilde t_\psi [ \gamma(x_1^+,0)+\gamma(x_1^-,0)],
\end{align}
where $\tilde t_\psi = -i \frac{t_{\psi}}{4v_e}e^{- i \pi h_\psi}$.
Combining with Eq.~\eqref{solB} and defining $e^{i\chi} = e^{i k_e(x_2-x_1)}$, one obtains
\begin{equation}
  \gamma(x_2^+,0)  = e^{i \chi} \frac{1+2 e^{-i \chi}\tilde t_\psi + \tilde t_\psi^2}{1+2e^{i \chi}\tilde t_\psi + \tilde t_\psi^2} \gamma(x_1^-,0).
  \label{solD}
\end{equation}
Finally, Eqs.~\eqref{solA}, \eqref{solC}, and \eqref{solD} together imply that $\gamma(x_3,0) = e^{i \phi_{\rm emergent}} \gamma(x_0,0)$ with
\begin{align}
  e^{i \phi_{\rm emergent}} = e^{i k_e(x_3-x_0)} \frac{1+2e^{-i \chi} \tilde t_\psi + \tilde t_\psi^2}{1+2e^{i \chi}\tilde t_\psi + \tilde t_\psi^2}.
  \label{gen_sol}
\end{align}

To make contact with Sec.~\ref{sec:interf_1psi} from the main text, we now set $x_3-x_0 = L_e$ and $x_2-x_1 = L_a$.  Additionally, we expand Eq.~\eqref{gen_sol} to first order in $\tilde t_\psi$ and use $i e^{- i \pi h_\psi} = 1$, leading to
\begin{align}
  e^{i \phi_{\rm emergent}} \approx e^{i k_e L_e}\left[ 1 + \frac{t_{\psi}}{2v_e}\left(-e^{-i k_e L_a} + e^{i k_e L_a}\right)\right].
\end{align}
The three terms in brackets respectively correspond to paths $(i)$, $(ii)$, and $(iii)$ discussed in Sec.~\ref{sec:interf_1psi}.  
From this explicit calculation we can trace the relative minus sign between the terms for paths $(ii)$ and $(iii)$ to the anticommutation relations obeyed by the Majorana fermions.
Comparing to Eqs.~\eqref{Acomplete} and \eqref{phi_emergent_def}, the weights $w_{ii}$ and $w_{iii}$ are given by Eq.~\eqref{wii_iii} with $\alpha_{ii} = - \alpha_{iii} = -1/2$, as quoted in Eq.~\eqref{alpha_def}.

%%%%%%%%%%%%%%%%%%%%%%%%%%%%%%%%%%%%%%%%%%%%%%%%
\section{Analysis of Ising-anyon tunneling across a constriction}
\label{app:Ising_tunneling}

In this Appendix, we continue to study the interferometer in Fig.~\ref{psidetector}, but now allowing only Ising anyons to tunnel across the constriction.
The geometry and coordinates used are again given in Fig.~\ref{coords_fig}. 
We will evaluate the transmission amplitude describing propagation of an emergent fermion from position to $x_0$ to $x_3$ perturbatively in Ising-anyon tunneling, assuming that the interferometer does not contain any nontrivial bulk quasiparticles as in Appendix~\ref{fermion_tunneling}.

%%%%%%%%%%%%%%%%%%%%%%%%%%%%%%%%%%%%%%%%%%%%%%%%
\subsection{Hamiltonian and conventions}

We set $t_\psi = 0$ but take $t_\sigma \neq 0$ in Eq.~\eqref{Htun}, so that the full Hamiltonian becomes 
\begin{align} \begin{aligned}
	\mathcal{H} &= \mathcal{H}_{0} + \mathcal{H}_{\rm tun} \,,
\\	\mathcal{H}_{\rm tun} &= e^{-i \pi h_\sigma} t_\sigma \sigma(x_2)\sigma(x_1) \,.
	\label{eq:appHtun}
\end{aligned} \end{align}
Precisely as in Appendix~\ref{fermion_tunneling}, $\mathcal{H}_{0}$ describes a chiral right-moving free Majorana fermion.
In the tunneling term $h_\sigma = 1/16$ is the conformal weight (spin) of the $\sigma$ field and $t_\sigma \in \mathbb{R}$ is the coupling coefficient.
For the remainder of this appendix we set the velocity $v_e = 1$.

We choose $\sigma$ to be Hermitian and normalized such that
\begin{align}
	\bigXp{ \sigma(x',t') \, \sigma(x,t) }_0 &= \frac{1}{ (it'-ix' - it+ix)^{1/8} } \,.
	\label{eq:sigma_corr}
\end{align}
The subscript of the correlation function $\braket{\cdots}_0$ indicates that the correlator is computed with respect to the free CFT Hamiltonian $\mathcal{H}_{0}$.
[The choice of phase on the right side of Eq.~\eqref{eq:sigma_corr} guarantees that the correlator $\bigXp{ \sigma(0,-i\beta) \sigma(0) } = \operatorname{Tr}[\sigma(0) e^{-\beta H} \sigma(0)] / \operatorname{Tr}e^{-\beta H}$ is positive---a necessary condition for $\sigma=\sigma^\dag$.]
It is straightforward to check that $\mathcal{H}_{\rm tun}$ is indeed Hermitian.
Likewise, we define the normalized Majorana field
\begin{align}
	\psi = \sqrt{4\pi}\gamma
\end{align}
such that 
\begin{equation}
  \bigXp{ \psi(x',t') \, \psi(x,t) }_0 = \frac{1}{(i t' - i x'-i t + i x)}. 
\end{equation}

%%%%%%%%%%%%%%%%%%%%%%%%%%%%%%%%%%%%%%%%%%%%%%%%
\subsection{Scattering states}

Next we specify the formalism used to compute the transmission amplitude.
We quantize the CFT along slices at fixed positions $x$; wavefunctions are written on `position slices' that live for all of time.
(One should contrast to the usual framework wherein states live on fixed time slices.)
Such a  reformulation is in principle applicable to all field theories, but is particular convenient for CFT's due to the symmetry between space and time coordinates%
	~\footnote{For a chiral CFT, one can transform between the `position slice' quantization and the `time slice' quantization via a Wick rotation (real time to imaginary time), a $90^\circ$ Euclidean rotation (exchange time and space), and then another Wick rotation (imaginary time back to real time).}%
.
An interaction term may manifest itself in different ways within this rotated frame.
For example, a point defect localized in space become a global (instaneous) quantum quench.

In the same spirit as Eq.~\eqref{eq:appHtun}, we decompose the action into its free part and an interaction part:
\begin{align}
	S &= \int\!P_0 \, dx + S_{\rm tun} .
	\label{eq:tunneling_action}
\end{align}
The operator $P_0$ generates spatial translations of the free CFT, while $S_{\rm tun}$ consists of a spatially nonlocal global term (spanning all of time),
\begin{align} \begin{split}
	S_{\rm tun} &= -\int_{-\infty}^{\infty} \!\!\! dt \, \mathcal{H}_{\rm tun}(t)
	\\	&= -e^{-i \pi h_\sigma} t_\sigma \int_{-\infty}^{\infty} \!\!\! dt \, \sigma(x_2,t) \, \sigma(x_1,t) .
\end{split} \end{align}
For the problem at hand, the tunnel junction `teleports' particles between positions $x_1$ and $x_2$ and can be interpreted as a wormhole allowing `time-travel' between the two positions.
(This nonlocality makes it difficult to write the corresponding momentum operator for the tunneling term.)
We choose to work in the interaction picture, where operators are related to those in the Schr\"odinger picture via
\begin{align}
	O(x,t) = e^{-i P_0 x} \, O(t) \, e^{i P_0 x} .
\end{align}

The state describing an incoming emergent fermion with positive frequency $\omega$ is written as 
\begin{align}
	\ket{\psi_\omega} &= \frac{1}{2\pi} \int_{-\infty}^\infty \! dt \, e^{-i\omega t} \psi(t) \ket{0} ,
\end{align}
where $\ket{0}$ is the ground state of $P_0$.
This state exhibits the normalization
\begin{align}
	\braket{\psi_{\omega'}|\psi_{\omega}} &= \delta(\omega-\omega')
\end{align}
and carries momentum
\begin{align}
	P_0 \ket{\psi_\omega} = \omega \ket{\psi_\omega} .
\end{align}
Recall that we set the pesky velocity to unity; hence here and below $\omega$ corresponds to $k_e$ from Sec.~\ref{sec:interf_1psi}.

Let $A(\omega; x_0,x_3)$ denote the amplitude for transmission of the Majorana fermion from position $x_0$ to $x_3$ at frequency $\omega$.
In terms of the field theory, $A$ is the quantum amplitude associated with the spatial strip $[x_0,x_3]$ with boundary conditions set by the incoming/outgoing states.
Formally, the amplitude is defined via
\begin{align}
	A(\omega; x_0,x_3)\delta(\omega-\omega') = \frac{\Braket{ \psi_{\omega'} | \mathcal{P} e^{iS\big|_{x_0}^{x_3}} | \psi_\omega }} {\Braket{ \mathcal{P} e^{iS\big|_{x_0}^{x_3}} }} ,
\end{align}
where $S\big|_x^y$ is the action [Eq.~\eqref{eq:tunneling_action}] restricted to the spatial interval $[x,y]$, and $\mathcal{P}$ denotes path ordering (of the position coordinates).
We expand $A$ in powers of $t_\sigma$,
\begin{align}
	A &= A^{(0)} + t_\sigma A^{(1)} + t_\sigma^2 A^{(2)} + \dots ,
\end{align}
so that $A^{(n)}$ captures the $n$\textsuperscript{th}-order correction in the perturbative series.

The zeroth-order piece follows from free propagation of the scattering states, i.e., evolving $\psi(x,t)$ with the $t_\sigma = 0$ Hamiltonian: 
\begin{align} \begin{aligned}
	A^{(0)}(\omega; x_3,x_0) \delta(\omega-\omega')
		&= \frac{\Braket{ \psi_{\omega'} | e^{iP_0(x_3-x_0)} | \psi_\omega }_0} {\Braket{e^{iP_0(x_3-x_0)}}_0}
	\\	&= e^{i\omega (x_3-x_0)} \braket{ \psi_{\omega'} | \psi_\omega }_0 \,.
\end{aligned} \end{align}
We thus obtain the expected result $A^{(0)}(\omega; x_3,x_0) = e^{i\omega(x_3-x_0)}$.

\begin{widetext}
The first-order correction to the amplitude is 
\begin{align} \begin{aligned}
	t_\sigma A^{(1)}(\omega; x_3,x_0) \delta(\omega-\omega')
		&= -i \Braket{ \psi_{\omega'}(x_3) | \int_{-\infty}^\infty \!\!\!dt' \Big[ \mathcal{H}_{\rm tun}(t') - \braket{ \mathcal{H}_{\rm tun} }_0 \Big] | \psi_\omega(x_0) }_0
	\\	&= -i t_\sigma e^{-i\pi h_\sigma} \Braket{ \psi_{\omega'}(x_3) | \int_{-\infty}^\infty \!\!\!dt' \Big[ \sigma(x_2,t') \, \sigma(x_1,t') - \Delta \Big] | \psi_\omega(x_0) }_0 .
\end{aligned} \end{align}
Here $\Delta$ is a constant, defined through $t_\sigma e^{-i\pi h_\sigma} \Delta = \braket{ \mathcal{H}_{\rm tun} }_0$, chosen to cancel off the phase correction to the vacuum.
We also let $\ket{\psi_\omega(x)} = \int \frac{dt}{2\pi} e^{-i\omega t}\psi(x,t)\ket{0}$ denote a (temporal) plane wave at position $x$.
Expanding the scattering states in terms of their integral definitions and the operators in the Schr\"odinger picture yields 
\begin{align}
	A^{(1)}(\omega;x_3,x_0) \delta(\omega-\omega')
	&= \frac{-i e^{-i\pi h_\sigma}}{(2\pi)^2} \int_{t_2,t_1,t'}\mkern-29mu
		e^{i(\omega't_2-\omega t_1)} \Braket{ \psi(t_2)  \, e^{iP(x_3-x_2)} \Big[ \sigma(t') e^{iPL_a} \sigma(t') - \Delta e^{iPL_a} \Big] e^{iP(x_1-x_0)} \, \psi(t_1) }_0 .
\end{align}
Above we used $x_2-x_1 = L_a$.
As the scattering states are eigenstates of the momentum operator, we can replace $e^{iP(x_3-x_2)}$ and $e^{iP(x_1-x_0)}$ with their respective eigenvalues $e^{i\omega'(x_3-x_2)}$ and $e^{i\omega(x_1-x_0)}$.
In addition, we can eliminate the factor $\delta(\omega-\omega')$ on left-hand side by integrating over $\omega$ on both sides; doing so fixes $t_1 = 0$ and eliminates one of the integrals on the right-hand side.
With some simple substitution of variables, the first-order amplitude correction can now be written as
\begin{align}
	A^{(1)}(\omega;x_3,x_0) = -\frac{i e^{-i\pi h_\sigma} e^{i\omega(x_3-x_0-L_a)}}{2\pi}
		\int_{t_2,t'}\!\!\! e^{i\omega t_2} \Braket{ \psi(t_2) \Big[ \sigma(t')\,e^{iPL_a}\,\sigma(t') - \Delta e^{iPL_a} \Big] \psi(0) }_0 .
	\label{eq:IsingT_2int}
\end{align}
From Eq.~\eqref{eq:sigma_corr} one finds $\Delta = \bigXp{\sigma(t')\,e^{iPL_a}\,\sigma(t')}_0 = (-iL_a)^{-1/8}$.

%%%%%%%%%%%%%%%%%%%%%%%%%%%%%%%%%%%%%%%%%%%%%%%%
\subsection{Evaluation of the first-order correction}

We will now evaluate the first-order contribution $A^{(1)}(\omega)$ by first integrating over $t'$ and then integrating over $t_2$.
Let
\begin{align} \begin{aligned}
	\mathcal{I}(t') &= \BigXp{ \psi(t_2-i\epsilon) \Big[ \sigma(t')\,e^{iPL_a}\,\sigma(t') - \Delta e^{iPL_a} \Big] \psi(i\epsilon) }_0
	\\	&= \bigXp{ \psi(t_2-L_a-i\epsilon) \, \sigma(t'-L_a) \, \sigma(t') \, \psi(i\epsilon) }_0 - \bigXp{ \sigma(t'-L_a) \, \sigma(t')  }_0 \bigXp{ \psi(t_2-L_a-i\epsilon) \, \psi(i\epsilon) }_0
\end{aligned} \end{align}
be the regulated form the integrand in Eq.~\eqref{eq:IsingT_2int} (without the $e^{i\omega t_2}$ factor).
We will take $\epsilon \to 0^+$ at the end of the calculation.
The correlation function $\mathcal{I}$ can be computed via standard CFT techniques~\cite{MooreRead, NayakWilczek:2nStatesQHPf:96}.
For instance, one can evaluate $\braket{\psi(z_1) \sigma(z_2) \sigma(z_3) \psi(z_4)}$ using a conformal transformation that maps the plane to a cylinder, placing the $\sigma$ fields at $t = \pm\infty$. 
The correlator then reduces to computing the two-point correlation function $\braket{ \psi(x,t) \psi(x',t') }_\text{cyl}$ with the fermion field $\psi$ having periodic boundary conditions.
Undoing the conformal transformation yields the desired result:
\begin{align}
	\mathcal{I}(t')
	&= \frac{(-iL_a)^{-1/8}} {2(2\epsilon+it_2-iL_a)} \left[
			\sqrt{\frac{ (\epsilon+it_2-it')(\epsilon+it') }{ (\epsilon+it_2-iL_a-it')(\epsilon-iL_a+it') }}
			+ \sqrt{\frac{ (\epsilon+it_2-iL_a-it')(\epsilon-iL_a+it') }{ (\epsilon+it_2-it')(\epsilon+it') }}
			- 2 \right] .
	\label{eq:IsingT_integrand}
\end{align}
Notice that $\int_{t'} \mathcal{I}$ is absolutely convergent since $\mathcal{I}(t')$ decays as $\mathcal{O}\big((t')^{-4}\big)$ for large $t'$.
This convergence results from the $\Delta$ subtraction (corresponding to the $-2$ term in brackets), which eliminates the leading contribution in $\mathcal{I}(t')$.

The function $\mathcal{I}(t')$ has four branch points at $t_2-L_a-i\epsilon$, $t_2-i\epsilon$, $i\epsilon$, and $L_a+i\epsilon$.
Observe that the first two sit below the real axis while the latter two sit above the real axis.
The $\int\! dt'$ integral is to be evaluated with a branch cut connecting the two upper branch points, and a branch cut connecting the two lower points, such that $\mathcal{I}$ is analytic along the real line.
To simplify the terms in brackets we introduce a shift of variables $t' \mapsto t'+\frac{t_2}{2}$:
\begin{align}
	\mathcal{I} \big(t'+\tfrac{t_2}{2}\big) &= \frac{1}{2 (-iL_a)^{1/8} (2\epsilon+it_2-iL_a)} \left[
			  \sqrt{\frac{(t')^2 + a^2}{(t')^2 + b^2}} + \sqrt{\frac{(t')^2 + b^2}{(t')^2 + a^2}} - 2 \right]
\end{align}
with $a = \epsilon + it_2/2$, $b = \epsilon + it_2/2 - iL_a$.
At this point we can utilize the integral identity in Eq.~\eqref{eq:eq:4branch_integral} from Appendix~\ref{app:ellipticintegral} to write
\begin{align} \begin{aligned}
	\int_{-\infty}^\infty \!\!\! dt' \, \mathcal{I}\left(t'+\tfrac{t_2}{2}\right)
	&= \frac{1}{2 (-iL_a)^{1/8} (2\epsilon+it_2-iL_a)} \times 4(a+b) \left[ K\Big(\frac{a-b}{a+b}\Big) - E\Big(\frac{a-b}{a+b}\Big) \right] .
\\	&=	\frac{2i^{1/8}}{L_a^{1/8}} \left[ K\bigg(\frac{L_a}{t_2-L_a-2i\epsilon}\bigg) - E\bigg(\frac{L_a}{t_2-L_a-2i\epsilon}\bigg) \right] ,
\end{aligned} \end{align}
where $K$ and $E$ are complete elliptic integrals of the first and second kind, respectively.
Inserting this result into Eq.~\eqref{eq:IsingT_2int} yields
\begin{align} \begin{aligned}
	A^{(1)}(\omega;x_3,x_0)
		&= \lim_{\epsilon\to0^+} -\frac{i e^{-i\pi h_\sigma} e^{i\omega(x_3-x_0-L_a)}}{2\pi}
			\int_{-\infty}^\infty \!\!\!dt_2 \, e^{i\omega t_2} \frac{2i^{1/8}}{L_a^{1/8}} \left[ K\bigg(\frac{L_a}{t_2-L_a-2i\epsilon}\bigg) - E\bigg(\frac{L_a}{t_2-L_a-2i\epsilon}\bigg) \right]
	\\	&= \lim_{\epsilon'\to0^+} 2i L_a^{7/8} e^{i\omega (x_3-x_0)} \int_{-\infty}^\infty \frac{d y}{2\pi} e^{i(\omega L_a)y} \left[ E\bigg(\frac{1}{y-i\epsilon'}\bigg) - K\bigg(\frac{1}{y-i\epsilon'}\bigg) \right] .
\end{aligned} \end{align}
\end{widetext}
From the first to the second line, we substituted $t_2 = L_a(y+1)$ and introduced $\epsilon' \propto \epsilon$ as the small parameter to be taken to zero.
Finally, we write the amplitude as
\begin{align}
	A^{(1)}(\omega;x_3,x_0) &= 2i L_a^{7/8} e^{i\omega(x_3-x_0)} g(\omega L_a) ,
\end{align}
where $g(u)$ is defined as a Fourier transform via
\begin{subequations} \begin{align}
	g(u) &= \lim_{\epsilon\to0^+} \int_{-\infty}^\infty \frac{dy}{2\pi} e^{iuy} \tilde{g}(y - i\epsilon) ,
\label{gdef} \\	
\tilde{g}(y) &= E\big(\tfrac{1}{y}\big) - K\big(\tfrac{1}{y}\big) .
\end{align} \end{subequations}

States $\ket{\psi_\omega}$ with negative frequencies do not exist in a chiral CFT, so strictly speaking the amplitude is ill-defined for $\omega < 0$, and hence $g(u<0)$ is ill-defined as well.
Nevertheless, it is convenient to now extend the domain of $g(u)$ defined in Eq.~\eqref{gdef} to all real $u$.  
[We caution that one should not confuse this continuation with that adopted later in Eq.~\eqref{Gc}, which serves a quite different purpose.]
The function $g(u)$ extended in this way vanishes for $u < 0$, since $\tilde{g}(y)$ has no singularities in the lower half plane ($\operatorname{Im}y < 0$).
In addition, $\lim_{\epsilon\to0^+} \operatorname{Re}\tilde{g}(y - i\epsilon)$ is symmetric in $y$ while $\lim_{\epsilon\to0^+} \operatorname{Im}\tilde{g}(y - i\epsilon)$ is antisymmetric.
Together these properties allow us to write $g(u)$ over the physical domain $u\geq 0$ in terms of simply the imaginary part of $\tilde{g}$:
\begin{subequations} \label{eq:IsingT_g_sin} \begin{align}
	g(u\geq0) &= \frac{2}{\pi} \int_0^\infty \!\!\! dy \, \sin(uy) \, \tilde{g}_i(y) ,
\\	\tilde{g}_i(y) &= -\lim_{\epsilon\to0^+} \operatorname{Im} \tilde{g}(y - i\epsilon) .
\end{align} \end{subequations}
For $y > 1$ or $y < -1$, $\tilde{g}(y)$ is purely real because $1/y$ lies within the interval $-1 < \tfrac{1}{y} < 1$ (cf.~Appendix~\ref{app:ellipticintegral}).
Using Eqs.~\eqref{eq:EllipticK_branch} and~\eqref{eq:EllipticE_branch}, the imaginary component $\tilde{g}_i$ can be written as
\begin{align}
	\tilde{g}_i(y) &= \begin{cases} 0 & |y| > 1 ,  \\  \tfrac{1}{y} E\Big(\sqrt{1-y^2}\Big) & |y| \leq 1 ,  \end{cases}
\end{align}
which is supported on the finite interval $-1 \leq y \leq 1$.
From Eqs.~\eqref{eq:IsingT_g_sin} we can now re-express $g(u\geq 0)$ as
\begin{align}
	g(u\geq0) &= \frac{2}{\pi} \int_0^1 \! dy \frac{\sin(uy)}{y} E\Big(\sqrt{1-y^2}\Big) .
\end{align}
Evidently $g$ is a purely real function.
As shown in Appendix~\ref{app:EllipticE_int}, it can be written in terms of a generalized hypergeometric function,
\begin{align}
	g(u\geq 0) = \dfrac{\pi u}{4} \HGpFqb{1}{2}{\tfrac{1}{2}}{1,2}{-\tfrac{1}{4}u^2}
	\label{eq:IsingT_g}
\end{align}
plotted in Fig.~\ref{gf_fig}(a).

Upon replacing $\omega \rightarrow k_e$ to match the notation from Sec.~\ref{sec:interf_1psi} and setting $x_3-x_0 = L_e$, the zeroth- and first-order terms in the transmission amplitude are
\begin{subequations} \begin{align}
	A^{(0)}(k_e) &= e^{i k_e L_e} ,
\\	t_\sigma A^{(1)}(k_e;x_3,x_0) &= e^{i k_e L_e} \times 2i t_\sigma L_a^{7/8} g(k_e L_a).
\end{align} \end{subequations}
Restoring $v_e$ factors recovers precisely the $t_\sigma$ correction quoted in Eqs.~\eqref{Ag} and \eqref{gHGF} in the $n_\psi = 0$ case.

%%%%%%%%%%%%%%%%%%%%%%%%%%%%%%%%%%%%%%%%%%%%%%%%
\section{A few facts regarding complete elliptic integrals}
\label{app:ellipticintegral}

Let $K(k)$ and $E(k)$ denote the complete elliptic integrals of the first and second kind, respectively.
For $|k| < 1$ they are defined as
\begin{subequations} \begin{align}
	K(k) &= \frac{1}{4} \int_0^{2\pi} \frac{d\theta}{\sqrt{1 - k^2\cos^2\theta}} \,,
\\	E(k) &= \frac{1}{4} \int_0^{2\pi} \!\!d\theta \sqrt{1 - k^2\cos^2\theta} \,;
\end{align} \end{subequations}
beyond the unit circle they are defined via analytical continuation.  
Both are even [e.g., $K(-k) = K(k)$] and have branch cuts along the real line at $k \leq -1$ and $k \geq 1$.

Define $\bar{x} = \sqrt{1-x^2}$.
For $0 < x < 1$, these functions satisfy the algebraic identities~\cite{EllipticK:08.02.04.0009.01, EllipticK:08.02.17.0001.01, EllipticE:08.01.04.0009.01, EllipticE:08.01.17.0002.01}
\begin{align}
	\lim_{\epsilon\to0^+} \frac{K\big(\tfrac{1}{x}+i\epsilon\big) - K\big(\tfrac{1}{x}-i\epsilon\big)}{2} &= i x \mkern1mu K(\bar{x}) ,
	\label{eq:EllipticK_branch}
\\	\lim_{\epsilon\to0^+} \frac{E\big(\tfrac{1}{x}+i\epsilon\big) - E\big(\tfrac{1}{x}-i\epsilon\big)}{2} &= i \big[x \mkern1mu K(\bar{x}) - \tfrac{1}{x}E(\bar{x})\big] 
	\label{eq:EllipticE_branch}
\end{align}
along with the differential and integral identities
\begin{subequations} \begin{align}
	\frac{d}{dx} E(\bar{x}) &= \frac{x}{1-x^2}\big[ K(\bar{x}) - E(\bar{x}) \big] ,
	\label{eq:EllipticE_d}
\\	\frac{d}{dx} \big[ E(\bar{x}) - K(\bar{x}) \big] &= \frac{1}{x} E(\bar{x}) ,
	\label{eq:EllipticEK_d}
\\	\int_0^1 \! dx \, E(\bar{x}) &= \int_0^1 \! dx \, \frac{x}{\bar{x}} E(x) = \frac{\pi^2}{8} .
	\label{eq:EllipticE_someint1}
\end{align} \end{subequations}

%%%%%%%%%%%%%%%%%%%%%%%%%%%%%%%%%%%%%%%%%%%%%%%%
\subsection{An integral identity}

Suppose that $a,b$ are complex numbers such that $\operatorname{Re}a > 0$ and $\operatorname{Re}b > 0$.  
Following hours of struggle with \texttt{Mathematica} one can show that
\begin{align} \begin{aligned}
&	\int_{-\infty}^\infty \!\!\! dz \left[ \sqrt{\frac{z^2 + a^2}{z^2 + b^2}} + \sqrt{\frac{z^2 + b^2}{z^2 + a^2}} - 2 \right]
\\&\quad = 4(a+b) \left[ K\Big(\frac{a-b}{a+b}\Big) - E\Big(\frac{a-b}{a+b}\Big) \right] .
	\label{eq:eq:4branch_integral}
\end{aligned} \end{align}
To be precise, the integrand has branch points at $\pm ia$ and $\pm ib$; the integral is evaluated assuming branch cuts between $ia \leftrightarrow ib$ and $-ia \leftrightarrow -ib$ with one pair above the real line and the other pair below.

%%%%%%%%%%%%%%%%%%%%%%%%%%%%%%%%%%%%%%%%%%%%%%%%
\subsection{A different integral identity}
\label{app:EllipticE_int}
Here we show that the integral
\begin{align}
	g(u) \overset{\text{def}}{=} \frac{2}{\pi} \int_0^1 \! dy \frac{\sin(uy)}{y} E\Big(\sqrt{1-y^2}\Big)
	\label{eq:EllipticE_int_g}
\end{align}
can be expressed as a generalized hypergeometric function
\begin{align}
	g(u) &\overset{?}{=} \frac{\pi u}{4} \HGpFqb{1}{2}{\tfrac{1}{2}}{1,2}{-\tfrac{1}{4}u^2}  \notag
	\\	&= \frac{\pi u}{4} \sum_{m=0}^\infty \frac{1}{m!} \frac{ \big(\tfrac12) \big(\tfrac32) \cdots \big(\tfrac{2m-1}{2}\big) }{ m! \, (m+1)! } \big({-}\tfrac{1}{4}u^2\big)^m  \notag
	\\	&= \frac{\pi u}{4} \sum_{m=0}^\infty \frac{ (-1)^m (2m-1)!! }{ 4^m m! \, (m+1)! \, (2m)!! } u^{2m} .
	\label{eq:EllipticE_int_1F2}
\end{align}
To do so we Taylor expand Eq.~\eqref{eq:EllipticE_int_g} in powers of $u$ and show that it takes the form of Eq.~\eqref{eq:EllipticE_int_1F2}.

Notice that terms with even powers vanish in \eqref{eq:EllipticE_int_g}.
The coefficients for odd powers $u^{2m+1}$ are given by
\begin{align}
	c_{2m+1} = \frac{2}{\pi} \frac{(-1)^m}{(2m+1)!}M_m
\end{align}
with 
\begin{align}
  M_m = \int_0^1 \! dy \, y^{2m} E\Big(\sqrt{1-y^2}\Big).
\end{align}
One can evaluate $M_m$ as follows.
Performing integration by parts twice and using Eqs.~\eqref{eq:EllipticE_d} and~\eqref{eq:EllipticEK_d} yields $M_m = (2m+1)^2M_m - (4m^2-1)M_{m-1}$, implying 
the recursion relation 
\begin{align}
  M_m = \frac{(2m+1)(2m-1)}{4m(m+1)}M_{m-1}.
\end{align}
Since $M_0 = \frac{\pi^2}{8}$ from Eq.~\eqref{eq:EllipticE_someint1}, we can deduce that 
  $4^m m! \, (m+1)! \, M_m = \frac{\pi^2}{8} (2m-1)!! \, (2m+1)!!$.
Therefore
\begin{align}
	c_{2m+1} &= \frac{\pi}{4} (-1)^m \frac{(2m-1)!!}{4^{m} m! \, (m+1)! \, (2m)!!},
\end{align}
which indeeds matches the coefficients in the series expansion in Eq.~\eqref{eq:EllipticE_int_1F2}.

%%%%%%%%%%%%%%%%%%%%%%%%%%%%%%%%%%%%%%%%%%%%%%%%
\section{Extraction of Ising-anyon tunneling weights}
\label{app:Ising_tunneling_inversion}

Here we will formally invert Eq.~\eqref{gu} to extract the scaling functions $f_{iv}$ and $f_v$ that quantify energy partitioning in Ising-anyon tunneling events.  
First we define
\begin{equation}
  G(u) = 2i g(u) u^{7/8}
\end{equation}
and
\begin{equation}
  F(y) = \begin{cases}
f_{v}(y),  &y>0\\
f_{iv}(-y), &-1<y<0\\
0, &y<-1
\end{cases}
\end{equation}
so that Eq.~\eqref{gu} can be compactly expressed as
\begin{align}
  G(u) &= \int_{-1}^\infty \! dy \, e^{i u y} F(y).
  \label{Gu}
\end{align}
Since $G(u)$ is defined only for $u>0$, one can not exploit standard plane-wave orthogonality to isolate $F(y)$.
To proceed we continue $G(u)$ to $u<0$.  
For clarity we denote the resulting function defined for all real $u$ by $G_c(u)$.  

Care must be taken in defining this continuation to ensure that $F(y<-1)$ remains zero as demanded by our physical system.
Consider the integral
\begin{equation}
  \mathcal{F}(y) \equiv \int_0^\infty \frac{du}{2\pi} e^{-i u y} e^{-\epsilon u}G(u),
\end{equation}
where we introduced a regulator with $\epsilon \rightarrow 0^+$ for convergence, which is necessary given that $G(u) \sim u^{7/8}$ as $u \rightarrow \infty$.  
Evaluating this integral gives a result in terms of a generalized hypergeometric function.  
Here we simply note that
\begin{subequations} \begin{align}
  |\mathcal{F}(y)| &= |\mathcal{F}(-y)| ,
  \label{Feven} \\
  \mathcal{F}(y>1) &= -e^{i \frac{\pi}{16}}|\mathcal{F}(y>1)| ,
  \label{Fg} \\
  \mathcal{F}(y<-1) &= e^{-i \frac{\pi}{16}}|\mathcal{F}(y<-1)|.
  \label{Fl}
\end{align} \end{subequations}
These crucial properties imply that the desired continuation is given by
\begin{equation}
  G_c(u) = \begin{cases}
G(u),  &u>0\\
e^{-i \frac{\pi}{8}}G(|u|), &u<0
\end{cases}.
\label{Gc}
\end{equation}
It follows that
\begin{align}
  \int_{-\infty}^\infty \frac{du}{2\pi} e^{-i u y}e^{-\epsilon |u|} G_c(u) = \mathcal{F}(y) + e^{-i \frac{\pi}{8}}\mathcal{F}(-y) = F(y),
  \label{final_inversion}
\end{align}
where again we introduced a regulator with $\epsilon \rightarrow 0^+$.  
By virtue of Eqs.~\eqref{Feven} through \eqref{Fl}, we see that $F(y<-1)$ indeed vanishes.  
Taking the $\epsilon \rightarrow 0^+$ limit allows us to explicitly deduce
\begin{align}
  &e^{i \frac{9\pi}{16}} f_{iv}(y) =\frac{ 2 \, \Gamma(\frac{15}{8}) }{\pi y^{15/8}} \HGpFqb{3}{2}{-\tfrac12,\tfrac12,\tfrac12}{-\tfrac{7}{16},\tfrac{1}{16}}{y^2}
			\nonumber \\	
			&~~~~\;\mathrel{-} \frac{7^2 \sqrt{\pi} \, \Gamma(\frac{1}{16})}{2^{25/8}\cdot15 \, \Gamma(\frac{9}{16})^3} \HGpFqb{3}{2}{\tfrac{7}{16},\tfrac{23}{16},\tfrac{23}{16}}{\tfrac{1}{2},\tfrac{31}{16}}{y^2}
			\nonumber \\	
			&~~~~\;\mathrel{+} \frac{2^{15/8}\cdot15^2 \sqrt{\pi} \, \Gamma(\frac{9}{16}) \, y}{7\cdot23 \, \Gamma(\frac{1}{16})^3} \HGpFqb{3}{2}{\tfrac{15}{16},\tfrac{31}{16},\tfrac{31}{16}}{\tfrac{3}{2},\tfrac{39}{16}}{y^2}
\end{align}
and
\begin{align}
  &	e^{i \frac{9\pi}{16}} f_v(y<1) =
  \nonumber \\
  &~~~~ \; \frac{ 2\cos(\frac{\pi}{8}) \, \Gamma(\frac{15}{8}) }{\pi y^{15/8}} \HGpFqb{3}{2}{-\tfrac12,\tfrac12,\tfrac12}{-\tfrac{7}{16},\tfrac{1}{16}}{y^2}
			\nonumber \\	
			& ~~~~\;\mathrel{-} \frac{7^2 \sqrt{\pi} \, \Gamma(\frac{1}{16})}{2^{25/8}\cdot15 \, \Gamma(\frac{9}{16})^3} \HGpFqb{3}{2}{\tfrac{7}{16},\tfrac{23}{16},\tfrac{23}{16}}{\tfrac{1}{2},\tfrac{31}{16}}{y^2}
			\nonumber \\	
			&~~~~\;\mathrel{+} \frac{2^{15/8}\cdot15^2 \sqrt{\pi} \, \Gamma(\frac{9}{16}) \, y}{7\cdot23 \, \Gamma(\frac{1}{16})^3} \HGpFqb{3}{2}{\tfrac{15}{16},\tfrac{31}{16},\tfrac{31}{16}}{\tfrac{3}{2},\tfrac{39}{16}}{y^2} ,
			\\
  &	e^{i \frac{9\pi}{16}} f_v(y>1) = \dfrac{\sin(\tfrac{\pi}{8}) \, \Gamma(\tfrac{23}{8})}{2 y^{23/8}} \HGpFqb{3}{2}{\tfrac12,\tfrac{23}{16},\tfrac{31}{16}}{1,2}{\tfrac{1}{y^2}}.		
\end{align}
[Recall that $f_{iv}(y)$ is defined for $0<y<1$ while $f_v(y)$ is defined for $y>0$.]
We have thus completed the desired inversion.
Figure~\ref{gf_fig}(b) from the main text plots the magnitude of the scaling functions $f_{iv}$ and $f_v$.

Some limits of $F(y)$ can be deduced from the asymptotics specified in Eqs.~\eqref{gsmall} and \eqref{glarge}.  
The small-$y$ singularities follow from the leading $G(u\gg 1)$ behavior and can be obtained by simply replacing $G(u) \rightarrow 2i u^{7/8}$ in Eq.~\eqref{final_inversion}.
One finds
\begin{align} \begin{split}
  F\big( |y|\ll1 \big) &\approx \frac{\Gamma\left(\frac{15}{8}\right)}{\pi}\left[\frac{e^{i \frac{\pi}{2}}}{(\epsilon+i y)^{15/8}} + \frac{e^{i \frac{3\pi}{8}}}{(\epsilon-i y)^{15/8}}\right] 
\\	&\mkern-10mu \approx \begin{cases}
  -i e^{-i \frac{\pi}{16}}\left[\frac{2}{\pi}\cos(\frac{\pi}{8})\Gamma(\tfrac{15}{8}) \right] y^{-15/8}  &y>0\\
  -i e^{-i \frac{\pi}{16}}\left[\frac{2}{\pi}\Gamma(\tfrac{15}{8})\right] |y|^{-15/8} &y<0
\end{cases}.
  \label{Fsmall}
\end{split} \end{align}
In the bottom lines we took the $\epsilon \rightarrow 0^+$ limit.  
We caution, however, that inserting the bottom lines of Eq.~\eqref{Fsmall} into Eq.~\eqref{Gu} would produce an unphysical infrared divergence; hence in the top line we explicitly displayed the regularization that circumvents this problem. 
Elsewhere we are free to send $\epsilon \rightarrow 0^+$ as no such issues arise.
The singularities at $y \rightarrow \pm 1^+$ instead follow from the subleading $G(u \gg 1)$ behavior; they can be captured by replacing $G(u) \rightarrow -2i u^{-1/8} (\cos u)$ in Eq.~\eqref{final_inversion}, yielding
\begin{equation}
  F(y \rightarrow \pm 1^+) \approx -i e^{-i \frac{\pi}{16}}\left[\frac{1}{\Gamma\left(\frac{1}{8}\right)}\right](y\mp 1)^{-7/8}.
\end{equation}
Finally, the asymptotic decay at $y \gg 1$ encodes the $G(u \ll 1)$ behavior and follows from replacing $G(u) \rightarrow i \frac{\pi}{2} u^{15/8}$ in Eq.~\eqref{Gu}:
\begin{equation}
  F(y \gg 1) \approx -i e^{-i \frac{\pi}{16}}\left[\frac{\pi}{2\,\Gamma\big({-}\frac{15}{8}\big)}\right] y^{-23/8}.
\end{equation}
Figure~\ref{gf_fig}(b) indicates the scaling behaviors captured above.

%%%%%%%%%%%%%%%%%%%%%%%%%%%%%%%%%%%%%%%%%%%%%%%%
\section{Conductance from Majorana phase accumulation}
\label{ConductanceApp}

For completeness, we will briefly review how electrical conductance follows from the relative phases acquired by Majorana fermions propagating in the circuits from Figs.~\ref{interference} and \ref{psidetector}.
We denote the part of the wavefunction describing an electron incident at the lower edge of the $\nu = 1$ quantum Hall system by
\begin{equation}
  \int_{-\infty}^{x_0} dx \, e^{i E x/u}c^\dagger(x)\ket{0}.
  \label{incoming}
\end{equation}
Here $\ket{0}$ is the ground state, $E$ is the incident energy, $u$ is the edge velocity in the region without induced pairing, $c^\dagger(x)$ adds the electron to position $x$ of the edge, and $x_0$ is the location at which the edge state meets the proximitizing superconductor.
(In this appendix we use coordinates consistent with those in Fig.~\ref{coords_fig}.)
Employing a Majorana representation via $c= \gamma_1+i \gamma_2$, Eq.~\eqref{incoming} equivalently becomes
\begin{equation}
  \int_{-\infty}^{x_0}dx \, e^{i E x/u}[\gamma_1(x)-i \gamma_2(x)]\ket{0}.
  \label{incoming2}
\end{equation}

Beyond position $x_0$, the constituent Majorana fermions $\gamma_1$ and $\gamma_2$ follow diverging paths that eventually recombine at position $x_3$ at the upper edge of the $\nu = 1$ quantum Hall system.
En route they generally acquire different phase factors; hence at the upper $\nu = 1$ edge the outgoing part of the wavefunction becomes
\begin{equation}
  e^{i \bar \phi(E)}\int_{x_3}^{-\infty} \mkern-10mu dx \, e^{i E x/u} \big[\gamma_1(x)-i e^{i \delta \phi(E)}\gamma_2(x)\big] \ket{0},
  \label{outgoing}
\end{equation}
where $\bar \phi(E)$ denotes a possible phase common to both Majorana fermions (which is unimportant here) and $\delta \phi(E)$ is the accumulated phase difference.
Reverting back to complex fermions by writing $\gamma_1 = (c + c^\dagger)/2$ and $\gamma_2 = -i(c-c^\dagger)/2$, Eq.~\eqref{outgoing} reads
\begin{align} \begin{split}
  &e^{i \bar \phi(E)}\int_{x_3}^{-\infty} \mkern-10mu dx \, e^{i E x/u}\bigg{[}\left(\frac{1+ e^{i \delta \phi(E)}}{2}\right)c^\dagger(x) 
  \\&\mkern150mu + \left(\frac{1- e^{i \delta \phi(E)}}{2}\right)c(x)\bigg{]}\ket{0}.
  \label{outgoing2}
\end{split} \end{align}
Thus with probability 
\begin{equation}
  P_A(E) = \left|\frac{1- e^{i \delta \phi(E)}}{2}\right|^2 = \frac{1-\cos[\delta \phi(E)]}{2}
\end{equation}
the incident electron returns at the upper edge as a hole---transmitting a Cooper pair into the superconductor.  
The conductance at bias voltage $V$ arising from such Andreev processes is
\begin{equation}
  G(V) = \frac{2e^2}{h} P_A(eV).
\end{equation}

\hbadness=10000	% suppress underfull hbox warnings from this point forward
\bibliography{references}

\begin{thebibliography}{123}%
\makeatletter
\providecommand \@ifxundefined [1]{%
 \@ifx{#1\undefined}
}%
\providecommand \@ifnum [1]{%
 \ifnum #1\expandafter \@firstoftwo
 \else \expandafter \@secondoftwo
 \fi
}%
\providecommand \@ifx [1]{%
 \ifx #1\expandafter \@firstoftwo
 \else \expandafter \@secondoftwo
 \fi
}%
\providecommand \natexlab [1]{#1}%
\providecommand \enquote  [1]{``#1''}%
\providecommand \bibnamefont  [1]{#1}%
\providecommand \bibfnamefont [1]{#1}%
\providecommand \citenamefont [1]{#1}%
\providecommand \href@noop [0]{\@secondoftwo}%
\providecommand \href [0]{\begingroup \@sanitize@url \@href}%
\providecommand \@href[1]{\@@startlink{#1}\@@href}%
\providecommand \@@href[1]{\endgroup#1\@@endlink}%
\providecommand \@sanitize@url [0]{\catcode `\\12\catcode `\$12\catcode
  `\&12\catcode `\#12\catcode `\^12\catcode `\_12\catcode `\%12\relax}%
\providecommand \@@startlink[1]{}%
\providecommand \@@endlink[0]{}%
\providecommand \url  [0]{\begingroup\@sanitize@url \@url }%
\providecommand \@url [1]{\endgroup\@href {#1}{\urlprefix }}%
\providecommand \urlprefix  [0]{URL }%
\providecommand \Eprint [0]{\href }%
\providecommand \doibase [0]{http://dx.doi.org/}%
\providecommand \selectlanguage [0]{\@gobble}%
\providecommand \bibinfo  [0]{\@secondoftwo}%
\providecommand \bibfield  [0]{\@secondoftwo}%
\providecommand \translation [1]{[#1]}%
\providecommand \BibitemOpen [0]{}%
\providecommand \bibitemStop [0]{}%
\providecommand \bibitemNoStop [0]{.\EOS\space}%
\providecommand \EOS [0]{\spacefactor3000\relax}%
\providecommand \BibitemShut  [1]{\csname bibitem#1\endcsname}%
\let\auto@bib@innerbib\@empty
%</preamble>
\bibitem [{\citenamefont {Kitaev}(2003)}]{Kitaev:2003}%
  \BibitemOpen
  \bibfield  {author} {\bibinfo {author} {\bibfnamefont {Alexei~Yu}\
  \bibnamefont {Kitaev}},\ }\bibfield  {title} {\enquote {\bibinfo {title}
  {{Fault-tolerant quantum computation by anyons}},}\ }\href {\doibase%
  10.1016/S0003-4916(02)00018-0} {\bibfield  {journal} {\bibinfo  {journal}
  {Ann.\ Phys.}\ }\textbf {\bibinfo {volume} {303}},\ \bibinfo {pages} {2--30}
  (\bibinfo {year} {2003})}\BibitemShut {NoStop}%
\bibitem [{\citenamefont {Nayak}\ \emph {et~al.}(2008)\citenamefont {Nayak},
  \citenamefont {Simon}, \citenamefont {Stern}, \citenamefont {Freedman},\ and\
  \citenamefont {Das~Sarma}}]{TQCreview}%
  \BibitemOpen
  \bibfield  {author} {\bibinfo {author} {\bibfnamefont {Chetan}\ \bibnamefont
  {Nayak}}, \bibinfo {author} {\bibfnamefont {Steven~H.}\ \bibnamefont
  {Simon}}, \bibinfo {author} {\bibfnamefont {Ady}\ \bibnamefont {Stern}},
  \bibinfo {author} {\bibfnamefont {Michael}\ \bibnamefont {Freedman}}, and\
  \bibinfo {author} {\bibfnamefont {Sankar}\ \bibnamefont {Das~Sarma}},\
  }\bibfield  {title} {\enquote {\bibinfo {title} {Non-{Abelian} anyons and
  topological quantum computation},}\ }\href {\doibase%
  10.1103/RevModPhys.80.1083} {\bibfield  {journal} {\bibinfo  {journal} {Rev.
  Mod. Phys.}\ }\textbf {\bibinfo {volume} {80}},\ \bibinfo {pages}
  {1083--1159} (\bibinfo {year} {2008})}\BibitemShut {NoStop}%
\bibitem [{\citenamefont {Willett}\ \emph {et~al.}(1987)\citenamefont
  {Willett}, \citenamefont {Eisenstein}, \citenamefont {St\"ormer},
  \citenamefont {Tsui}, \citenamefont {Gossard},\ and\ \citenamefont
  {English}}]{Willett1987}%
  \BibitemOpen
  \bibfield  {author} {\bibinfo {author} {\bibfnamefont {R.}~\bibnamefont
  {Willett}}, \bibinfo {author} {\bibfnamefont {J.~P.}\ \bibnamefont
  {Eisenstein}}, \bibinfo {author} {\bibfnamefont {H.~L.}\ \bibnamefont
  {St\"ormer}}, \bibinfo {author} {\bibfnamefont {D.~C.}\ \bibnamefont {Tsui}},
  \bibinfo {author} {\bibfnamefont {A.~C.}\ \bibnamefont {Gossard}}, and\
  \bibinfo {author} {\bibfnamefont {J.~H.}\ \bibnamefont {English}},\
  }\bibfield  {title} {\enquote {\bibinfo {title} {Observation of an
  even-denominator quantum number in the fractional quantum {Hall} effect},}\
  }\href {\doibase 10.1103/PhysRevLett.59.1776} {\bibfield  {journal} {\bibinfo
   {journal} {Phys. Rev. Lett.}\ }\textbf {\bibinfo {volume} {59}},\ \bibinfo
  {pages} {1776--1779} (\bibinfo {year} {1987})}\BibitemShut {NoStop}%
\bibitem [{\citenamefont {Moore}\ and\ \citenamefont {Read}(1991)}]{MooreRead}%
  \BibitemOpen
  \bibfield  {author} {\bibinfo {author} {\bibfnamefont {Gregory}\ \bibnamefont
  {Moore}}\ and\ \bibinfo {author} {\bibfnamefont {Nicholas}\ \bibnamefont
  {Read}},\ }\bibfield  {title} {\enquote {\bibinfo {title} {Nonabelions in the
  fractional quantum {H}all effect},}\ }\href {\doibase%
  10.1016/0550-3213(91)90407-O} {\bibfield  {journal} {\bibinfo  {journal}
  {Nucl. Phys. B}\ }\textbf {\bibinfo {volume} {360}},\ \bibinfo {pages}
  {362--396} (\bibinfo {year} {1991})}\BibitemShut {NoStop}%
\bibitem [{\citenamefont {Levin}\ \emph {et~al.}(2007)\citenamefont {Levin},
  \citenamefont {Halperin},\ and\ \citenamefont {Rosenow}}]{Levin2007}%
  \BibitemOpen
  \bibfield  {author} {\bibinfo {author} {\bibfnamefont {Michael}\ \bibnamefont
  {Levin}}, \bibinfo {author} {\bibfnamefont {Bertrand~I.}\ \bibnamefont
  {Halperin}}, and\ \bibinfo {author} {\bibfnamefont {Bernd}\ \bibnamefont
  {Rosenow}},\ }\bibfield  {title} {\enquote {\bibinfo {title} {Particle-hole
  symmetry and the {Pfaffian} state},}\ }\href {\doibase%
  10.1103/PhysRevLett.99.236806} {\bibfield  {journal} {\bibinfo  {journal}
  {Phys. Rev. Lett.}\ }\textbf {\bibinfo {volume} {99}},\ \bibinfo {pages}
  {236806} (\bibinfo {year} {2007})}\BibitemShut {NoStop}%
\bibitem [{\citenamefont {Lee}\ \emph {et~al.}(2007)\citenamefont {Lee},
  \citenamefont {Ryu}, \citenamefont {Nayak},\ and\ \citenamefont
  {Fisher}}]{Lee2007}%
  \BibitemOpen
  \bibfield  {author} {\bibinfo {author} {\bibfnamefont {Sung-Sik}\
  \bibnamefont {Lee}}, \bibinfo {author} {\bibfnamefont {Shinsei}\ \bibnamefont
  {Ryu}}, \bibinfo {author} {\bibfnamefont {Chetan}\ \bibnamefont {Nayak}},\
  and\ \bibinfo {author} {\bibfnamefont {Matthew P.~A.}\ \bibnamefont
  {Fisher}},\ }\bibfield  {title} {\enquote {\bibinfo {title} {Particle-hole
  symmetry and the $\ensuremath{\nu}=\frac{5}{2}$ quantum {Hall} state},}\
  }\href {\doibase 10.1103/PhysRevLett.99.236807} {\bibfield  {journal}
  {\bibinfo  {journal} {Phys. Rev. Lett.}\ }\textbf {\bibinfo {volume} {99}},\
  \bibinfo {pages} {236807} (\bibinfo {year} {2007})}\BibitemShut {NoStop}%
\bibitem [{\citenamefont {Son}(2015)}]{SonDiracCFL}%
  \BibitemOpen
  \bibfield  {author} {\bibinfo {author} {\bibfnamefont {Dam~Thanh}\
  \bibnamefont {Son}},\ }\bibfield  {title} {\enquote {\bibinfo {title} {Is the
  composite fermion a {Dirac} particle?}}\ }\href {\doibase%
  10.1103/PhysRevX.5.031027} {\bibfield  {journal} {\bibinfo  {journal} {Phys.
  Rev. X}\ }\textbf {\bibinfo {volume} {5}},\ \bibinfo {pages} {031027}
  (\bibinfo {year} {2015})}\BibitemShut {NoStop}%
\bibitem [{\citenamefont {Banerjee}\ \emph
  {et~al.}(2018{\natexlab{a}})\citenamefont {Banerjee}, \citenamefont
  {Heiblum}, \citenamefont {Umansky}, \citenamefont {Feldman}, \citenamefont
  {Oreg},\ and\ \citenamefont {Stern}}]{Banerjee}%
  \BibitemOpen
  \bibfield  {author} {\bibinfo {author} {\bibfnamefont {Mitali}\ \bibnamefont
  {Banerjee}}, \bibinfo {author} {\bibfnamefont {Moty}\ \bibnamefont
  {Heiblum}}, \bibinfo {author} {\bibfnamefont {Vladimir}\ \bibnamefont
  {Umansky}}, \bibinfo {author} {\bibfnamefont {Dima~E.}\ \bibnamefont
  {Feldman}}, \bibinfo {author} {\bibfnamefont {Yuval}\ \bibnamefont {Oreg}},\
  and\ \bibinfo {author} {\bibfnamefont {Ady}\ \bibnamefont {Stern}},\
  }\bibfield  {title} {\enquote {\bibinfo {title} {Observation of half-integer
  thermal {Hall} conductance},}\ }\href {\doibase 10.1038/s41586-018-0184-1}
  {\bibfield  {journal} {\bibinfo  {journal} {Nature}\ }\textbf {\bibinfo
  {volume} {559}},\ \bibinfo {pages} {205} (\bibinfo {year}
  {2018}{\natexlab{a}})}\BibitemShut {NoStop}%
\bibitem [{\citenamefont {Das~Sarma}\ \emph {et~al.}(2005)\citenamefont
  {Das~Sarma}, \citenamefont {Freedman},\ and\ \citenamefont
  {Nayak}}]{Nayak2005}%
  \BibitemOpen
  \bibfield  {author} {\bibinfo {author} {\bibfnamefont {Sankar}\ \bibnamefont
  {Das~Sarma}}, \bibinfo {author} {\bibfnamefont {Michael}\ \bibnamefont
  {Freedman}}, and\ \bibinfo {author} {\bibfnamefont {Chetan}\ \bibnamefont
  {Nayak}},\ }\bibfield  {title} {\enquote {\bibinfo {title} {Topologically
  protected qubits from a possible non-{A}belian fractional quantum {H}all
  state},}\ }\href {\doibase 10.1103/PhysRevLett.94.166802} {\bibfield
  {journal} {\bibinfo  {journal} {Phys. Rev. Lett.}\ }\textbf {\bibinfo
  {volume} {94}},\ \bibinfo {pages} {166802} (\bibinfo {year}
  {2005})}\BibitemShut {NoStop}%
\bibitem [{\citenamefont {Stern}\ and\ \citenamefont
  {Halperin}(2006)}]{Stern2006}%
  \BibitemOpen
  \bibfield  {author} {\bibinfo {author} {\bibfnamefont {Ady}\ \bibnamefont
  {Stern}}\ and\ \bibinfo {author} {\bibfnamefont {Bertrand~I.}\ \bibnamefont
  {Halperin}},\ }\bibfield  {title} {\enquote {\bibinfo {title} {Proposed
  experiments to probe the non-{A}belian $\ensuremath{\nu}=5/2$ quantum {H}all
  state},}\ }\href {\doibase 10.1103/PhysRevLett.96.016802} {\bibfield
  {journal} {\bibinfo  {journal} {Phys. Rev. Lett.}\ }\textbf {\bibinfo
  {volume} {96}},\ \bibinfo {pages} {016802} (\bibinfo {year}
  {2006})}\BibitemShut {NoStop}%
\bibitem [{\citenamefont {Bonderson}\ \emph {et~al.}(2006)\citenamefont
  {Bonderson}, \citenamefont {Kitaev},\ and\ \citenamefont
  {Shtengel}}]{Bonderson2006}%
  \BibitemOpen
  \bibfield  {author} {\bibinfo {author} {\bibfnamefont {Parsa}\ \bibnamefont
  {Bonderson}}, \bibinfo {author} {\bibfnamefont {Alexei}\ \bibnamefont
  {Kitaev}}, and\ \bibinfo {author} {\bibfnamefont {Kirill}\ \bibnamefont
  {Shtengel}},\ }\bibfield  {title} {\enquote {\bibinfo {title} {Detecting
  non-{A}belian statistics in the $\ensuremath{\nu}=5/2$ fractional quantum
  {H}all state},}\ }\href {\doibase 10.1103/PhysRevLett.96.016803} {\bibfield
  {journal} {\bibinfo  {journal} {Phys. Rev. Lett.}\ }\textbf {\bibinfo
  {volume} {96}},\ \bibinfo {pages} {016803} (\bibinfo {year}
  {2006})}\BibitemShut {NoStop}%
\bibitem [{\citenamefont {Willett}\ \emph {et~al.}(2019)\citenamefont
  {Willett}, \citenamefont {Shtengel}, \citenamefont {Nayak}, \citenamefont
  {Pfeiffer}, \citenamefont {Chung}, \citenamefont {Peabody}, \citenamefont
  {Baldwin},\ and\ \citenamefont {West}}]{Willett2019}%
  \BibitemOpen
  \bibfield  {author} {\bibinfo {author} {\bibfnamefont {R.~L.}\ \bibnamefont
  {Willett}}, \bibinfo {author} {\bibfnamefont {K.}~\bibnamefont {Shtengel}},
  \bibinfo {author} {\bibfnamefont {C.}~\bibnamefont {Nayak}}, \bibinfo
  {author} {\bibfnamefont {L.~N.}\ \bibnamefont {Pfeiffer}}, \bibinfo {author}
  {\bibfnamefont {Y.~J.}\ \bibnamefont {Chung}}, \bibinfo {author}
  {\bibfnamefont {M.~L.}\ \bibnamefont {Peabody}}, \bibinfo {author}
  {\bibfnamefont {K.~W.}\ \bibnamefont {Baldwin}}, and\ \bibinfo {author}
  {\bibfnamefont {K.~W.}\ \bibnamefont {West}},\ }\href@noop {} {\enquote
  {\bibinfo {title} {Interference measurements of non-{Abelian} $e/4$ and
  {Abelian} $e/2$ quasiparticle braiding},}\ } (\bibinfo {year} {2019}),\
  \bibinfo {note} {unpublished},\ \Eprint {http://arxiv.org/abs/1905.10248}
  {arXiv:1905.10248} \BibitemShut {NoStop}%
\bibitem [{\citenamefont {Read}\ and\ \citenamefont {Green}(2000)}]{ReadGreen}%
  \BibitemOpen
  \bibfield  {author} {\bibinfo {author} {\bibfnamefont {N.}~\bibnamefont
  {Read}}\ and\ \bibinfo {author} {\bibfnamefont {Dmitry}\ \bibnamefont
  {Green}},\ }\bibfield  {title} {\enquote {\bibinfo {title} {Paired states of
  fermions in two dimensions with breaking of parity and time-reversal
  symmetries and the fractional quantum {H}all effect},}\ }\href {\doibase%
  10.1103/PhysRevB.61.10267} {\bibfield  {journal} {\bibinfo  {journal} {Phys.
  Rev. B}\ }\textbf {\bibinfo {volume} {61}},\ \bibinfo {pages} {10267--10297}
  (\bibinfo {year} {2000})}\BibitemShut {NoStop}%
\bibitem [{\citenamefont {Kitaev}(2001)}]{Kitaev:2001}%
  \BibitemOpen
  \bibfield  {author} {\bibinfo {author} {\bibfnamefont {Alexei~Yu}\
  \bibnamefont {Kitaev}},\ }\bibfield  {title} {\enquote {\bibinfo {title}
  {{Unpaired Majorana fermions in quantum wires}},}\ }\href {\doibase%
  10.1070/1063-7869/44/10S/S29} {\bibfield  {journal} {\bibinfo  {journal}
  {Sov. Phys.--Uspeki}\ }\textbf {\bibinfo {volume} {44}},\ \bibinfo {pages}
  {131} (\bibinfo {year} {2001})}\BibitemShut {NoStop}%
\bibitem [{\citenamefont {Hasan}\ and\ \citenamefont {Kane}(2010)}]{HasanKane}%
  \BibitemOpen
  \bibfield  {author} {\bibinfo {author} {\bibfnamefont {M.~Z.}\ \bibnamefont
  {Hasan}}\ and\ \bibinfo {author} {\bibfnamefont {C.~L.}\ \bibnamefont
  {Kane}},\ }\bibfield  {title} {\enquote {\bibinfo {title} {Colloquium:
  Topological insulators},}\ }\href {\doibase 10.1103/RevModPhys.82.3045}
  {\bibfield  {journal} {\bibinfo  {journal} {Rev. Mod. Phys.}\ }\textbf
  {\bibinfo {volume} {82}},\ \bibinfo {pages} {3045--3067} (\bibinfo {year}
  {2010})}\BibitemShut {NoStop}%
\bibitem [{\citenamefont {Qi}\ and\ \citenamefont {Zhang}(2011)}]{QiZhang}%
  \BibitemOpen
  \bibfield  {author} {\bibinfo {author} {\bibfnamefont {Xiao-Liang}\
  \bibnamefont {Qi}}\ and\ \bibinfo {author} {\bibfnamefont {Shou-Cheng}\
  \bibnamefont {Zhang}},\ }\bibfield  {title} {\enquote {\bibinfo {title}
  {Topological insulators and superconductors},}\ }\href {\doibase%
  10.1103/RevModPhys.83.1057} {\bibfield  {journal} {\bibinfo  {journal} {Rev.
  Mod. Phys.}\ }\textbf {\bibinfo {volume} {83}},\ \bibinfo {pages}
  {1057--1110} (\bibinfo {year} {2011})}\BibitemShut {NoStop}%
\bibitem [{\citenamefont {Beenakker}(2013)}]{BeenakkerReview}%
  \BibitemOpen
  \bibfield  {author} {\bibinfo {author} {\bibfnamefont {C.~W.~J.}\
  \bibnamefont {Beenakker}},\ }\bibfield  {title} {\enquote {\bibinfo {title}
  {Search for {Majorana} fermions in superconductors},}\ }\href {\doibase%
  10.1146/annurev-conmatphys-030212-184337} {\bibfield  {journal} {\bibinfo
  {journal} {Annu. Rev. Con. Mat. Phys.}\ }\textbf {\bibinfo {volume} {4}},\
  \bibinfo {pages} {113--136} (\bibinfo {year} {2013})}\BibitemShut {NoStop}%
\bibitem [{\citenamefont {Alicea}(2012)}]{AliceaReview}%
  \BibitemOpen
  \bibfield  {author} {\bibinfo {author} {\bibfnamefont {Jason}\ \bibnamefont
  {Alicea}},\ }\bibfield  {title} {\enquote {\bibinfo {title} {New directions
  in the pursuit of {Majorana} fermions in solid state systems},}\ }\href
  {\doibase 10.1088/0034-4885/75/7/076501} {\bibfield  {journal} {\bibinfo
  {journal} {Reports on Progress in Physics}\ }\textbf {\bibinfo {volume}
  {75}},\ \bibinfo {pages} {076501} (\bibinfo {year} {2012})}\BibitemShut
  {NoStop}%
\bibitem [{\citenamefont {Leijnse}\ and\ \citenamefont
  {Flensberg}(2012)}]{FlensbergReview}%
  \BibitemOpen
  \bibfield  {author} {\bibinfo {author} {\bibfnamefont {Martin}\ \bibnamefont
  {Leijnse}}\ and\ \bibinfo {author} {\bibfnamefont {Karsten}\ \bibnamefont
  {Flensberg}},\ }\bibfield  {title} {\enquote {\bibinfo {title} {Introduction
  to topological superconductivity and {Majorana} fermions},}\ }\href {\doibase%
  10.1088/0268-1242/27/12/124003} {\bibfield  {journal} {\bibinfo  {journal}
  {Semicond. Sci. Technol.}\ }\textbf {\bibinfo {volume} {27}},\ \bibinfo
  {pages} {124003} (\bibinfo {year} {2012})}\BibitemShut {NoStop}%
\bibitem [{\citenamefont {Stanescu}\ and\ \citenamefont
  {Tewari}(2013)}]{TewariReview}%
  \BibitemOpen
  \bibfield  {author} {\bibinfo {author} {\bibfnamefont {Tudor~D.}\
  \bibnamefont {Stanescu}}\ and\ \bibinfo {author} {\bibfnamefont {Sumanta}\
  \bibnamefont {Tewari}},\ }\bibfield  {title} {\enquote {\bibinfo {title}
  {{Majorana} fermions in semiconductor nanowires: Fundamentals, modeling, and
  experiment},}\ }\href {\doibase 10.1088/0953-8984/25/23/233201} {\bibfield
  {journal} {\bibinfo  {journal} {J. Phys.: Condens. Matter}\ }\textbf
  {\bibinfo {volume} {25}},\ \bibinfo {pages} {233201} (\bibinfo {year}
  {2013})}\BibitemShut {NoStop}%
\bibitem [{\citenamefont {Elliott}\ and\ \citenamefont
  {Franz}(2015)}]{FranzReview}%
  \BibitemOpen
  \bibfield  {author} {\bibinfo {author} {\bibfnamefont {Steven~R.}\
  \bibnamefont {Elliott}}\ and\ \bibinfo {author} {\bibfnamefont {Marcel}\
  \bibnamefont {Franz}},\ }\bibfield  {title} {\enquote {\bibinfo {title}
  {\textit{Colloquium} : {Majorana} fermions in nuclear, particle, and
  solid-state physics},}\ }\href {\doibase 10.1103/RevModPhys.87.137}
  {\bibfield  {journal} {\bibinfo  {journal} {Rev. Mod. Phys.}\ }\textbf
  {\bibinfo {volume} {87}},\ \bibinfo {pages} {137--163} (\bibinfo {year}
  {2015})}\BibitemShut {NoStop}%
\bibitem [{\citenamefont {{Das Sarma}}\ \emph {et~al.}(2015)\citenamefont {{Das
  Sarma}}, \citenamefont {Freedman},\ and\ \citenamefont
  {Nayak}}]{ChetanReview}%
  \BibitemOpen
  \bibfield  {author} {\bibinfo {author} {\bibfnamefont {Sankar}\ \bibnamefont
  {{Das Sarma}}}, \bibinfo {author} {\bibfnamefont {Michael}\ \bibnamefont
  {Freedman}}, and\ \bibinfo {author} {\bibfnamefont {Chetan}\ \bibnamefont
  {Nayak}},\ }\bibfield  {title} {\enquote {\bibinfo {title} {{Majorana} zero
  modes and topological quantum computation},}\ }\href {\doibase%
  10.1038/npjqi.2015.1} {\bibfield  {journal} {\bibinfo  {journal} {npj Quantum
  Information}\ }\textbf {\bibinfo {volume} {1}},\ \bibinfo {pages} {15001}
  (\bibinfo {year} {2015})}\BibitemShut {NoStop}%
\bibitem [{\citenamefont {Sato}\ and\ \citenamefont
  {Fujimoto}(2016)}]{SatoReview}%
  \BibitemOpen
  \bibfield  {author} {\bibinfo {author} {\bibfnamefont {Masatoshi}\
  \bibnamefont {Sato}}\ and\ \bibinfo {author} {\bibfnamefont {Satoshi}\
  \bibnamefont {Fujimoto}},\ }\bibfield  {title} {\enquote {\bibinfo {title}
  {Majorana fermions and topology in superconductors},}\ }\href {\doibase%
  10.7566/JPSJ.85.072001} {\bibfield  {journal} {\bibinfo  {journal} {Journal
  of the Physical Society of Japan}\ }\textbf {\bibinfo {volume} {85}},\
  \bibinfo {pages} {072001} (\bibinfo {year} {2016})}\BibitemShut {NoStop}%
\bibitem [{\citenamefont {Aguado}(2017)}]{AguadoReview}%
  \BibitemOpen
  \bibfield  {author} {\bibinfo {author} {\bibfnamefont {Ramon}\ \bibnamefont
  {Aguado}},\ }\bibfield  {title} {\enquote {\bibinfo {title} {Majorana
  quasiparticles in condensed matter},}\ }\href {\doibase%
  10.1393/ncr/i2017-10141-9} {\bibfield  {journal} {\bibinfo  {journal} {Riv.
  Nuovo Cimento}\ }\textbf {\bibinfo {volume} {40}},\ \bibinfo {pages} {523}
  (\bibinfo {year} {2017})}\BibitemShut {NoStop}%
\bibitem [{\citenamefont {Lutchyn}\ \emph {et~al.}(2018)\citenamefont
  {Lutchyn}, \citenamefont {Bakkers}, \citenamefont {Kouwenhoven},
  \citenamefont {Krogstrup}, \citenamefont {Marcus},\ and\ \citenamefont
  {Oreg}}]{LutchynReview}%
  \BibitemOpen
  \bibfield  {author} {\bibinfo {author} {\bibfnamefont {R.~M.}\ \bibnamefont
  {Lutchyn}}, \bibinfo {author} {\bibfnamefont {E.~P. A.~M.}\ \bibnamefont
  {Bakkers}}, \bibinfo {author} {\bibfnamefont {L.~P.}\ \bibnamefont
  {Kouwenhoven}}, \bibinfo {author} {\bibfnamefont {P.}~\bibnamefont
  {Krogstrup}}, \bibinfo {author} {\bibfnamefont {C.~M.}\ \bibnamefont
  {Marcus}}, and\ \bibinfo {author} {\bibfnamefont {Y.}~\bibnamefont
  {Oreg}},\ }\bibfield  {title} {\enquote {\bibinfo {title} {Majorana zero
  modes in superconductor/semiconductor heterostructures},}\ }\href {\doibase%
  10.1038/s41578-018-0003-1} {\bibfield  {journal} {\bibinfo  {journal} {Nature
  Reviews Materials}\ }\textbf {\bibinfo {volume} {3}},\ \bibinfo {pages} {52}
  (\bibinfo {year} {2018})}\BibitemShut {NoStop}%
\bibitem [{\citenamefont {Fu}\ and\ \citenamefont
  {Kane}(2009)}]{FuKaneInterferometer}%
  \BibitemOpen
  \bibfield  {author} {\bibinfo {author} {\bibfnamefont {Liang}\ \bibnamefont
  {Fu}}\ and\ \bibinfo {author} {\bibfnamefont {C.~L.}\ \bibnamefont {Kane}},\
  }\bibfield  {title} {\enquote {\bibinfo {title} {Probing neutral {M}ajorana
  fermion edge modes with charge transport},}\ }\href {\doibase%
  10.1103/PhysRevLett.102.216403} {\bibfield  {journal} {\bibinfo  {journal}
  {Phys. Rev. Lett.}\ }\textbf {\bibinfo {volume} {102}},\ \bibinfo {pages}
  {216403} (\bibinfo {year} {2009})}\BibitemShut {NoStop}%
\bibitem [{\citenamefont {Akhmerov}\ \emph {et~al.}(2009)\citenamefont
  {Akhmerov}, \citenamefont {Nilsson},\ and\ \citenamefont
  {Beenakker}}]{AkhmerovInterferometer}%
  \BibitemOpen
  \bibfield  {author} {\bibinfo {author} {\bibfnamefont {A.~R.}\ \bibnamefont
  {Akhmerov}}, \bibinfo {author} {\bibfnamefont {Johan}\ \bibnamefont
  {Nilsson}}, and\ \bibinfo {author} {\bibfnamefont {C.~W.~J.}\ \bibnamefont
  {Beenakker}},\ }\bibfield  {title} {\enquote {\bibinfo {title} {Electrically
  detected interferometry of {M}ajorana fermions in a topological insulator},}\
  }\href {\doibase 10.1103/PhysRevLett.102.216404} {\bibfield  {journal}
  {\bibinfo  {journal} {Phys. Rev. Lett.}\ }\textbf {\bibinfo {volume} {102}},\
  \bibinfo {pages} {216404} (\bibinfo {year} {2009})}\BibitemShut {NoStop}%
\bibitem [{\citenamefont {Law}\ \emph {et~al.}(2009)\citenamefont {Law},
  \citenamefont {Lee},\ and\ \citenamefont {Ng}}]{Law}%
  \BibitemOpen
  \bibfield  {author} {\bibinfo {author} {\bibfnamefont {K.~T.}\ \bibnamefont
  {Law}}, \bibinfo {author} {\bibfnamefont {Patrick~A.}\ \bibnamefont {Lee}},\
  and\ \bibinfo {author} {\bibfnamefont {T.~K.}\ \bibnamefont {Ng}},\
  }\bibfield  {title} {\enquote {\bibinfo {title} {{Majorana} fermion induced
  resonant {Andreev} reflection},}\ }\href {\doibase%
  10.1103/PhysRevLett.103.237001} {\bibfield  {journal} {\bibinfo  {journal}
  {Phys. Rev. Lett.}\ }\textbf {\bibinfo {volume} {103}},\ \bibinfo {pages}
  {237001} (\bibinfo {year} {2009})}\BibitemShut {NoStop}%
\bibitem [{\citenamefont {Fu}(2010)}]{FuTeleportation}%
  \BibitemOpen
  \bibfield  {author} {\bibinfo {author} {\bibfnamefont {Liang}\ \bibnamefont
  {Fu}},\ }\bibfield  {title} {\enquote {\bibinfo {title} {Electron
  teleportation via {Majorana} bound states in a mesoscopic superconductor},}\
  }\href {\doibase 10.1103/PhysRevLett.104.056402} {\bibfield  {journal}
  {\bibinfo  {journal} {Phys. Rev. Lett.}\ }\textbf {\bibinfo {volume} {104}},\
  \bibinfo {pages} {056402} (\bibinfo {year} {2010})}\BibitemShut {NoStop}%
\bibitem [{\citenamefont {Chung}\ \emph {et~al.}(2011)\citenamefont {Chung},
  \citenamefont {Qi}, \citenamefont {Maciejko},\ and\ \citenamefont
  {Zhang}}]{Chung2011}%
  \BibitemOpen
  \bibfield  {author} {\bibinfo {author} {\bibfnamefont {Suk~Bum}\ \bibnamefont
  {Chung}}, \bibinfo {author} {\bibfnamefont {Xiao-Liang}\ \bibnamefont {Qi}},
  \bibinfo {author} {\bibfnamefont {Joseph}\ \bibnamefont {Maciejko}}, and\
  \bibinfo {author} {\bibfnamefont {Shou-Cheng}\ \bibnamefont {Zhang}},\
  }\bibfield  {title} {\enquote {\bibinfo {title} {Conductance and noise
  signatures of {M}ajorana backscattering},}\ }\href {\doibase%
  10.1103/PhysRevB.83.100512} {\bibfield  {journal} {\bibinfo  {journal} {Phys.
  Rev. B}\ }\textbf {\bibinfo {volume} {83}},\ \bibinfo {pages} {100512}
  (\bibinfo {year} {2011})}\BibitemShut {NoStop}%
\bibitem [{\citenamefont {Karzig}\ \emph {et~al.}(2017)\citenamefont {Karzig},
  \citenamefont {Knapp}, \citenamefont {Lutchyn}, \citenamefont {Bonderson},
  \citenamefont {Hastings}, \citenamefont {Nayak}, \citenamefont {Alicea},
  \citenamefont {Flensberg}, \citenamefont {Plugge}, \citenamefont {Oreg},
  \citenamefont {Marcus},\ and\ \citenamefont {Freedman}}]{Karzig2017}%
  \BibitemOpen
  \bibfield  {author} {\bibinfo {author} {\bibfnamefont {Torsten}\ \bibnamefont
  {Karzig}}, \bibinfo {author} {\bibfnamefont {Christina}\ \bibnamefont
  {Knapp}}, \bibinfo {author} {\bibfnamefont {Roman~M.}\ \bibnamefont
  {Lutchyn}}, \bibinfo {author} {\bibfnamefont {Parsa}\ \bibnamefont
  {Bonderson}}, \bibinfo {author} {\bibfnamefont {Matthew~B.}\ \bibnamefont
  {Hastings}}, \bibinfo {author} {\bibfnamefont {Chetan}\ \bibnamefont
  {Nayak}}, \bibinfo {author} {\bibfnamefont {Jason}\ \bibnamefont {Alicea}},
  \bibinfo {author} {\bibfnamefont {Karsten}\ \bibnamefont {Flensberg}},
  \bibinfo {author} {\bibfnamefont {Stephan}\ \bibnamefont {Plugge}}, \bibinfo
  {author} {\bibfnamefont {Yuval}\ \bibnamefont {Oreg}}, \bibinfo {author}
  {\bibfnamefont {Charles~M.}\ \bibnamefont {Marcus}}, and\ \bibinfo {author}
  {\bibfnamefont {Michael~H.}\ \bibnamefont {Freedman}},\ }\bibfield  {title}
  {\enquote {\bibinfo {title} {Scalable designs for
  quasiparticle-poisoning-protected topological quantum computation with
  {M}ajorana zero modes},}\ }\href {\doibase 10.1103/PhysRevB.95.235305}
  {\bibfield  {journal} {\bibinfo  {journal} {Phys. Rev. B}\ }\textbf {\bibinfo
  {volume} {95}},\ \bibinfo {pages} {235305} (\bibinfo {year}
  {2017})}\BibitemShut {NoStop}%
\bibitem [{\citenamefont {Plumb}\ \emph {et~al.}(2014)\citenamefont {Plumb},
  \citenamefont {Clancy}, \citenamefont {Sandilands}, \citenamefont {Shankar},
  \citenamefont {Hu}, \citenamefont {Burch}, \citenamefont {Kee},\ and\
  \citenamefont {Kim}}]{Plumb2014}%
  \BibitemOpen
  \bibfield  {author} {\bibinfo {author} {\bibfnamefont {K.~W.}\ \bibnamefont
  {Plumb}}, \bibinfo {author} {\bibfnamefont {J.~P.}\ \bibnamefont {Clancy}},
  \bibinfo {author} {\bibfnamefont {L.~J.}\ \bibnamefont {Sandilands}},
  \bibinfo {author} {\bibfnamefont {V.~Vijay}\ \bibnamefont {Shankar}},
  \bibinfo {author} {\bibfnamefont {Y.~F.}\ \bibnamefont {Hu}}, \bibinfo
  {author} {\bibfnamefont {K.~S.}\ \bibnamefont {Burch}}, \bibinfo {author}
  {\bibfnamefont {Hae-Young}\ \bibnamefont {Kee}}, and\ \bibinfo {author}
  {\bibfnamefont {Young-June}\ \bibnamefont {Kim}},\ }\bibfield  {title}
  {\enquote {\bibinfo {title} {$\alpha$-$\mathrm{RuCl}_3$: A spin-orbit
  assisted {M}ott insulator on a honeycomb lattice},}\ }\href {\doibase%
  10.1103/PhysRevB.90.041112} {\bibfield  {journal} {\bibinfo  {journal} {Phys.
  Rev. B}\ }\textbf {\bibinfo {volume} {90}},\ \bibinfo {pages} {041112}
  (\bibinfo {year} {2014})}\BibitemShut {NoStop}%
\bibitem [{\citenamefont {Kim}\ \emph {et~al.}(2015)\citenamefont {Kim},
  \citenamefont {V.}, \citenamefont {Catuneanu},\ and\ \citenamefont
  {Kee}}]{Kim2015}%
  \BibitemOpen
  \bibfield  {author} {\bibinfo {author} {\bibfnamefont {Heung-Sik}\
  \bibnamefont {Kim}}, \bibinfo {author} {\bibfnamefont {Vijay~Shankar}\
  \bibnamefont {V.}}, \bibinfo {author} {\bibfnamefont {Andrei}\ \bibnamefont
  {Catuneanu}}, and\ \bibinfo {author} {\bibfnamefont {Hae-Young}\
  \bibnamefont {Kee}},\ }\bibfield  {title} {\enquote {\bibinfo {title} {Kitaev
  magnetism in honeycomb $\alpha$-$\mathrm{RuCl}_3$ with intermediate
  spin-orbit coupling},}\ }\href {\doibase 10.1103/PhysRevB.91.241110}
  {\bibfield  {journal} {\bibinfo  {journal} {Phys. Rev. B}\ }\textbf {\bibinfo
  {volume} {91}},\ \bibinfo {pages} {241110} (\bibinfo {year}
  {2015})}\BibitemShut {NoStop}%
\bibitem [{\citenamefont {Kitaev}(2006)}]{Kitaev2006}%
  \BibitemOpen
  \bibfield  {author} {\bibinfo {author} {\bibfnamefont {Alexei}\ \bibnamefont
  {Kitaev}},\ }\bibfield  {title} {\enquote {\bibinfo {title} {Anyons in an
  exactly solved model and beyond},}\ }\href {\doibase%
  10.1016/j.aop.2005.10.005} {\bibfield  {journal} {\bibinfo  {journal} {Annals
  of Physics}\ }\textbf {\bibinfo {volume} {321}},\ \bibinfo {pages} {2 -- 111}
  (\bibinfo {year} {2006})}\BibitemShut {NoStop}%
\bibitem [{\citenamefont {Jackeli}\ and\ \citenamefont
  {Khaliullin}(2009)}]{Jackeli2009}%
  \BibitemOpen
  \bibfield  {author} {\bibinfo {author} {\bibfnamefont {G.}~\bibnamefont
  {Jackeli}}\ and\ \bibinfo {author} {\bibfnamefont {G.}~\bibnamefont
  {Khaliullin}},\ }\bibfield  {title} {\enquote {\bibinfo {title} {Mott
  insulators in the strong spin-orbit coupling limit: From {H}eisenberg to a
  quantum compass and {K}itaev models},}\ }\href {\doibase%
  10.1103/PhysRevLett.102.017205} {\bibfield  {journal} {\bibinfo  {journal}
  {Phys. Rev. Lett.}\ }\textbf {\bibinfo {volume} {102}},\ \bibinfo {pages}
  {017205} (\bibinfo {year} {2009})}\BibitemShut {NoStop}%
\bibitem [{\citenamefont {Winter}\ \emph {et~al.}(2017)\citenamefont {Winter},
  \citenamefont {Tsirlin}, \citenamefont {Daghofer}, \citenamefont {van~den
  Brink}, \citenamefont {Singh}, \citenamefont {Gegenwart},\ and\ \citenamefont
  {Valent{\'{\i}}}}]{Winter2017}%
  \BibitemOpen
  \bibfield  {author} {\bibinfo {author} {\bibfnamefont {Stephen~M}\
  \bibnamefont {Winter}}, \bibinfo {author} {\bibfnamefont {Alexander~A}\
  \bibnamefont {Tsirlin}}, \bibinfo {author} {\bibfnamefont {Maria}\
  \bibnamefont {Daghofer}}, \bibinfo {author} {\bibfnamefont {Jeroen}\
  \bibnamefont {van~den Brink}}, \bibinfo {author} {\bibfnamefont {Yogesh}\
  \bibnamefont {Singh}}, \bibinfo {author} {\bibfnamefont {Philipp}\
  \bibnamefont {Gegenwart}}, and\ \bibinfo {author} {\bibfnamefont {Roser}\
  \bibnamefont {Valent{\'{\i}}}},\ }\bibfield  {title} {\enquote {\bibinfo
  {title} {Models and materials for generalized {K}itaev magnetism},}\ }\href
  {\doibase 10.1088/1361-648x/aa8cf5} {\bibfield  {journal} {\bibinfo
  {journal} {Journal of Physics: Condensed Matter}\ }\textbf {\bibinfo {volume}
  {29}},\ \bibinfo {pages} {493002} (\bibinfo {year} {2017})}\BibitemShut
  {NoStop}%
\bibitem [{\citenamefont {Trebst}(2017)}]{Trebst2017}%
  \BibitemOpen
  \bibfield  {author} {\bibinfo {author} {\bibfnamefont {Simon}\ \bibnamefont
  {Trebst}},\ }\href@noop {} {\enquote {\bibinfo {title} {Kitaev materials},}\
  } (\bibinfo {year} {2017}),\ \bibinfo {note} {unpublished},\ \Eprint
  {http://arxiv.org/abs/1701.07056} {arXiv:1701.07056} \BibitemShut {NoStop}%
\bibitem [{\citenamefont {Janssen}\ and\ \citenamefont
  {Vojta}(2019)}]{Janssen2019}%
  \BibitemOpen
  \bibfield  {author} {\bibinfo {author} {\bibfnamefont {Lukas}\ \bibnamefont
  {Janssen}}\ and\ \bibinfo {author} {\bibfnamefont {Matthias}\ \bibnamefont
  {Vojta}},\ }\bibfield  {title} {\enquote {\bibinfo {title}
  {Heisenberg{\textendash}{Kitaev} physics in magnetic fields},}\ }\href
  {\doibase 10.1088/1361-648x/ab283e} {\bibfield  {journal} {\bibinfo
  {journal} {Journal of Physics: Condensed Matter}\ }\textbf {\bibinfo {volume}
  {31}},\ \bibinfo {pages} {423002} (\bibinfo {year} {2019})}\BibitemShut
  {NoStop}%
\bibitem [{\citenamefont {Motome}\ and\ \citenamefont
  {Nasu}(2019)}]{Motome2019}%
  \BibitemOpen
  \bibfield  {author} {\bibinfo {author} {\bibfnamefont {Yukitoshi}\
  \bibnamefont {Motome}}\ and\ \bibinfo {author} {\bibfnamefont {Joji}\
  \bibnamefont {Nasu}},\ }\href@noop {} {\enquote {\bibinfo {title} {Hunting
  {Majorana} fermions in {Kitaev} magnets},}\ } (\bibinfo {year} {2019}),\
  \bibinfo {note} {unpublished},\ \Eprint {http://arxiv.org/abs/1909.02234}
  {arXiv:1909.02234} \BibitemShut {NoStop}%
\bibitem [{\citenamefont {Sears}\ \emph {et~al.}(2015)\citenamefont {Sears},
  \citenamefont {Songvilay}, \citenamefont {Plumb}, \citenamefont {Clancy},
  \citenamefont {Qiu}, \citenamefont {Zhao}, \citenamefont {Parshall},\ and\
  \citenamefont {Kim}}]{Sears2015}%
  \BibitemOpen
  \bibfield  {author} {\bibinfo {author} {\bibfnamefont {J.~A.}\ \bibnamefont
  {Sears}}, \bibinfo {author} {\bibfnamefont {M.}~\bibnamefont {Songvilay}},
  \bibinfo {author} {\bibfnamefont {K.~W.}\ \bibnamefont {Plumb}}, \bibinfo
  {author} {\bibfnamefont {J.~P.}\ \bibnamefont {Clancy}}, \bibinfo {author}
  {\bibfnamefont {Y.}~\bibnamefont {Qiu}}, \bibinfo {author} {\bibfnamefont
  {Y.}~\bibnamefont {Zhao}}, \bibinfo {author} {\bibfnamefont {D.}~\bibnamefont
  {Parshall}}, and\ \bibinfo {author} {\bibfnamefont {Young-June}\
  \bibnamefont {Kim}},\ }\bibfield  {title} {\enquote {\bibinfo {title}
  {Magnetic order in $\alpha$-$\mathrm{RuCl}_3$: A honeycomb-lattice quantum
  magnet with strong spin-orbit coupling},}\ }\href {\doibase%
  10.1103/PhysRevB.91.144420} {\bibfield  {journal} {\bibinfo  {journal} {Phys.
  Rev. B}\ }\textbf {\bibinfo {volume} {91}},\ \bibinfo {pages} {144420}
  (\bibinfo {year} {2015})}\BibitemShut {NoStop}%
\bibitem [{\citenamefont {Chaloupka}\ \emph {et~al.}(2013)\citenamefont
  {Chaloupka}, \citenamefont {Jackeli},\ and\ \citenamefont
  {Khaliullin}}]{Chaloupka2013}%
  \BibitemOpen
  \bibfield  {author} {\bibinfo {author} {\bibfnamefont {J.}~\bibnamefont
  {Chaloupka}}, \bibinfo {author} {\bibfnamefont {George}\ \bibnamefont
  {Jackeli}}, and\ \bibinfo {author} {\bibfnamefont {Giniyat}\ \bibnamefont
  {Khaliullin}},\ }\bibfield  {title} {\enquote {\bibinfo {title} {Zigzag
  magnetic order in the iridium oxide $\mathrm{Na}_{2}\mathrm{IrO}_{3}$},}\
  }\href {\doibase 10.1103/PhysRevLett.110.097204} {\bibfield  {journal}
  {\bibinfo  {journal} {Phys. Rev. Lett.}\ }\textbf {\bibinfo {volume} {110}},\
  \bibinfo {pages} {097204} (\bibinfo {year} {2013})}\BibitemShut {NoStop}%
\bibitem [{\citenamefont {Sandilands}\ \emph {et~al.}(2015)\citenamefont
  {Sandilands}, \citenamefont {Tian}, \citenamefont {Plumb}, \citenamefont
  {Kim},\ and\ \citenamefont {Burch}}]{Sandilands2015}%
  \BibitemOpen
  \bibfield  {author} {\bibinfo {author} {\bibfnamefont {Luke~J.}\ \bibnamefont
  {Sandilands}}, \bibinfo {author} {\bibfnamefont {Yao}\ \bibnamefont {Tian}},
  \bibinfo {author} {\bibfnamefont {Kemp~W.}\ \bibnamefont {Plumb}}, \bibinfo
  {author} {\bibfnamefont {Young-June}\ \bibnamefont {Kim}}, and\ \bibinfo
  {author} {\bibfnamefont {Kenneth~S.}\ \bibnamefont {Burch}},\ }\bibfield
  {title} {\enquote {\bibinfo {title} {Scattering continuum and possible
  fractionalized excitations in $\alpha$-$\mathrm{RuCl}_3$},}\ }\href {\doibase%
  10.1103/PhysRevLett.114.147201} {\bibfield  {journal} {\bibinfo  {journal}
  {Phys. Rev. Lett.}\ }\textbf {\bibinfo {volume} {114}},\ \bibinfo {pages}
  {147201} (\bibinfo {year} {2015})}\BibitemShut {NoStop}%
\bibitem [{\citenamefont {Nasu}\ \emph {et~al.}(2016)\citenamefont {Nasu},
  \citenamefont {Knolle}, \citenamefont {Kovrizhin}, \citenamefont {Motome},\
  and\ \citenamefont {Moessner}}]{Nasu2016}%
  \BibitemOpen
  \bibfield  {author} {\bibinfo {author} {\bibfnamefont {J.}~\bibnamefont
  {Nasu}}, \bibinfo {author} {\bibfnamefont {J.}~\bibnamefont {Knolle}},
  \bibinfo {author} {\bibfnamefont {D.~L.}\ \bibnamefont {Kovrizhin}}, \bibinfo
  {author} {\bibfnamefont {Y.}~\bibnamefont {Motome}}, and\ \bibinfo {author}
  {\bibfnamefont {R.}~\bibnamefont {Moessner}},\ }\bibfield  {title} {\enquote
  {\bibinfo {title} {Fermionic response from fractionalization in an insulating
  two-dimensional magnet},}\ }\href {\doibase 10.1038/nphys3809} {\bibfield
  {journal} {\bibinfo  {journal} {Nature Physics}\ }\textbf {\bibinfo {volume}
  {12}},\ \bibinfo {pages} {912--915} (\bibinfo {year} {2016})}\BibitemShut
  {NoStop}%
\bibitem [{\citenamefont {Banerjee}\ \emph {et~al.}(2016)\citenamefont
  {Banerjee}, \citenamefont {Bridges}, \citenamefont {Yan}, \citenamefont
  {Aczel}, \citenamefont {Li}, \citenamefont {Stone}, \citenamefont {Granroth},
  \citenamefont {Lumsden}, \citenamefont {Yiu}, \citenamefont {Knolle},
  \citenamefont {Bhattacharjee}, \citenamefont {Kovrizhin}, \citenamefont
  {Moessner}, \citenamefont {Tennant}, \citenamefont {Mandrus},\ and\
  \citenamefont {Nagler}}]{Banerjee2016}%
  \BibitemOpen
  \bibfield  {author} {\bibinfo {author} {\bibfnamefont {A.}~\bibnamefont
  {Banerjee}}, \bibinfo {author} {\bibfnamefont {C.~A.}\ \bibnamefont
  {Bridges}}, \bibinfo {author} {\bibfnamefont {J.-Q.}\ \bibnamefont {Yan}},
  \bibinfo {author} {\bibfnamefont {A.~A.}\ \bibnamefont {Aczel}}, \bibinfo
  {author} {\bibfnamefont {L.}~\bibnamefont {Li}}, \bibinfo {author}
  {\bibfnamefont {M.~B.}\ \bibnamefont {Stone}}, \bibinfo {author}
  {\bibfnamefont {G.~E.}\ \bibnamefont {Granroth}}, \bibinfo {author}
  {\bibfnamefont {M.~D.}\ \bibnamefont {Lumsden}}, \bibinfo {author}
  {\bibfnamefont {Y.}~\bibnamefont {Yiu}}, \bibinfo {author} {\bibfnamefont
  {J.}~\bibnamefont {Knolle}}, \bibinfo {author} {\bibfnamefont
  {S.}~\bibnamefont {Bhattacharjee}}, \bibinfo {author} {\bibfnamefont {D.~L.}\
  \bibnamefont {Kovrizhin}}, \bibinfo {author} {\bibfnamefont {R.}~\bibnamefont
  {Moessner}}, \bibinfo {author} {\bibfnamefont {D.~A.}\ \bibnamefont
  {Tennant}}, \bibinfo {author} {\bibfnamefont {D.~G.}\ \bibnamefont
  {Mandrus}}, and\ \bibinfo {author} {\bibfnamefont {S.~E.}\ \bibnamefont
  {Nagler}},\ }\bibfield  {title} {\enquote {\bibinfo {title} {Proximate
  {Kitaev} quantum spin liquid behaviour in a honeycomb magnet},}\ }\href
  {\doibase 10.1038/nmat4604} {\bibfield  {journal} {\bibinfo  {journal}
  {Nature Materials}\ }\textbf {\bibinfo {volume} {15}},\ \bibinfo {pages}
  {733} (\bibinfo {year} {2016})}\BibitemShut {NoStop}%
\bibitem [{\citenamefont {Banerjee}\ \emph {et~al.}(2017)\citenamefont
  {Banerjee}, \citenamefont {Yan}, \citenamefont {Knolle}, \citenamefont
  {Bridges}, \citenamefont {Stone}, \citenamefont {Lumsden}, \citenamefont
  {Mandrus}, \citenamefont {Tennant}, \citenamefont {Moessner},\ and\
  \citenamefont {Nagler}}]{Banerjee2017}%
  \BibitemOpen
  \bibfield  {author} {\bibinfo {author} {\bibfnamefont {Arnab}\ \bibnamefont
  {Banerjee}}, \bibinfo {author} {\bibfnamefont {Jiaqiang}\ \bibnamefont
  {Yan}}, \bibinfo {author} {\bibfnamefont {Johannes}\ \bibnamefont {Knolle}},
  \bibinfo {author} {\bibfnamefont {Craig~A.}\ \bibnamefont {Bridges}},
  \bibinfo {author} {\bibfnamefont {Matthew~B.}\ \bibnamefont {Stone}},
  \bibinfo {author} {\bibfnamefont {Mark~D.}\ \bibnamefont {Lumsden}}, \bibinfo
  {author} {\bibfnamefont {David~G.}\ \bibnamefont {Mandrus}}, \bibinfo
  {author} {\bibfnamefont {David~A.}\ \bibnamefont {Tennant}}, \bibinfo
  {author} {\bibfnamefont {Roderich}\ \bibnamefont {Moessner}}, and\ \bibinfo
  {author} {\bibfnamefont {Stephen~E.}\ \bibnamefont {Nagler}},\ }\bibfield
  {title} {\enquote {\bibinfo {title} {Neutron scattering in the proximate
  quantum spin liquid $\alpha$-{RuCl}$_3$},}\ }\href {\doibase%
  10.1126/science.aah6015} {\bibfield  {journal} {\bibinfo  {journal}
  {Science}\ }\textbf {\bibinfo {volume} {356}},\ \bibinfo {pages} {1055--1059}
  (\bibinfo {year} {2017})}\BibitemShut {NoStop}%
\bibitem [{\citenamefont {Do}\ \emph {et~al.}(2017)\citenamefont {Do},
  \citenamefont {Park}, \citenamefont {Yoshitake}, \citenamefont {Nasu},
  \citenamefont {Motome}, \citenamefont {Kwon}, \citenamefont {Adroja},
  \citenamefont {Voneshen}, \citenamefont {Kim}, \citenamefont {Jang},
  \citenamefont {Park}, \citenamefont {Choi},\ and\ \citenamefont
  {Ji}}]{Do2017}%
  \BibitemOpen
  \bibfield  {author} {\bibinfo {author} {\bibfnamefont {Seung-Hwan}\
  \bibnamefont {Do}}, \bibinfo {author} {\bibfnamefont {Sang-Youn}\
  \bibnamefont {Park}}, \bibinfo {author} {\bibfnamefont {Junki}\ \bibnamefont
  {Yoshitake}}, \bibinfo {author} {\bibfnamefont {Joji}\ \bibnamefont {Nasu}},
  \bibinfo {author} {\bibfnamefont {Yukitoshi}\ \bibnamefont {Motome}},
  \bibinfo {author} {\bibfnamefont {Yong~Seung}\ \bibnamefont {Kwon}}, \bibinfo
  {author} {\bibfnamefont {D.~T.}\ \bibnamefont {Adroja}}, \bibinfo {author}
  {\bibfnamefont {D.~J.}\ \bibnamefont {Voneshen}}, \bibinfo {author}
  {\bibfnamefont {Kyoo}\ \bibnamefont {Kim}}, \bibinfo {author} {\bibfnamefont
  {T.-H.}\ \bibnamefont {Jang}}, \bibinfo {author} {\bibfnamefont {J.-H.}\
  \bibnamefont {Park}}, \bibinfo {author} {\bibfnamefont {Kwang-Yong}\
  \bibnamefont {Choi}}, and\ \bibinfo {author} {\bibfnamefont {Sungdae}\
  \bibnamefont {Ji}},\ }\bibfield  {title} {\enquote {\bibinfo {title}
  {Majorana fermions in the {Kitaev} quantum spin system
  $\alpha$-$\mathrm{RuCl}_3$},}\ }\href {\doibase 10.1038/nphys4264} {\bibfield
   {journal} {\bibinfo  {journal} {Nature Physics}\ }\textbf {\bibinfo {volume}
  {13}},\ \bibinfo {pages} {1079} (\bibinfo {year} {2017})}\BibitemShut
  {NoStop}%
\bibitem [{\citenamefont {Kasahara}\ \emph
  {et~al.}(2018{\natexlab{a}})\citenamefont {Kasahara}, \citenamefont {Sugii},
  \citenamefont {Ohnishi}, \citenamefont {Shimozawa}, \citenamefont
  {Yamashita}, \citenamefont {Kurita}, \citenamefont {Tanaka}, \citenamefont
  {Nasu}, \citenamefont {Motome}, \citenamefont {Shibauchi},\ and\
  \citenamefont {Matsuda}}]{Kasahara2017}%
  \BibitemOpen
  \bibfield  {author} {\bibinfo {author} {\bibfnamefont {Y.}~\bibnamefont
  {Kasahara}}, \bibinfo {author} {\bibfnamefont {K.}~\bibnamefont {Sugii}},
  \bibinfo {author} {\bibfnamefont {T.}~\bibnamefont {Ohnishi}}, \bibinfo
  {author} {\bibfnamefont {M.}~\bibnamefont {Shimozawa}}, \bibinfo {author}
  {\bibfnamefont {M.}~\bibnamefont {Yamashita}}, \bibinfo {author}
  {\bibfnamefont {N.}~\bibnamefont {Kurita}}, \bibinfo {author} {\bibfnamefont
  {H.}~\bibnamefont {Tanaka}}, \bibinfo {author} {\bibfnamefont
  {J.}~\bibnamefont {Nasu}}, \bibinfo {author} {\bibfnamefont {Y.}~\bibnamefont
  {Motome}}, \bibinfo {author} {\bibfnamefont {T.}~\bibnamefont {Shibauchi}},\
  and\ \bibinfo {author} {\bibfnamefont {Y.}~\bibnamefont {Matsuda}},\
  }\bibfield  {title} {\enquote {\bibinfo {title} {Unusual thermal {H}all
  effect in a {K}itaev spin liquid candidate $\alpha$-$\mathrm{RuCl}_3$},}\
  }\href {\doibase 10.1103/PhysRevLett.120.217205} {\bibfield  {journal}
  {\bibinfo  {journal} {Phys. Rev. Lett.}\ }\textbf {\bibinfo {volume} {120}},\
  \bibinfo {pages} {217205} (\bibinfo {year} {2018}{\natexlab{a}})}\BibitemShut
  {NoStop}%
\bibitem [{\citenamefont {Wellm}\ \emph {et~al.}(2018)\citenamefont {Wellm},
  \citenamefont {Zeisner}, \citenamefont {Alfonsov}, \citenamefont {Wolter},
  \citenamefont {Roslova}, \citenamefont {Isaeva}, \citenamefont {Doert},
  \citenamefont {Vojta}, \citenamefont {B\"uchner},\ and\ \citenamefont
  {Kataev}}]{Wellm2018}%
  \BibitemOpen
  \bibfield  {author} {\bibinfo {author} {\bibfnamefont {C.}~\bibnamefont
  {Wellm}}, \bibinfo {author} {\bibfnamefont {J.}~\bibnamefont {Zeisner}},
  \bibinfo {author} {\bibfnamefont {A.}~\bibnamefont {Alfonsov}}, \bibinfo
  {author} {\bibfnamefont {A.~U.~B.}\ \bibnamefont {Wolter}}, \bibinfo {author}
  {\bibfnamefont {M.}~\bibnamefont {Roslova}}, \bibinfo {author} {\bibfnamefont
  {A.}~\bibnamefont {Isaeva}}, \bibinfo {author} {\bibfnamefont
  {T.}~\bibnamefont {Doert}}, \bibinfo {author} {\bibfnamefont
  {M.}~\bibnamefont {Vojta}}, \bibinfo {author} {\bibfnamefont
  {B.}~\bibnamefont {B\"uchner}}, and\ \bibinfo {author} {\bibfnamefont
  {V.}~\bibnamefont {Kataev}},\ }\bibfield  {title} {\enquote {\bibinfo {title}
  {Signatures of low-energy fractionalized excitations in
  $\alpha$-$\mathrm{RuCl}_3$ from field-dependent microwave absorption},}\
  }\href {\doibase 10.1103/PhysRevB.98.184408} {\bibfield  {journal} {\bibinfo
  {journal} {Phys. Rev. B}\ }\textbf {\bibinfo {volume} {98}},\ \bibinfo
  {pages} {184408} (\bibinfo {year} {2018})}\BibitemShut {NoStop}%
\bibitem [{\citenamefont {Wang}\ \emph {et~al.}(2018)\citenamefont {Wang},
  \citenamefont {Osterhoudt}, \citenamefont {Tian}, \citenamefont
  {Lampen-Kelley}, \citenamefont {Banerjee}, \citenamefont {Goldstein},
  \citenamefont {Yan}, \citenamefont {Knolle}, \citenamefont {Ji},
  \citenamefont {Cava}, \citenamefont {Nasu}, \citenamefont {Motome},
  \citenamefont {Nagler}, \citenamefont {Mandrus},\ and\ \citenamefont
  {Burch}}]{Wang2018}%
  \BibitemOpen
  \bibfield  {author} {\bibinfo {author} {\bibfnamefont {Yiping}\ \bibnamefont
  {Wang}}, \bibinfo {author} {\bibfnamefont {Gavin~B.}\ \bibnamefont
  {Osterhoudt}}, \bibinfo {author} {\bibfnamefont {Yao}\ \bibnamefont {Tian}},
  \bibinfo {author} {\bibfnamefont {Paige}\ \bibnamefont {Lampen-Kelley}},
  \bibinfo {author} {\bibfnamefont {Arnab}\ \bibnamefont {Banerjee}}, \bibinfo
  {author} {\bibfnamefont {Thomas}\ \bibnamefont {Goldstein}}, \bibinfo
  {author} {\bibfnamefont {Jun}\ \bibnamefont {Yan}}, \bibinfo {author}
  {\bibfnamefont {Johannes}\ \bibnamefont {Knolle}}, \bibinfo {author}
  {\bibfnamefont {Huiwen}\ \bibnamefont {Ji}}, \bibinfo {author} {\bibfnamefont
  {Robert~J.}\ \bibnamefont {Cava}}, \bibinfo {author} {\bibfnamefont {Joji}\
  \bibnamefont {Nasu}}, \bibinfo {author} {\bibfnamefont {Yukitoshi}\
  \bibnamefont {Motome}}, \bibinfo {author} {\bibfnamefont {Stephen~E.}\
  \bibnamefont {Nagler}}, \bibinfo {author} {\bibfnamefont {David}\
  \bibnamefont {Mandrus}}, and\ \bibinfo {author} {\bibfnamefont
  {Kenneth~S.}\ \bibnamefont {Burch}},\ }\href@noop {} {\enquote {\bibinfo
  {title} {Direct evidence for {Fermi Statistics} from proximity to the
  {Kitaev} spin liquid in $\mathrm{RuCl}_3$},}\ } (\bibinfo {year} {2018}),\
  \bibinfo {note} {unpublished},\ \Eprint {http://arxiv.org/abs/1809.07782}
  {arXiv:1809.07782} \BibitemShut {NoStop}%
\bibitem [{\citenamefont {Jansa}\ \emph {et~al.}(2018)\citenamefont {Jansa},
  \citenamefont {Zorko}, \citenamefont {Gomilsek}, \citenamefont {Pregelj},
  \citenamefont {Kramer}, \citenamefont {Biner}, \citenamefont {Biffin},
  \citenamefont {Ruegg},\ and\ \citenamefont {Klanjsek}}]{Jansa2018}%
  \BibitemOpen
  \bibfield  {author} {\bibinfo {author} {\bibfnamefont {Nejc}\ \bibnamefont
  {Jansa}}, \bibinfo {author} {\bibfnamefont {Andrej}\ \bibnamefont {Zorko}},
  \bibinfo {author} {\bibfnamefont {M.}~\bibnamefont {Gomilsek}}, \bibinfo
  {author} {\bibfnamefont {Matej}\ \bibnamefont {Pregelj}}, \bibinfo {author}
  {\bibfnamefont {Karl~W.}\ \bibnamefont {Kramer}}, \bibinfo {author}
  {\bibfnamefont {Daniel}\ \bibnamefont {Biner}}, \bibinfo {author}
  {\bibfnamefont {Alun}\ \bibnamefont {Biffin}}, \bibinfo {author}
  {\bibfnamefont {Christian}\ \bibnamefont {Ruegg}}, and\ \bibinfo {author}
  {\bibfnamefont {Martin}\ \bibnamefont {Klanjsek}},\ }\bibfield  {title}
  {\enquote {\bibinfo {title} {Observation of two types of fractional
  excitation in the {Kitaev} honeycomb magnet},}\ }\href {\doibase%
  10.1038/s41567-018-0129-5} {\bibfield  {journal} {\bibinfo  {journal} {Nature
  Physics}\ }\textbf {\bibinfo {volume} {14}},\ \bibinfo {pages} {786}
  (\bibinfo {year} {2018})}\BibitemShut {NoStop}%
\bibitem [{\citenamefont {Widmann}\ \emph {et~al.}(2019)\citenamefont
  {Widmann}, \citenamefont {Tsurkan}, \citenamefont {Prishchenko},
  \citenamefont {Mazurenko}, \citenamefont {Tsirlin},\ and\ \citenamefont
  {Loidl}}]{Widmann2019}%
  \BibitemOpen
  \bibfield  {author} {\bibinfo {author} {\bibfnamefont {S.}~\bibnamefont
  {Widmann}}, \bibinfo {author} {\bibfnamefont {V.}~\bibnamefont {Tsurkan}},
  \bibinfo {author} {\bibfnamefont {D.~A.}\ \bibnamefont {Prishchenko}},
  \bibinfo {author} {\bibfnamefont {V.~G.}\ \bibnamefont {Mazurenko}}, \bibinfo
  {author} {\bibfnamefont {A.~A.}\ \bibnamefont {Tsirlin}}, and\ \bibinfo
  {author} {\bibfnamefont {A.}~\bibnamefont {Loidl}},\ }\bibfield  {title}
  {\enquote {\bibinfo {title} {Thermodynamic evidence of fractionalized
  excitations in $\alpha$-$\mathrm{RuCl}_3$},}\ }\href {\doibase%
  10.1103/PhysRevB.99.094415} {\bibfield  {journal} {\bibinfo  {journal} {Phys.
  Rev. B}\ }\textbf {\bibinfo {volume} {99}},\ \bibinfo {pages} {094415}
  (\bibinfo {year} {2019})}\BibitemShut {NoStop}%
\bibitem [{\citenamefont {Zhang}\ \emph {et~al.}(2019)\citenamefont {Zhang},
  \citenamefont {Kim}, \citenamefont {Kim}, \citenamefont {Kee},\ and\
  \citenamefont {Yang}}]{Zhang2019}%
  \BibitemOpen
  \bibfield  {author} {\bibinfo {author} {\bibfnamefont {Haochen}\ \bibnamefont
  {Zhang}}, \bibinfo {author} {\bibfnamefont {Subin}\ \bibnamefont {Kim}},
  \bibinfo {author} {\bibfnamefont {Young-June}\ \bibnamefont {Kim}}, \bibinfo
  {author} {\bibfnamefont {Hae-Young}\ \bibnamefont {Kee}}, and\ \bibinfo
  {author} {\bibfnamefont {Luyi}\ \bibnamefont {Yang}},\ }\href@noop {}
  {\enquote {\bibinfo {title} {Ultrafast dynamics of fractional particles in
  $\alpha$-$\mathrm{RuCl}_3$},}\ } (\bibinfo {year} {2019}),\ \bibinfo {note}
  {unpublished},\ \Eprint {http://arxiv.org/abs/1908.04807} {arXiv:1908.04807}
  \BibitemShut {NoStop}%
\bibitem [{\citenamefont {Johnson}\ \emph {et~al.}(2015)\citenamefont
  {Johnson}, \citenamefont {Williams}, \citenamefont {Haghighirad},
  \citenamefont {Singleton}, \citenamefont {Zapf}, \citenamefont {Manuel},
  \citenamefont {Mazin}, \citenamefont {Li}, \citenamefont {Jeschke},
  \citenamefont {Valent\'{\i}},\ and\ \citenamefont {Coldea}}]{Johnson2015}%
  \BibitemOpen
  \bibfield  {author} {\bibinfo {author} {\bibfnamefont {R.~D.}\ \bibnamefont
  {Johnson}}, \bibinfo {author} {\bibfnamefont {S.~C.}\ \bibnamefont
  {Williams}}, \bibinfo {author} {\bibfnamefont {A.~A.}\ \bibnamefont
  {Haghighirad}}, \bibinfo {author} {\bibfnamefont {J.}~\bibnamefont
  {Singleton}}, \bibinfo {author} {\bibfnamefont {V.}~\bibnamefont {Zapf}},
  \bibinfo {author} {\bibfnamefont {P.}~\bibnamefont {Manuel}}, \bibinfo
  {author} {\bibfnamefont {I.~I.}\ \bibnamefont {Mazin}}, \bibinfo {author}
  {\bibfnamefont {Y.}~\bibnamefont {Li}}, \bibinfo {author} {\bibfnamefont
  {H.~O.}\ \bibnamefont {Jeschke}}, \bibinfo {author} {\bibfnamefont
  {R.}~\bibnamefont {Valent\'{\i}}}, and\ \bibinfo {author} {\bibfnamefont
  {R.}~\bibnamefont {Coldea}},\ }\bibfield  {title} {\enquote {\bibinfo {title}
  {Monoclinic crystal structure of $\alpha$-$\mathrm{RuCl}_3$ and the zigzag
  antiferromagnetic ground state},}\ }\href {\doibase%
  10.1103/PhysRevB.92.235119} {\bibfield  {journal} {\bibinfo  {journal} {Phys.
  Rev. B}\ }\textbf {\bibinfo {volume} {92}},\ \bibinfo {pages} {235119}
  (\bibinfo {year} {2015})}\BibitemShut {NoStop}%
\bibitem [{\citenamefont {Baek}\ \emph {et~al.}(2017)\citenamefont {Baek},
  \citenamefont {Do}, \citenamefont {Choi}, \citenamefont {Kwon}, \citenamefont
  {Wolter}, \citenamefont {Nishimoto}, \citenamefont {van~den Brink},\ and\
  \citenamefont {B\"uchner}}]{Baek2017}%
  \BibitemOpen
  \bibfield  {author} {\bibinfo {author} {\bibfnamefont {S.-H.}\ \bibnamefont
  {Baek}}, \bibinfo {author} {\bibfnamefont {S.-H.}\ \bibnamefont {Do}},
  \bibinfo {author} {\bibfnamefont {K.-Y.}\ \bibnamefont {Choi}}, \bibinfo
  {author} {\bibfnamefont {Y.~S.}\ \bibnamefont {Kwon}}, \bibinfo {author}
  {\bibfnamefont {A.~U.~B.}\ \bibnamefont {Wolter}}, \bibinfo {author}
  {\bibfnamefont {S.}~\bibnamefont {Nishimoto}}, \bibinfo {author}
  {\bibfnamefont {Jeroen}\ \bibnamefont {van~den Brink}}, and\ \bibinfo
  {author} {\bibfnamefont {B.}~\bibnamefont {B\"uchner}},\ }\bibfield  {title}
  {\enquote {\bibinfo {title} {Evidence for a field-induced quantum spin liquid
  in $\alpha$-$\mathrm{RuCl}_3$},}\ }\href {\doibase%
  10.1103/PhysRevLett.119.037201} {\bibfield  {journal} {\bibinfo  {journal}
  {Phys. Rev. Lett.}\ }\textbf {\bibinfo {volume} {119}},\ \bibinfo {pages}
  {037201} (\bibinfo {year} {2017})}\BibitemShut {NoStop}%
\bibitem [{\citenamefont {Sears}\ \emph {et~al.}(2017)\citenamefont {Sears},
  \citenamefont {Zhao}, \citenamefont {Xu}, \citenamefont {Lynn},\ and\
  \citenamefont {Kim}}]{Sears2017}%
  \BibitemOpen
  \bibfield  {author} {\bibinfo {author} {\bibfnamefont {J.~A.}\ \bibnamefont
  {Sears}}, \bibinfo {author} {\bibfnamefont {Y.}~\bibnamefont {Zhao}},
  \bibinfo {author} {\bibfnamefont {Z.}~\bibnamefont {Xu}}, \bibinfo {author}
  {\bibfnamefont {J.~W.}\ \bibnamefont {Lynn}}, and\ \bibinfo {author}
  {\bibfnamefont {Young-June}\ \bibnamefont {Kim}},\ }\bibfield  {title}
  {\enquote {\bibinfo {title} {Phase diagram of $\alpha$-$\mathrm{RuCl}_3$ in
  an in-plane magnetic field},}\ }\href {\doibase 10.1103/PhysRevB.95.180411}
  {\bibfield  {journal} {\bibinfo  {journal} {Phys. Rev. B}\ }\textbf {\bibinfo
  {volume} {95}},\ \bibinfo {pages} {180411} (\bibinfo {year}
  {2017})}\BibitemShut {NoStop}%
\bibitem [{\citenamefont {Wolter}\ \emph {et~al.}(2017)\citenamefont {Wolter},
  \citenamefont {Corredor}, \citenamefont {Janssen}, \citenamefont {Nenkov},
  \citenamefont {Sch\"onecker}, \citenamefont {Do}, \citenamefont {Choi},
  \citenamefont {Albrecht}, \citenamefont {Hunger}, \citenamefont {Doert},
  \citenamefont {Vojta},\ and\ \citenamefont {B\"uchner}}]{Wolter2017}%
  \BibitemOpen
  \bibfield  {author} {\bibinfo {author} {\bibfnamefont {A.~U.~B.}\
  \bibnamefont {Wolter}}, \bibinfo {author} {\bibfnamefont {L.~T.}\
  \bibnamefont {Corredor}}, \bibinfo {author} {\bibfnamefont {L.}~\bibnamefont
  {Janssen}}, \bibinfo {author} {\bibfnamefont {K.}~\bibnamefont {Nenkov}},
  \bibinfo {author} {\bibfnamefont {S.}~\bibnamefont {Sch\"onecker}}, \bibinfo
  {author} {\bibfnamefont {S.-H.}\ \bibnamefont {Do}}, \bibinfo {author}
  {\bibfnamefont {K.-Y.}\ \bibnamefont {Choi}}, \bibinfo {author}
  {\bibfnamefont {R.}~\bibnamefont {Albrecht}}, \bibinfo {author}
  {\bibfnamefont {J.}~\bibnamefont {Hunger}}, \bibinfo {author} {\bibfnamefont
  {T.}~\bibnamefont {Doert}}, \bibinfo {author} {\bibfnamefont
  {M.}~\bibnamefont {Vojta}}, and\ \bibinfo {author} {\bibfnamefont
  {B.}~\bibnamefont {B\"uchner}},\ }\bibfield  {title} {\enquote {\bibinfo
  {title} {Field-induced quantum criticality in the {Kitaev} system
  $\alpha$-$\mathrm{RuCl}_3$},}\ }\href {\doibase 10.1103/PhysRevB.96.041405}
  {\bibfield  {journal} {\bibinfo  {journal} {Phys. Rev. B}\ }\textbf {\bibinfo
  {volume} {96}},\ \bibinfo {pages} {041405} (\bibinfo {year}
  {2017})}\BibitemShut {NoStop}%
\bibitem [{\citenamefont {Leahy}\ \emph {et~al.}(2017)\citenamefont {Leahy},
  \citenamefont {Pocs}, \citenamefont {Siegfried}, \citenamefont {Graf},
  \citenamefont {Do}, \citenamefont {Choi}, \citenamefont {Normand},\ and\
  \citenamefont {Lee}}]{Leahy2017}%
  \BibitemOpen
  \bibfield  {author} {\bibinfo {author} {\bibfnamefont {Ian~A.}\ \bibnamefont
  {Leahy}}, \bibinfo {author} {\bibfnamefont {Christopher~A.}\ \bibnamefont
  {Pocs}}, \bibinfo {author} {\bibfnamefont {Peter~E.}\ \bibnamefont
  {Siegfried}}, \bibinfo {author} {\bibfnamefont {David}\ \bibnamefont {Graf}},
  \bibinfo {author} {\bibfnamefont {S.-H.}\ \bibnamefont {Do}}, \bibinfo
  {author} {\bibfnamefont {Kwang-Yong}\ \bibnamefont {Choi}}, \bibinfo {author}
  {\bibfnamefont {B.}~\bibnamefont {Normand}}, and\ \bibinfo {author}
  {\bibfnamefont {Minhyea}\ \bibnamefont {Lee}},\ }\bibfield  {title} {\enquote
  {\bibinfo {title} {Anomalous thermal conductivity and magnetic torque
  response in the honeycomb magnet $\alpha$-$\mathrm{RuCl}_3$},}\ }\href
  {\doibase 10.1103/PhysRevLett.118.187203} {\bibfield  {journal} {\bibinfo
  {journal} {Phys. Rev. Lett.}\ }\textbf {\bibinfo {volume} {118}},\ \bibinfo
  {pages} {187203} (\bibinfo {year} {2017})}\BibitemShut {NoStop}%
\bibitem [{\citenamefont {Banerjee}\ \emph
  {et~al.}(2018{\natexlab{b}})\citenamefont {Banerjee}, \citenamefont
  {Lampen-Kelley}, \citenamefont {Knolle}, \citenamefont {Balz}, \citenamefont
  {Aczel}, \citenamefont {Winn}, \citenamefont {Liu}, \citenamefont
  {Pajerowski}, \citenamefont {Yan}, \citenamefont {Bridges}, \citenamefont
  {Savici}, \citenamefont {Chakoumakos}, \citenamefont {Lumsden}, \citenamefont
  {Tennant}, \citenamefont {Moessner}, \citenamefont {Mandrus},\ and\
  \citenamefont {Nagler}}]{Banerjee2018}%
  \BibitemOpen
  \bibfield  {author} {\bibinfo {author} {\bibfnamefont {Arnab}\ \bibnamefont
  {Banerjee}}, \bibinfo {author} {\bibfnamefont {Paula}\ \bibnamefont
  {Lampen-Kelley}}, \bibinfo {author} {\bibfnamefont {Johannes}\ \bibnamefont
  {Knolle}}, \bibinfo {author} {\bibfnamefont {Christian}\ \bibnamefont
  {Balz}}, \bibinfo {author} {\bibfnamefont {Adam~Anthony}\ \bibnamefont
  {Aczel}}, \bibinfo {author} {\bibfnamefont {Barry}\ \bibnamefont {Winn}},
  \bibinfo {author} {\bibfnamefont {Yaohua}\ \bibnamefont {Liu}}, \bibinfo
  {author} {\bibfnamefont {Daniel}\ \bibnamefont {Pajerowski}}, \bibinfo
  {author} {\bibfnamefont {Jiaqiang}\ \bibnamefont {Yan}}, \bibinfo {author}
  {\bibfnamefont {Craig~A.}\ \bibnamefont {Bridges}}, \bibinfo {author}
  {\bibfnamefont {Andrei~T.}\ \bibnamefont {Savici}}, \bibinfo {author}
  {\bibfnamefont {Bryan~C.}\ \bibnamefont {Chakoumakos}}, \bibinfo {author}
  {\bibfnamefont {Mark~D.}\ \bibnamefont {Lumsden}}, \bibinfo {author}
  {\bibfnamefont {David~Alan}\ \bibnamefont {Tennant}}, \bibinfo {author}
  {\bibfnamefont {Roderich}\ \bibnamefont {Moessner}}, \bibinfo {author}
  {\bibfnamefont {David~G.}\ \bibnamefont {Mandrus}}, and\ \bibinfo {author}
  {\bibfnamefont {Stephen~E.}\ \bibnamefont {Nagler}},\ }\bibfield  {title}
  {\enquote {\bibinfo {title} {Excitations in the field-induced quantum spin
  liquid state of $\alpha$-$\mathrm{RuCl}_3$},}\ }\href {\doibase%
  10.1038/s41535-018-0079-2} {\bibfield  {journal} {\bibinfo  {journal} {npj
  Quantum Materials}\ }\textbf {\bibinfo {volume} {3}},\ \bibinfo {pages} {8}
  (\bibinfo {year} {2018}{\natexlab{b}})}\BibitemShut {NoStop}%
\bibitem [{\citenamefont {Hentrich}\ \emph {et~al.}(2018)\citenamefont
  {Hentrich}, \citenamefont {Wolter}, \citenamefont {Zotos}, \citenamefont
  {Brenig}, \citenamefont {Nowak}, \citenamefont {Isaeva}, \citenamefont
  {Doert}, \citenamefont {Banerjee}, \citenamefont {Lampen-Kelley},
  \citenamefont {Mandrus}, \citenamefont {Nagler}, \citenamefont {Sears},
  \citenamefont {Kim}, \citenamefont {B\"uchner},\ and\ \citenamefont
  {Hess}}]{Hentrich2018}%
  \BibitemOpen
  \bibfield  {author} {\bibinfo {author} {\bibfnamefont {Richard}\ \bibnamefont
  {Hentrich}}, \bibinfo {author} {\bibfnamefont {Anja U.~B.}\ \bibnamefont
  {Wolter}}, \bibinfo {author} {\bibfnamefont {Xenophon}\ \bibnamefont
  {Zotos}}, \bibinfo {author} {\bibfnamefont {Wolfram}\ \bibnamefont {Brenig}},
  \bibinfo {author} {\bibfnamefont {Domenic}\ \bibnamefont {Nowak}}, \bibinfo
  {author} {\bibfnamefont {Anna}\ \bibnamefont {Isaeva}}, \bibinfo {author}
  {\bibfnamefont {Thomas}\ \bibnamefont {Doert}}, \bibinfo {author}
  {\bibfnamefont {Arnab}\ \bibnamefont {Banerjee}}, \bibinfo {author}
  {\bibfnamefont {Paula}\ \bibnamefont {Lampen-Kelley}}, \bibinfo {author}
  {\bibfnamefont {David~G.}\ \bibnamefont {Mandrus}}, \bibinfo {author}
  {\bibfnamefont {Stephen~E.}\ \bibnamefont {Nagler}}, \bibinfo {author}
  {\bibfnamefont {Jennifer}\ \bibnamefont {Sears}}, \bibinfo {author}
  {\bibfnamefont {Young-June}\ \bibnamefont {Kim}}, \bibinfo {author}
  {\bibfnamefont {Bernd}\ \bibnamefont {B\"uchner}}, and\ \bibinfo {author}
  {\bibfnamefont {Christian}\ \bibnamefont {Hess}},\ }\bibfield  {title}
  {\enquote {\bibinfo {title} {Unusual phonon heat transport in
  $\alpha$-$\mathrm{RuCl}_3$: Strong spin-phonon scattering and field-induced
  spin gap},}\ }\href {\doibase 10.1103/PhysRevLett.120.117204} {\bibfield
  {journal} {\bibinfo  {journal} {Phys. Rev. Lett.}\ }\textbf {\bibinfo
  {volume} {120}},\ \bibinfo {pages} {117204} (\bibinfo {year}
  {2018})}\BibitemShut {NoStop}%
\bibitem [{\citenamefont {Kasahara}\ \emph
  {et~al.}(2018{\natexlab{b}})\citenamefont {Kasahara}, \citenamefont
  {Ohnishi}, \citenamefont {Mizukami}, \citenamefont {Tanaka}, \citenamefont
  {Ma}, \citenamefont {Sugii}, \citenamefont {Kurita}, \citenamefont {Tanaka},
  \citenamefont {Nasu}, \citenamefont {Motome}, \citenamefont {Shibauchi},\
  and\ \citenamefont {Matsuda}}]{Kasahara2018}%
  \BibitemOpen
  \bibfield  {author} {\bibinfo {author} {\bibfnamefont {Y}~\bibnamefont
  {Kasahara}}, \bibinfo {author} {\bibfnamefont {T}~\bibnamefont {Ohnishi}},
  \bibinfo {author} {\bibfnamefont {Y}~\bibnamefont {Mizukami}}, \bibinfo
  {author} {\bibfnamefont {O}~\bibnamefont {Tanaka}}, \bibinfo {author}
  {\bibfnamefont {Sixiao}\ \bibnamefont {Ma}}, \bibinfo {author} {\bibfnamefont
  {K}~\bibnamefont {Sugii}}, \bibinfo {author} {\bibfnamefont {N}~\bibnamefont
  {Kurita}}, \bibinfo {author} {\bibfnamefont {H}~\bibnamefont {Tanaka}},
  \bibinfo {author} {\bibfnamefont {J}~\bibnamefont {Nasu}}, \bibinfo {author}
  {\bibfnamefont {Y}~\bibnamefont {Motome}}, \bibinfo {author} {\bibfnamefont
  {T}~\bibnamefont {Shibauchi}}, and\ \bibinfo {author} {\bibfnamefont
  {Y}~\bibnamefont {Matsuda}},\ }\bibfield  {title} {\enquote {\bibinfo {title}
  {{M}ajorana quantization and half-integer thermal quantum {H}all effect in a
  {K}itaev spin liquid},}\ }\href {\doibase 10.1038/s41586-018-0274-0}
  {\bibfield  {journal} {\bibinfo  {journal} {Nature}\ }\textbf {\bibinfo
  {volume} {559}},\ \bibinfo {pages} {227--231} (\bibinfo {year}
  {2018}{\natexlab{b}})}\BibitemShut {NoStop}%
\bibitem [{\citenamefont {Balz}\ \emph {et~al.}(2019)\citenamefont {Balz},
  \citenamefont {Lampen-Kelley}, \citenamefont {Banerjee}, \citenamefont {Yan},
  \citenamefont {Lu}, \citenamefont {Hu}, \citenamefont {Yadav}, \citenamefont
  {Takano}, \citenamefont {Liu}, \citenamefont {Tennant}, \citenamefont
  {Lumsden}, \citenamefont {Mandrus},\ and\ \citenamefont {Nagler}}]{Balz2019}%
  \BibitemOpen
  \bibfield  {author} {\bibinfo {author} {\bibfnamefont {Christian}\
  \bibnamefont {Balz}}, \bibinfo {author} {\bibfnamefont {Paula}\ \bibnamefont
  {Lampen-Kelley}}, \bibinfo {author} {\bibfnamefont {Arnab}\ \bibnamefont
  {Banerjee}}, \bibinfo {author} {\bibfnamefont {Jiaqiang}\ \bibnamefont
  {Yan}}, \bibinfo {author} {\bibfnamefont {Zhilun}\ \bibnamefont {Lu}},
  \bibinfo {author} {\bibfnamefont {Xinzhe}\ \bibnamefont {Hu}}, \bibinfo
  {author} {\bibfnamefont {Swapnil~M.}\ \bibnamefont {Yadav}}, \bibinfo
  {author} {\bibfnamefont {Yasu}\ \bibnamefont {Takano}}, \bibinfo {author}
  {\bibfnamefont {Yaohua}\ \bibnamefont {Liu}}, \bibinfo {author}
  {\bibfnamefont {D.~Alan}\ \bibnamefont {Tennant}}, \bibinfo {author}
  {\bibfnamefont {Mark~D.}\ \bibnamefont {Lumsden}}, \bibinfo {author}
  {\bibfnamefont {David}\ \bibnamefont {Mandrus}}, and\ \bibinfo {author}
  {\bibfnamefont {Stephen~E.}\ \bibnamefont {Nagler}},\ }\bibfield  {title}
  {\enquote {\bibinfo {title} {Finite field regime for a quantum spin liquid in
  $\alpha$-$\mathrm{RuCl}_3$},}\ }\href {\doibase 10.1103/PhysRevB.100.060405}
  {\bibfield  {journal} {\bibinfo  {journal} {Phys. Rev. B}\ }\textbf {\bibinfo
  {volume} {100}},\ \bibinfo {pages} {060405} (\bibinfo {year}
  {2019})}\BibitemShut {NoStop}%
\bibitem [{\citenamefont {Ye}\ \emph {et~al.}(2018)\citenamefont {Ye},
  \citenamefont {Hal\'asz}, \citenamefont {Savary},\ and\ \citenamefont
  {Balents}}]{Ye2018}%
  \BibitemOpen
  \bibfield  {author} {\bibinfo {author} {\bibfnamefont {Mengxing}\
  \bibnamefont {Ye}}, \bibinfo {author} {\bibfnamefont {G\'abor~B.}\
  \bibnamefont {Hal\'asz}}, \bibinfo {author} {\bibfnamefont {Lucile}\
  \bibnamefont {Savary}}, and\ \bibinfo {author} {\bibfnamefont {Leon}\
  \bibnamefont {Balents}},\ }\bibfield  {title} {\enquote {\bibinfo {title}
  {Quantization of the thermal {H}all conductivity at small {H}all angles},}\
  }\href {\doibase 10.1103/PhysRevLett.121.147201} {\bibfield  {journal}
  {\bibinfo  {journal} {Phys. Rev. Lett.}\ }\textbf {\bibinfo {volume} {121}},\
  \bibinfo {pages} {147201} (\bibinfo {year} {2018})}\BibitemShut {NoStop}%
\bibitem [{\citenamefont {Vinkler-Aviv}\ and\ \citenamefont
  {Rosch}(2018)}]{Vinkler2018}%
  \BibitemOpen
  \bibfield  {author} {\bibinfo {author} {\bibfnamefont {Yuval}\ \bibnamefont
  {Vinkler-Aviv}}\ and\ \bibinfo {author} {\bibfnamefont {Achim}\ \bibnamefont
  {Rosch}},\ }\bibfield  {title} {\enquote {\bibinfo {title} {Approximately
  quantized thermal {H}all effect of chiral liquids coupled to phonons},}\
  }\href {\doibase 10.1103/PhysRevX.8.031032} {\bibfield  {journal} {\bibinfo
  {journal} {Phys. Rev. X}\ }\textbf {\bibinfo {volume} {8}},\ \bibinfo {pages}
  {031032} (\bibinfo {year} {2018})}\BibitemShut {NoStop}%
\bibitem [{\citenamefont {Yokoi}\ \emph {et~al.}(2020)\citenamefont {Yokoi},
  \citenamefont {Ma}, \citenamefont {Kasahara}, \citenamefont {Kasahara},
  \citenamefont {Shibauchi}, \citenamefont {Kurita}, \citenamefont {Tanaka},
  \citenamefont {Nasu}, \citenamefont {Motome}, \citenamefont {Hickey},
  \citenamefont {Trebst},\ and\ \citenamefont {Matsuda}}]{Tokoi}%
  \BibitemOpen
  \bibfield  {author} {\bibinfo {author} {\bibfnamefont {T.}~\bibnamefont
  {Yokoi}}, \bibinfo {author} {\bibfnamefont {S.}~\bibnamefont {Ma}}, \bibinfo
  {author} {\bibfnamefont {Y.}~\bibnamefont {Kasahara}}, \bibinfo {author}
  {\bibfnamefont {S.}~\bibnamefont {Kasahara}}, \bibinfo {author}
  {\bibfnamefont {T.}~\bibnamefont {Shibauchi}}, \bibinfo {author}
  {\bibfnamefont {N.}~\bibnamefont {Kurita}}, \bibinfo {author} {\bibfnamefont
  {H.}~\bibnamefont {Tanaka}}, \bibinfo {author} {\bibfnamefont
  {J.}~\bibnamefont {Nasu}}, \bibinfo {author} {\bibfnamefont {Y.}~\bibnamefont
  {Motome}}, \bibinfo {author} {\bibfnamefont {C.}~\bibnamefont {Hickey}},
  \bibinfo {author} {\bibfnamefont {S.}~\bibnamefont {Trebst}}, and\ \bibinfo
  {author} {\bibfnamefont {Y.}~\bibnamefont {Matsuda}},\ }\href@noop {}
  {\enquote {\bibinfo {title} {Half-integer quantized anomalous thermal {Hall}
  effect in the {Kitaev} material $\alpha$-$\mathrm{RuCl}_3$},}\ } (\bibinfo
  {year} {2020}),\ \bibinfo {note} {unpublished},\ \Eprint
  {http://arxiv.org/abs/2001.01899} {arXiv:2001.01899} \BibitemShut {NoStop}%
\bibitem [{Note1()}]{Note1}%
  \BibitemOpen
  \bibinfo {note} {Any topological quantum computing platform would ideally be
  run at the lowest accessible temperatures. A large gap is nevertheless
  desirable for suppressing errors.}\BibitemShut {Stop}%
\bibitem [{\citenamefont {Zhou}\ \emph {et~al.}(2018)\citenamefont {Zhou},
  \citenamefont {Balgley}, \citenamefont {Lampen-Kelley}, \citenamefont {Yan},
  \citenamefont {Mandrus},\ and\ \citenamefont {Henriksen}}]{Zhou2018}%
  \BibitemOpen
  \bibfield  {author} {\bibinfo {author} {\bibfnamefont {Boyi}\ \bibnamefont
  {Zhou}}, \bibinfo {author} {\bibfnamefont {J.}~\bibnamefont {Balgley}},
  \bibinfo {author} {\bibfnamefont {P.}~\bibnamefont {Lampen-Kelley}}, \bibinfo
  {author} {\bibfnamefont {J.-Q.}\ \bibnamefont {Yan}}, \bibinfo {author}
  {\bibfnamefont {D.~G.}\ \bibnamefont {Mandrus}}, and\ \bibinfo {author}
  {\bibfnamefont {E.~A.}\ \bibnamefont {Henriksen}},\ }\href@noop {} {\enquote
  {\bibinfo {title} {Gate-tuned charge-doping and magnetism in
  graphene/$\alpha$-$\mathrm{RuCl}_3$ heterostructures},}\ } (\bibinfo {year}
  {2018}),\ \bibinfo {note} {unpublished},\ \Eprint
  {http://arxiv.org/abs/1811.04838} {arXiv:1811.04838} \BibitemShut {NoStop}%
\bibitem [{\citenamefont {Zhou}\ \emph {et~al.}(2019)\citenamefont {Zhou},
  \citenamefont {Wang}, \citenamefont {Osterhoudt}, \citenamefont
  {Lampen-Kelley}, \citenamefont {Mandrus}, \citenamefont {He}, \citenamefont
  {Burch},\ and\ \citenamefont {Henriksen}}]{Zhou2019}%
  \BibitemOpen
  \bibfield  {author} {\bibinfo {author} {\bibfnamefont {Boyi}\ \bibnamefont
  {Zhou}}, \bibinfo {author} {\bibfnamefont {Yiping}\ \bibnamefont {Wang}},
  \bibinfo {author} {\bibfnamefont {Gavin~B.}\ \bibnamefont {Osterhoudt}},
  \bibinfo {author} {\bibfnamefont {Paula}\ \bibnamefont {Lampen-Kelley}},
  \bibinfo {author} {\bibfnamefont {David}\ \bibnamefont {Mandrus}}, \bibinfo
  {author} {\bibfnamefont {Rui}\ \bibnamefont {He}}, \bibinfo {author}
  {\bibfnamefont {Kenneth~S.}\ \bibnamefont {Burch}}, and\ \bibinfo {author}
  {\bibfnamefont {Erik~A.}\ \bibnamefont {Henriksen}},\ }\bibfield  {title}
  {\enquote {\bibinfo {title} {Possible structural transformation and enhanced
  magnetic fluctuations in exfoliated $\alpha$-$\mathrm{RuCl}_3$},}\ }\href
  {\doibase https://doi.org/10.1016/j.jpcs.2018.01.026} {\bibfield  {journal}
  {\bibinfo  {journal} {Journal of Physics and Chemistry of Solids}\ }\textbf
  {\bibinfo {volume} {128}},\ \bibinfo {pages} {291 -- 295} (\bibinfo {year}
  {2019})}\BibitemShut {NoStop}%
\bibitem [{\citenamefont {Mashhadi}\ \emph {et~al.}(2018)\citenamefont
  {Mashhadi}, \citenamefont {Weber}, \citenamefont {Schoop}, \citenamefont
  {Schulz}, \citenamefont {Lotsch}, \citenamefont {Burghard},\ and\
  \citenamefont {Kern}}]{Mashhadi2018}%
  \BibitemOpen
  \bibfield  {author} {\bibinfo {author} {\bibfnamefont {Soudabeh}\
  \bibnamefont {Mashhadi}}, \bibinfo {author} {\bibfnamefont {Daniel}\
  \bibnamefont {Weber}}, \bibinfo {author} {\bibfnamefont {Leslie~M.}\
  \bibnamefont {Schoop}}, \bibinfo {author} {\bibfnamefont {Armin}\
  \bibnamefont {Schulz}}, \bibinfo {author} {\bibfnamefont {Bettina~V.}\
  \bibnamefont {Lotsch}}, \bibinfo {author} {\bibfnamefont {Marko}\
  \bibnamefont {Burghard}}, and\ \bibinfo {author} {\bibfnamefont {Klaus}\
  \bibnamefont {Kern}},\ }\bibfield  {title} {\enquote {\bibinfo {title}
  {Electrical transport signature of the magnetic fluctuation-structure
  relation in $\alpha$-$\mathrm{RuCl}_3$ nanoflakes},}\ }\href {\doibase%
  10.1021/acs.nanolett.8b00926} {\bibfield  {journal} {\bibinfo  {journal}
  {Nano Letters}\ }\textbf {\bibinfo {volume} {18}},\ \bibinfo {pages} {3203}
  (\bibinfo {year} {2018})}\BibitemShut {NoStop}%
\bibitem [{\citenamefont {Mashhadi}\ \emph {et~al.}(2019)\citenamefont
  {Mashhadi}, \citenamefont {Kim}, \citenamefont {Kim}, \citenamefont {Weber},
  \citenamefont {Taniguchi}, \citenamefont {Watanabe}, \citenamefont {Park},
  \citenamefont {Lotsch}, \citenamefont {Smet}, \citenamefont {Burghard},\ and\
  \citenamefont {Kern}}]{Mashhadi2019}%
  \BibitemOpen
  \bibfield  {author} {\bibinfo {author} {\bibfnamefont {Soudabeh}\
  \bibnamefont {Mashhadi}}, \bibinfo {author} {\bibfnamefont {Youngwook}\
  \bibnamefont {Kim}}, \bibinfo {author} {\bibfnamefont {Jeongwoo}\
  \bibnamefont {Kim}}, \bibinfo {author} {\bibfnamefont {Daniel}\ \bibnamefont
  {Weber}}, \bibinfo {author} {\bibfnamefont {Takashi}\ \bibnamefont
  {Taniguchi}}, \bibinfo {author} {\bibfnamefont {Kenji}\ \bibnamefont
  {Watanabe}}, \bibinfo {author} {\bibfnamefont {Noejung}\ \bibnamefont
  {Park}}, \bibinfo {author} {\bibfnamefont {Bettina}\ \bibnamefont {Lotsch}},
  \bibinfo {author} {\bibfnamefont {Jurgen~H.}\ \bibnamefont {Smet}}, \bibinfo
  {author} {\bibfnamefont {Marko}\ \bibnamefont {Burghard}}, and\ \bibinfo
  {author} {\bibfnamefont {Klaus}\ \bibnamefont {Kern}},\ }\bibfield  {title}
  {\enquote {\bibinfo {title} {Spin-split band hybridization in graphene
  proximitized with $\alpha$-$\mathrm{RuCl}_3$ nanosheets},}\ }\href {\doibase%
  10.1021/acs.nanolett.9b01691} {\bibfield  {journal} {\bibinfo  {journal}
  {Nano Letters}\ }\textbf {\bibinfo {volume} {19}},\ \bibinfo {pages} {4659}
  (\bibinfo {year} {2019})}\BibitemShut {NoStop}%
\bibitem [{\citenamefont {Willans}\ \emph {et~al.}(2011)\citenamefont
  {Willans}, \citenamefont {Chalker},\ and\ \citenamefont
  {Moessner}}]{HoneycombVacancies}%
  \BibitemOpen
  \bibfield  {author} {\bibinfo {author} {\bibfnamefont {A.~J.}\ \bibnamefont
  {Willans}}, \bibinfo {author} {\bibfnamefont {J.~T.}\ \bibnamefont
  {Chalker}}, and\ \bibinfo {author} {\bibfnamefont {R.}~\bibnamefont
  {Moessner}},\ }\bibfield  {title} {\enquote {\bibinfo {title} {Site dilution
  in the {K}itaev honeycomb model},}\ }\href {\doibase%
  10.1103/PhysRevB.84.115146} {\bibfield  {journal} {\bibinfo  {journal} {Phys.
  Rev. B}\ }\textbf {\bibinfo {volume} {84}},\ \bibinfo {pages} {115146}
  (\bibinfo {year} {2011})}\BibitemShut {NoStop}%
\bibitem [{\citenamefont {Dhochak}\ \emph {et~al.}(2010)\citenamefont
  {Dhochak}, \citenamefont {Shankar},\ and\ \citenamefont
  {Tripathi}}]{HoneycombImpurity}%
  \BibitemOpen
  \bibfield  {author} {\bibinfo {author} {\bibfnamefont {Kusum}\ \bibnamefont
  {Dhochak}}, \bibinfo {author} {\bibfnamefont {R.}~\bibnamefont {Shankar}},\
  and\ \bibinfo {author} {\bibfnamefont {V.}~\bibnamefont {Tripathi}},\
  }\bibfield  {title} {\enquote {\bibinfo {title} {Magnetic impurities in the
  honeycomb {Kitaev} model},}\ }\href {\doibase 10.1103/PhysRevLett.105.117201}
  {\bibfield  {journal} {\bibinfo  {journal} {Phys. Rev. Lett.}\ }\textbf
  {\bibinfo {volume} {105}},\ \bibinfo {pages} {117201} (\bibinfo {year}
  {2010})}\BibitemShut {NoStop}%
\bibitem [{\citenamefont {Vojta}\ \emph {et~al.}(2016)\citenamefont {Vojta},
  \citenamefont {Mitchell},\ and\ \citenamefont
  {Zschocke}}]{HoneycombImpurity2}%
  \BibitemOpen
  \bibfield  {author} {\bibinfo {author} {\bibfnamefont {Matthias}\
  \bibnamefont {Vojta}}, \bibinfo {author} {\bibfnamefont {Andrew~K.}\
  \bibnamefont {Mitchell}}, and\ \bibinfo {author} {\bibfnamefont {Fabian}\
  \bibnamefont {Zschocke}},\ }\bibfield  {title} {\enquote {\bibinfo {title}
  {Kondo impurities in the {Kitaev} spin liquid: Numerical renormalization
  group solution and gauge-flux-driven screening},}\ }\href {\doibase%
  10.1103/PhysRevLett.117.037202} {\bibfield  {journal} {\bibinfo  {journal}
  {Phys. Rev. Lett.}\ }\textbf {\bibinfo {volume} {117}},\ \bibinfo {pages}
  {037202} (\bibinfo {year} {2016})}\BibitemShut {NoStop}%
\bibitem [{\citenamefont {Barkeshli}\ \emph {et~al.}(2014)\citenamefont
  {Barkeshli}, \citenamefont {Berg},\ and\ \citenamefont
  {Kivelson}}]{Barkeshli2014}%
  \BibitemOpen
  \bibfield  {author} {\bibinfo {author} {\bibfnamefont {Maissam}\ \bibnamefont
  {Barkeshli}}, \bibinfo {author} {\bibfnamefont {Erez}\ \bibnamefont {Berg}},
  and\ \bibinfo {author} {\bibfnamefont {Steven}\ \bibnamefont {Kivelson}},\
  }\bibfield  {title} {\enquote {\bibinfo {title} {Coherent transmutation of
  electrons into fractionalized anyons},}\ }\href {\doibase%
  10.1126/science.1253251} {\bibfield  {journal} {\bibinfo  {journal}
  {Science}\ }\textbf {\bibinfo {volume} {346}},\ \bibinfo {pages} {722--725}
  (\bibinfo {year} {2014})}\BibitemShut {NoStop}%
\bibitem [{\citenamefont {Barkeshli}\ and\ \citenamefont
  {Nayak}(2015)}]{Barkeshli2015}%
  \BibitemOpen
  \bibfield  {author} {\bibinfo {author} {\bibfnamefont {Maissam}\ \bibnamefont
  {Barkeshli}}\ and\ \bibinfo {author} {\bibfnamefont {Chetan}\ \bibnamefont
  {Nayak}},\ }\href@noop {} {\enquote {\bibinfo {title} {Superconductivity
  induced topological phase transition at the edge of even denominator
  fractional quantum {Hall} states},}\ } (\bibinfo {year} {2015}),\ \bibinfo
  {note} {unpublished},\ \Eprint {http://arxiv.org/abs/1507.06305}
  {arXiv:1507.06305} \BibitemShut {NoStop}%
\bibitem [{Note2()}]{Note2}%
  \BibitemOpen
  \bibinfo {note} {Reference~\protect \rev@citealp {Barkeshli2015} also briefly
  discusses applications to the non-Abelian spin liquid in Kitaev's honeycomb
  model, though their approach is very different from the one developed
  here.}\BibitemShut {Stop}%
\bibitem [{\citenamefont {Aasen}\ \emph {et~al.}(2017)\citenamefont {Aasen},
  \citenamefont {Lake},\ and\ \citenamefont {Walker}}]{AasenFC}%
  \BibitemOpen
  \bibfield  {author} {\bibinfo {author} {\bibfnamefont {David}\ \bibnamefont
  {Aasen}}, \bibinfo {author} {\bibfnamefont {Ethan}\ \bibnamefont {Lake}},\
  and\ \bibinfo {author} {\bibfnamefont {Kevin}\ \bibnamefont {Walker}},\
  }\href@noop {} {\enquote {\bibinfo {title} {Fermion condensation and super
  pivotal categories},}\ } (\bibinfo {year} {2017}),\ \bibinfo {note}
  {unpublished},\ \Eprint {http://arxiv.org/abs/1709.01941} {arXiv:1709.01941}
  \BibitemShut {NoStop}%
\bibitem [{\citenamefont {Rahmani}\ and\ \citenamefont
  {Franz}(2019)}]{Rahmani2019}%
  \BibitemOpen
  \bibfield  {author} {\bibinfo {author} {\bibfnamefont {Armin}\ \bibnamefont
  {Rahmani}}\ and\ \bibinfo {author} {\bibfnamefont {Marcel}\ \bibnamefont
  {Franz}},\ }\bibfield  {title} {\enquote {\bibinfo {title} {Interacting
  {M}ajorana fermions},}\ }\href {\doibase 10.1088/1361-6633/ab28ef} {\bibfield
   {journal} {\bibinfo  {journal} {Reports on Progress in Physics}\ }\textbf
  {\bibinfo {volume} {82}},\ \bibinfo {pages} {084501} (\bibinfo {year}
  {2019})}\BibitemShut {NoStop}%
\bibitem [{Note3()}]{Note3}%
  \BibitemOpen
  \bibinfo {note} {Obtaining physical spin wavefunctions still requires
  enforcing the local constraint $D_{\protect \bf r} \equiv b^x_{\protect \bf
  r} b^y_{\protect \bf r} b^z_{\protect \bf r} c_{\protect \bf r} = +1$ for all
  ${\protect \bf r}$, which can be done by applying a projector $P = \DOTSB
  \prod@ \slimits@ _{\protect \bf r} \left (\protect \frac {1+D_{\protect \bf
  r}}{2}\right )$ to many-body fermion states. Although $[\protect \mathaccentV
  {hat}05E{u}_{\protect \bf r r'},D_{\protect \bf r''}] \not =0$,
  gauge-invariant quantities (e.g., the energy) can nevertheless be exactly
  extracted from the free-fermion limit of Eq.~\protect \textup {\hbox
  {\mathsurround \z@ \protect \normalfont (\ignorespaces \ref
  {hamfermion}\unskip \@@italiccorr )}} with fixed $\protect \mathaccentV
  {hat}05E{u}_{\protect \bf r r'}$ values.}\BibitemShut {Stop}%
\bibitem [{\citenamefont {Lieb}(1994)}]{LiebFlux}%
  \BibitemOpen
  \bibfield  {author} {\bibinfo {author} {\bibfnamefont {Elliott~H.}\
  \bibnamefont {Lieb}},\ }\bibfield  {title} {\enquote {\bibinfo {title} {Flux
  phase of the half-filled band},}\ }\href {\doibase%
  10.1103/PhysRevLett.73.2158} {\bibfield  {journal} {\bibinfo  {journal}
  {Phys. Rev. Lett.}\ }\textbf {\bibinfo {volume} {73}},\ \bibinfo {pages}
  {2158--2161} (\bibinfo {year} {1994})}\BibitemShut {NoStop}%
\bibitem [{\citenamefont {Song}\ \emph {et~al.}(2016)\citenamefont {Song},
  \citenamefont {You},\ and\ \citenamefont {Balents}}]{Balents2016}%
  \BibitemOpen
  \bibfield  {author} {\bibinfo {author} {\bibfnamefont {Xue-Yang}\
  \bibnamefont {Song}}, \bibinfo {author} {\bibfnamefont {Yi-Zhuang}\
  \bibnamefont {You}}, and\ \bibinfo {author} {\bibfnamefont {Leon}\
  \bibnamefont {Balents}},\ }\bibfield  {title} {\enquote {\bibinfo {title}
  {Low-energy spin dynamics of the honeycomb spin liquid beyond the {K}itaev
  limit},}\ }\href {\doibase 10.1103/PhysRevLett.117.037209} {\bibfield
  {journal} {\bibinfo  {journal} {Phys. Rev. Lett.}\ }\textbf {\bibinfo
  {volume} {117}},\ \bibinfo {pages} {037209} (\bibinfo {year}
  {2016})}\BibitemShut {NoStop}%
\bibitem [{\citenamefont {Haldane}(1988)}]{Haldane}%
  \BibitemOpen
  \bibfield  {author} {\bibinfo {author} {\bibfnamefont {F.~D.~M.}\
  \bibnamefont {Haldane}},\ }\bibfield  {title} {\enquote {\bibinfo {title}
  {Model for a quantum {Hall} effect without {Landau} levels: Condensed-matter
  realization of the `parity anomaly'},}\ }\href {\doibase%
  10.1103/PhysRevLett.61.2015} {\bibfield  {journal} {\bibinfo  {journal}
  {Phys. Rev. Lett.}\ }\textbf {\bibinfo {volume} {61}},\ \bibinfo {pages}
  {2015--2018} (\bibinfo {year} {1988})}\BibitemShut {NoStop}%
\bibitem [{\citenamefont {{Volovik}}(1990)}]{Volovik}%
  \BibitemOpen
  \bibfield  {author} {\bibinfo {author} {\bibfnamefont {G.~E.}\ \bibnamefont
  {{Volovik}}},\ }\bibfield  {title} {\enquote {\bibinfo {title} {{The
  gravitational topological {Chern-Simons} term in a film of superfluid
  $^{3}$He-A}},}\ }\href@noop {} {\bibfield  {journal} {\bibinfo  {journal}
  {JETP Letters}\ }\textbf {\bibinfo {volume} {51}},\ \bibinfo {pages} {125}
  (\bibinfo {year} {1990})}\BibitemShut {NoStop}%
\bibitem [{\citenamefont {Kane}\ and\ \citenamefont
  {Fisher}(1997)}]{KaneFisherThermal}%
  \BibitemOpen
  \bibfield  {author} {\bibinfo {author} {\bibfnamefont {C.~L.}\ \bibnamefont
  {Kane}}\ and\ \bibinfo {author} {\bibfnamefont {Matthew P.~A.}\ \bibnamefont
  {Fisher}},\ }\bibfield  {title} {\enquote {\bibinfo {title} {Quantized
  thermal transport in the fractional quantum {Hall} effect},}\ }\href
  {\doibase 10.1103/PhysRevB.55.15832} {\bibfield  {journal} {\bibinfo
  {journal} {Phys. Rev. B}\ }\textbf {\bibinfo {volume} {55}},\ \bibinfo
  {pages} {15832--15837} (\bibinfo {year} {1997})}\BibitemShut {NoStop}%
\bibitem [{\citenamefont {Cappelli}\ \emph {et~al.}(2002)\citenamefont
  {Cappelli}, \citenamefont {Huerta},\ and\ \citenamefont {Zemba}}]{Cappelli}%
  \BibitemOpen
  \bibfield  {author} {\bibinfo {author} {\bibfnamefont {Andrea}\ \bibnamefont
  {Cappelli}}, \bibinfo {author} {\bibfnamefont {Marina}\ \bibnamefont
  {Huerta}}, and\ \bibinfo {author} {\bibfnamefont {Guillermo~R.}\
  \bibnamefont {Zemba}},\ }\bibfield  {title} {\enquote {\bibinfo {title}
  {Thermal transport in chiral conformal theories and hierarchical quantum
  {Hall} states},}\ }\href {\doibase%
  https://doi.org/10.1016/S0550-3213(02)00340-1} {\bibfield  {journal}
  {\bibinfo  {journal} {Nuclear Physics B}\ }\textbf {\bibinfo {volume}
  {636}},\ \bibinfo {pages} {568 -- 582} (\bibinfo {year} {2002})}\BibitemShut
  {NoStop}%
\bibitem [{Note4()}]{Note4}%
  \BibitemOpen
  \bibinfo {note} {In general the edge velocity is expected to depend on
  details of the boundary and need not be spatially uniform, but for simplicity
  we ignore such complications in this paper.}\BibitemShut {Stop}%
\bibitem [{\citenamefont {Teo}\ and\ \citenamefont {Kane}(2014)}]{TeoKane}%
  \BibitemOpen
  \bibfield  {author} {\bibinfo {author} {\bibfnamefont {Jeffrey C.~Y.}\
  \bibnamefont {Teo}}\ and\ \bibinfo {author} {\bibfnamefont {C.~L.}\
  \bibnamefont {Kane}},\ }\bibfield  {title} {\enquote {\bibinfo {title} {From
  {L}uttinger liquid to non-{A}belian quantum {H}all states},}\ }\href
  {\doibase 10.1103/PhysRevB.89.085101} {\bibfield  {journal} {\bibinfo
  {journal} {Phys. Rev. B}\ }\textbf {\bibinfo {volume} {89}},\ \bibinfo
  {pages} {085101} (\bibinfo {year} {2014})}\BibitemShut {NoStop}%
\bibitem [{\citenamefont {O'Brien}\ and\ \citenamefont
  {Fendley}(2018)}]{Fendley2018}%
  \BibitemOpen
  \bibfield  {author} {\bibinfo {author} {\bibfnamefont {Edward}\ \bibnamefont
  {O'Brien}}\ and\ \bibinfo {author} {\bibfnamefont {Paul}\ \bibnamefont
  {Fendley}},\ }\bibfield  {title} {\enquote {\bibinfo {title} {Lattice
  supersymmetry and order-disorder coexistence in the tricritical {I}sing
  model},}\ }\href {\doibase 10.1103/PhysRevLett.120.206403} {\bibfield
  {journal} {\bibinfo  {journal} {Phys. Rev. Lett.}\ }\textbf {\bibinfo
  {volume} {120}},\ \bibinfo {pages} {206403} (\bibinfo {year}
  {2018})}\BibitemShut {NoStop}%
\bibitem [{\citenamefont {Sannomiya}\ and\ \citenamefont
  {Katsura}(2019)}]{Sannomiya2019}%
  \BibitemOpen
  \bibfield  {author} {\bibinfo {author} {\bibfnamefont {Noriaki}\ \bibnamefont
  {Sannomiya}}\ and\ \bibinfo {author} {\bibfnamefont {Hosho}\ \bibnamefont
  {Katsura}},\ }\bibfield  {title} {\enquote {\bibinfo {title} {Supersymmetry
  breaking and {N}ambu-{G}oldstone fermions in interacting {M}ajorana
  chains},}\ }\href {\doibase 10.1103/PhysRevD.99.045002} {\bibfield  {journal}
  {\bibinfo  {journal} {Phys. Rev. D}\ }\textbf {\bibinfo {volume} {99}},\
  \bibinfo {pages} {045002} (\bibinfo {year} {2019})}\BibitemShut {NoStop}%
\bibitem [{\citenamefont {Fendley}(2019)}]{Fendley2019}%
  \BibitemOpen
  \bibfield  {author} {\bibinfo {author} {\bibfnamefont {Paul}\ \bibnamefont
  {Fendley}},\ }\bibfield  {title} {\enquote {\bibinfo {title} {Free fermions
  in disguise},}\ }\href@noop {} {\bibfield  {journal} {\bibinfo  {journal}
  {arXiv preprint arXiv:1901.08078}\ } (\bibinfo {year} {2019})}\BibitemShut
  {NoStop}%
\bibitem [{Note5()}]{Note5}%
  \BibitemOpen
  \bibinfo {note} {In the purely 1D setting under consideration, $T$ symmetry
  is anomalous because it changes the sign of the total fermion parity operator
  $P = \DOTSB \prod@ \slimits@ _a (i \gamma _{2a}\gamma _{2a+1})$.}\BibitemShut
  {Stop}%
\bibitem [{Note6()}]{Note6}%
  \BibitemOpen
  \bibinfo {note} {At $t' = 0$ the chain also preserves a unitary reflection
  symmetry that sends $\gamma _a \to (-1)^a\gamma _{-a}$, but this symmetry
  will not play a role in our discussion.}\BibitemShut {Stop}%
\bibitem [{Note7()}]{Note7}%
  \BibitemOpen
  \bibinfo {note} {We caution, however, that physical fermions governed by the
  1D lattice model do not realize Ising non-Abelian anyons in the same sense as
  the spin liquid. In particular, explicitly breaking the anomalous translation
  symmetry $T$ in the 1D model generically confines the domain walls, whereas
  in the spin liquid interface no symmetry is required for their
  deconfinement.}\BibitemShut {Stop}%
\bibitem [{\citenamefont {Rahmani}\ \emph
  {et~al.}(2015{\natexlab{a}})\citenamefont {Rahmani}, \citenamefont {Zhu},
  \citenamefont {Franz},\ and\ \citenamefont {Affleck}}]{Rahmani1}%
  \BibitemOpen
  \bibfield  {author} {\bibinfo {author} {\bibfnamefont {Armin}\ \bibnamefont
  {Rahmani}}, \bibinfo {author} {\bibfnamefont {Xiaoyu}\ \bibnamefont {Zhu}},
  \bibinfo {author} {\bibfnamefont {Marcel}\ \bibnamefont {Franz}}, and\
  \bibinfo {author} {\bibfnamefont {Ian}\ \bibnamefont {Affleck}},\ }\bibfield
  {title} {\enquote {\bibinfo {title} {Emergent supersymmetry from strongly
  interacting {M}ajorana zero modes},}\ }\href {\doibase%
  10.1103/PhysRevLett.115.166401} {\bibfield  {journal} {\bibinfo  {journal}
  {Phys. Rev. Lett.}\ }\textbf {\bibinfo {volume} {115}},\ \bibinfo {pages}
  {166401} (\bibinfo {year} {2015}{\natexlab{a}})}\BibitemShut {NoStop}%
\bibitem [{\citenamefont {Rahmani}\ \emph
  {et~al.}(2015{\natexlab{b}})\citenamefont {Rahmani}, \citenamefont {Zhu},
  \citenamefont {Franz},\ and\ \citenamefont {Affleck}}]{Rahmani2}%
  \BibitemOpen
  \bibfield  {author} {\bibinfo {author} {\bibfnamefont {Armin}\ \bibnamefont
  {Rahmani}}, \bibinfo {author} {\bibfnamefont {Xiaoyu}\ \bibnamefont {Zhu}},
  \bibinfo {author} {\bibfnamefont {Marcel}\ \bibnamefont {Franz}}, and\
  \bibinfo {author} {\bibfnamefont {Ian}\ \bibnamefont {Affleck}},\ }\bibfield
  {title} {\enquote {\bibinfo {title} {Phase diagram of the interacting
  {M}ajorana chain model},}\ }\href {\doibase 10.1103/PhysRevB.92.235123}
  {\bibfield  {journal} {\bibinfo  {journal} {Phys. Rev. B}\ }\textbf {\bibinfo
  {volume} {92}},\ \bibinfo {pages} {235123} (\bibinfo {year}
  {2015}{\natexlab{b}})}\BibitemShut {NoStop}%
\bibitem [{Note8()}]{Note8}%
  \BibitemOpen
  \bibinfo {note} {Spin-orbit coupling facilitates such processes. For
  instance, spin-orbit interactions in the superconductor generically yield a
  triplet component at the interface, enabling injection and removal of Cooper
  pairs from the quantum Hall system even if the quantum Hall edge state
  exhibits perfect spin polarization.}\BibitemShut {Stop}%
\bibitem [{\citenamefont {Gamayun}\ \emph {et~al.}(2017)\citenamefont
  {Gamayun}, \citenamefont {Hutasoit},\ and\ \citenamefont
  {Cheianov}}]{Gamayun2017}%
  \BibitemOpen
  \bibfield  {author} {\bibinfo {author} {\bibfnamefont {Oleksandr}\
  \bibnamefont {Gamayun}}, \bibinfo {author} {\bibfnamefont {Jimmy~A.}\
  \bibnamefont {Hutasoit}}, and\ \bibinfo {author} {\bibfnamefont {Vadim~V.}\
  \bibnamefont {Cheianov}},\ }\bibfield  {title} {\enquote {\bibinfo {title}
  {Two-terminal transport along a proximity-induced superconducting quantum
  {H}all edge},}\ }\href {\doibase 10.1103/PhysRevB.96.241104} {\bibfield
  {journal} {\bibinfo  {journal} {Phys. Rev. B}\ }\textbf {\bibinfo {volume}
  {96}},\ \bibinfo {pages} {241104} (\bibinfo {year} {2017})}\BibitemShut
  {NoStop}%
\bibitem [{\citenamefont {Wan}\ and\ \citenamefont {Wang}(2017)}]{Wan2017}%
  \BibitemOpen
  \bibfield  {author} {\bibinfo {author} {\bibfnamefont {Yidun}\ \bibnamefont
  {Wan}}\ and\ \bibinfo {author} {\bibfnamefont {Chenjie}\ \bibnamefont
  {Wang}},\ }\bibfield  {title} {\enquote {\bibinfo {title} {Fermion
  condensation and gapped domain walls in topological orders},}\ }\href
  {\doibase 10.1007/JHEP03(2017)172} {\bibfield  {journal} {\bibinfo  {journal}
  {Journal of High Energy Physics}\ }\textbf {\bibinfo {volume} {2017}},\
  \bibinfo {pages} {172} (\bibinfo {year} {2017})}\BibitemShut {NoStop}%
\bibitem [{Note9()}]{Note9}%
  \BibitemOpen
  \bibinfo {note} {We stress the importance of a $\nu = 1$ edge in the
  arguments presented here. For a $\nu = 2$ quantum Hall system, by contrast,
  Andreev processes do \protect \emph {not} freeze out at low energies. In this
  alternative setting, Fermi statistics allows a pairing term $\Delta (\psi
  _{R1} \psi _{R2} + H.c.)$, where $\psi _{R1}$ and $\psi _{R2}$ describe
  fermions in the two edge channels at the $\nu = 2$ boundary. Such a pairing
  term does not vanish at zero momentum. As a corollary, at zero incident
  electron energy the momenta for the modes beneath the superconductor need not
  vanish---allowing a finite $\delta \phi $ even at asymptotically low
  energies.}\BibitemShut {Stop}%
\bibitem [{\citenamefont {Wang}\ \emph {et~al.}(2015)\citenamefont {Wang},
  \citenamefont {Zhou}, \citenamefont {Lian},\ and\ \citenamefont
  {Zhang}}]{Wang2015}%
  \BibitemOpen
  \bibfield  {author} {\bibinfo {author} {\bibfnamefont {Jing}\ \bibnamefont
  {Wang}}, \bibinfo {author} {\bibfnamefont {Quan}\ \bibnamefont {Zhou}},
  \bibinfo {author} {\bibfnamefont {Biao}\ \bibnamefont {Lian}}, and\
  \bibinfo {author} {\bibfnamefont {Shou-Cheng}\ \bibnamefont {Zhang}},\
  }\bibfield  {title} {\enquote {\bibinfo {title} {Chiral topological
  superconductor and half-integer conductance plateau from quantum anomalous
  {H}all plateau transition},}\ }\href {\doibase 10.1103/PhysRevB.92.064520}
  {\bibfield  {journal} {\bibinfo  {journal} {Phys. Rev. B}\ }\textbf {\bibinfo
  {volume} {92}},\ \bibinfo {pages} {064520} (\bibinfo {year}
  {2015})}\BibitemShut {NoStop}%
\bibitem [{\citenamefont {Lian}\ \emph {et~al.}(2016)\citenamefont {Lian},
  \citenamefont {Wang},\ and\ \citenamefont {Zhang}}]{Lian2016}%
  \BibitemOpen
  \bibfield  {author} {\bibinfo {author} {\bibfnamefont {Biao}\ \bibnamefont
  {Lian}}, \bibinfo {author} {\bibfnamefont {Jing}\ \bibnamefont {Wang}},\
  and\ \bibinfo {author} {\bibfnamefont {Shou-Cheng}\ \bibnamefont {Zhang}},\
  }\bibfield  {title} {\enquote {\bibinfo {title} {Edge-state-induced {A}ndreev
  oscillation in quantum anomalous {H}all insulator-superconductor
  junctions},}\ }\href {\doibase 10.1103/PhysRevB.93.161401} {\bibfield
  {journal} {\bibinfo  {journal} {Phys. Rev. B}\ }\textbf {\bibinfo {volume}
  {93}},\ \bibinfo {pages} {161401} (\bibinfo {year} {2016})}\BibitemShut
  {NoStop}%
\bibitem [{\citenamefont {Chen}\ \emph {et~al.}(2017)\citenamefont {Chen},
  \citenamefont {He}, \citenamefont {Xu},\ and\ \citenamefont {Law}}]{ChenLaw}%
  \BibitemOpen
  \bibfield  {author} {\bibinfo {author} {\bibfnamefont {Chui-Zhen}\
  \bibnamefont {Chen}}, \bibinfo {author} {\bibfnamefont {James~Jun}\
  \bibnamefont {He}}, \bibinfo {author} {\bibfnamefont {Dong-Hui}\ \bibnamefont
  {Xu}}, and\ \bibinfo {author} {\bibfnamefont {K.~T.}\ \bibnamefont {Law}},\
  }\bibfield  {title} {\enquote {\bibinfo {title} {Effects of domain walls in
  quantum anomalous {H}all insulator/superconductor heterostructures},}\ }\href
  {\doibase 10.1103/PhysRevB.96.041118} {\bibfield  {journal} {\bibinfo
  {journal} {Phys. Rev. B}\ }\textbf {\bibinfo {volume} {96}},\ \bibinfo
  {pages} {041118} (\bibinfo {year} {2017})}\BibitemShut {NoStop}%
\bibitem [{\citenamefont {Chen}\ \emph {et~al.}(2018)\citenamefont {Chen},
  \citenamefont {He}, \citenamefont {Xu},\ and\ \citenamefont
  {Law}}]{Chen2018}%
  \BibitemOpen
  \bibfield  {author} {\bibinfo {author} {\bibfnamefont {Chui-Zhen}\
  \bibnamefont {Chen}}, \bibinfo {author} {\bibfnamefont {James~Jun}\
  \bibnamefont {He}}, \bibinfo {author} {\bibfnamefont {Dong-Hui}\ \bibnamefont
  {Xu}}, and\ \bibinfo {author} {\bibfnamefont {K.~T.}\ \bibnamefont {Law}},\
  }\bibfield  {title} {\enquote {\bibinfo {title} {Emergent josephson current
  of $n=1$ chiral topological superconductor in quantum anomalous {H}all
  insulator/superconductor heterostructures},}\ }\href {\doibase%
  10.1103/PhysRevB.98.165439} {\bibfield  {journal} {\bibinfo  {journal} {Phys.
  Rev. B}\ }\textbf {\bibinfo {volume} {98}},\ \bibinfo {pages} {165439}
  (\bibinfo {year} {2018})}\BibitemShut {NoStop}%
\bibitem [{\citenamefont {Lian}\ and\ \citenamefont {Wang}(2019)}]{LianWang}%
  \BibitemOpen
  \bibfield  {author} {\bibinfo {author} {\bibfnamefont {Biao}\ \bibnamefont
  {Lian}}\ and\ \bibinfo {author} {\bibfnamefont {Jing}\ \bibnamefont {Wang}},\
  }\bibfield  {title} {\enquote {\bibinfo {title} {Distribution of conductances
  in chiral topological superconductor junctions},}\ }\href {\doibase%
  10.1103/PhysRevB.99.041404} {\bibfield  {journal} {\bibinfo  {journal} {Phys.
  Rev. B}\ }\textbf {\bibinfo {volume} {99}},\ \bibinfo {pages} {041404}
  (\bibinfo {year} {2019})}\BibitemShut {NoStop}%
\bibitem [{\citenamefont {He}\ \emph {et~al.}(2017)\citenamefont {He},
  \citenamefont {Pan}, \citenamefont {Stern}, \citenamefont {Burks},
  \citenamefont {Che}, \citenamefont {Yin}, \citenamefont {Wang}, \citenamefont
  {Lian}, \citenamefont {Zhou}, \citenamefont {Choi}, \citenamefont {Murata},
  \citenamefont {Kou}, \citenamefont {Chen}, \citenamefont {Nie}, \citenamefont
  {Shao}, \citenamefont {Fan}, \citenamefont {Zhang}, \citenamefont {Liu},
  \citenamefont {Xia},\ and\ \citenamefont {Wang}}]{WangMajorana}%
  \BibitemOpen
  \bibfield  {author} {\bibinfo {author} {\bibfnamefont {Qing~Lin}\
  \bibnamefont {He}}, \bibinfo {author} {\bibfnamefont {Lei}\ \bibnamefont
  {Pan}}, \bibinfo {author} {\bibfnamefont {Alexander~L.}\ \bibnamefont
  {Stern}}, \bibinfo {author} {\bibfnamefont {Edward~C.}\ \bibnamefont
  {Burks}}, \bibinfo {author} {\bibfnamefont {Xiaoyu}\ \bibnamefont {Che}},
  \bibinfo {author} {\bibfnamefont {Gen}\ \bibnamefont {Yin}}, \bibinfo
  {author} {\bibfnamefont {Jing}\ \bibnamefont {Wang}}, \bibinfo {author}
  {\bibfnamefont {Biao}\ \bibnamefont {Lian}}, \bibinfo {author} {\bibfnamefont
  {Quan}\ \bibnamefont {Zhou}}, \bibinfo {author} {\bibfnamefont {Eun~Sang}\
  \bibnamefont {Choi}}, \bibinfo {author} {\bibfnamefont {Koichi}\ \bibnamefont
  {Murata}}, \bibinfo {author} {\bibfnamefont {Xufeng}\ \bibnamefont {Kou}},
  \bibinfo {author} {\bibfnamefont {Zhijie}\ \bibnamefont {Chen}}, \bibinfo
  {author} {\bibfnamefont {Tianxiao}\ \bibnamefont {Nie}}, \bibinfo {author}
  {\bibfnamefont {Qiming}\ \bibnamefont {Shao}}, \bibinfo {author}
  {\bibfnamefont {Yabin}\ \bibnamefont {Fan}}, \bibinfo {author} {\bibfnamefont
  {Shou-Cheng}\ \bibnamefont {Zhang}}, \bibinfo {author} {\bibfnamefont {Kai}\
  \bibnamefont {Liu}}, \bibinfo {author} {\bibfnamefont {Jing}\ \bibnamefont
  {Xia}}, and\ \bibinfo {author} {\bibfnamefont {Kang~L.}\ \bibnamefont
  {Wang}},\ }\bibfield  {title} {\enquote {\bibinfo {title} {Chiral {M}ajorana
  fermion modes in a quantum anomalous {H}all
  insulator{\textendash}superconductor structure},}\ }\href {\doibase%
  10.1126/science.aag2792} {\bibfield  {journal} {\bibinfo  {journal}
  {Science}\ }\textbf {\bibinfo {volume} {357}},\ \bibinfo {pages} {294--299}
  (\bibinfo {year} {2017})}\BibitemShut {NoStop}%
\bibitem [{\citenamefont {Ji}\ and\ \citenamefont {Wen}(2018)}]{JiWen}%
  \BibitemOpen
  \bibfield  {author} {\bibinfo {author} {\bibfnamefont {Wenjie}\ \bibnamefont
  {Ji}}\ and\ \bibinfo {author} {\bibfnamefont {Xiao-Gang}\ \bibnamefont
  {Wen}},\ }\bibfield  {title} {\enquote {\bibinfo {title}
  {$\frac{1}{2}({e}^{2}/h)$ conductance plateau without 1{D} chiral {M}ajorana
  fermions},}\ }\href {\doibase 10.1103/PhysRevLett.120.107002} {\bibfield
  {journal} {\bibinfo  {journal} {Phys. Rev. Lett.}\ }\textbf {\bibinfo
  {volume} {120}},\ \bibinfo {pages} {107002} (\bibinfo {year}
  {2018})}\BibitemShut {NoStop}%
\bibitem [{\citenamefont {Huang}\ \emph {et~al.}(2018)\citenamefont {Huang},
  \citenamefont {Setiawan},\ and\ \citenamefont {Sau}}]{Huang2018}%
  \BibitemOpen
  \bibfield  {author} {\bibinfo {author} {\bibfnamefont {Yingyi}\ \bibnamefont
  {Huang}}, \bibinfo {author} {\bibfnamefont {F.}~\bibnamefont {Setiawan}},\
  and\ \bibinfo {author} {\bibfnamefont {Jay~D.}\ \bibnamefont {Sau}},\
  }\bibfield  {title} {\enquote {\bibinfo {title} {Disorder-induced
  half-integer quantized conductance plateau in quantum anomalous {H}all
  insulator-superconductor structures},}\ }\href {\doibase%
  10.1103/PhysRevB.97.100501} {\bibfield  {journal} {\bibinfo  {journal} {Phys.
  Rev. B}\ }\textbf {\bibinfo {volume} {97}},\ \bibinfo {pages} {100501}
  (\bibinfo {year} {2018})}\BibitemShut {NoStop}%
\bibitem [{\citenamefont {Kayyalha}\ \emph {et~al.}(2020)\citenamefont
  {Kayyalha}, \citenamefont {Xiao}, \citenamefont {Zhang}, \citenamefont
  {Shin}, \citenamefont {Jiang}, \citenamefont {Wang}, \citenamefont {Zhao},
  \citenamefont {Xiao}, \citenamefont {Zhang}, \citenamefont {Fijalkowski},
  \citenamefont {Mandal}, \citenamefont {Winnerlein}, \citenamefont {Gould},
  \citenamefont {Li}, \citenamefont {Molenkamp}, \citenamefont {Chan},
  \citenamefont {Samarth},\ and\ \citenamefont {Chang}}]{Kayyalha}%
  \BibitemOpen
  \bibfield  {author} {\bibinfo {author} {\bibfnamefont {Morteza}\ \bibnamefont
  {Kayyalha}}, \bibinfo {author} {\bibfnamefont {Di}~\bibnamefont {Xiao}},
  \bibinfo {author} {\bibfnamefont {Ruoxi}\ \bibnamefont {Zhang}}, \bibinfo
  {author} {\bibfnamefont {Jaeho}\ \bibnamefont {Shin}}, \bibinfo {author}
  {\bibfnamefont {Jue}\ \bibnamefont {Jiang}}, \bibinfo {author} {\bibfnamefont
  {Fei}\ \bibnamefont {Wang}}, \bibinfo {author} {\bibfnamefont {Yi-Fan}\
  \bibnamefont {Zhao}}, \bibinfo {author} {\bibfnamefont {Run}\ \bibnamefont
  {Xiao}}, \bibinfo {author} {\bibfnamefont {Ling}\ \bibnamefont {Zhang}},
  \bibinfo {author} {\bibfnamefont {Kajetan~M.}\ \bibnamefont {Fijalkowski}},
  \bibinfo {author} {\bibfnamefont {Pankaj}\ \bibnamefont {Mandal}}, \bibinfo
  {author} {\bibfnamefont {Martin}\ \bibnamefont {Winnerlein}}, \bibinfo
  {author} {\bibfnamefont {Charles}\ \bibnamefont {Gould}}, \bibinfo {author}
  {\bibfnamefont {Qi}~\bibnamefont {Li}}, \bibinfo {author} {\bibfnamefont
  {Laurens~W.}\ \bibnamefont {Molenkamp}}, \bibinfo {author} {\bibfnamefont
  {Moses H.~W.}\ \bibnamefont {Chan}}, \bibinfo {author} {\bibfnamefont
  {Nitin}\ \bibnamefont {Samarth}}, and\ \bibinfo {author} {\bibfnamefont
  {Cui-Zu}\ \bibnamefont {Chang}},\ }\bibfield  {title} {\enquote {\bibinfo
  {title} {Absence of evidence for chiral {M}ajorana modes in quantum anomalous
  {H}all-superconductor devices},}\ }\href {\doibase 10.1126/science.aax6361}
  {\bibfield  {journal} {\bibinfo  {journal} {Science}\ }\textbf {\bibinfo
  {volume} {367}},\ \bibinfo {pages} {64--67} (\bibinfo {year}
  {2020})}\BibitemShut {NoStop}%
\bibitem [{\citenamefont {Lian}\ \emph {et~al.}(2018)\citenamefont {Lian},
  \citenamefont {Wang}, \citenamefont {Sun}, \citenamefont {Vaezi},\ and\
  \citenamefont {Zhang}}]{LianVaezi}%
  \BibitemOpen
  \bibfield  {author} {\bibinfo {author} {\bibfnamefont {Biao}\ \bibnamefont
  {Lian}}, \bibinfo {author} {\bibfnamefont {Jing}\ \bibnamefont {Wang}},
  \bibinfo {author} {\bibfnamefont {Xiao-Qi}\ \bibnamefont {Sun}}, \bibinfo
  {author} {\bibfnamefont {Abolhassan}\ \bibnamefont {Vaezi}}, and\ \bibinfo
  {author} {\bibfnamefont {Shou-Cheng}\ \bibnamefont {Zhang}},\ }\bibfield
  {title} {\enquote {\bibinfo {title} {Quantum phase transition of chiral
  {M}ajorana fermions in the presence of disorder},}\ }\href {\doibase%
  10.1103/PhysRevB.97.125408} {\bibfield  {journal} {\bibinfo  {journal} {Phys.
  Rev. B}\ }\textbf {\bibinfo {volume} {97}},\ \bibinfo {pages} {125408}
  (\bibinfo {year} {2018})}\BibitemShut {NoStop}%
\bibitem [{\citenamefont {Fendley}\ \emph {et~al.}(2009)\citenamefont
  {Fendley}, \citenamefont {Fisher},\ and\ \citenamefont
  {Nayak}}]{Fendley2009}%
  \BibitemOpen
  \bibfield  {author} {\bibinfo {author} {\bibfnamefont {Paul}\ \bibnamefont
  {Fendley}}, \bibinfo {author} {\bibfnamefont {Matthew P.~A.}\ \bibnamefont
  {Fisher}}, and\ \bibinfo {author} {\bibfnamefont {Chetan}\ \bibnamefont
  {Nayak}},\ }\bibfield  {title} {\enquote {\bibinfo {title} {Boundary
  conformal field theory and tunneling of edge quasiparticles in non-{A}belian
  topological states},}\ }\href {\doibase 10.1016/j.aop.2009.03.005} {\bibfield
   {journal} {\bibinfo  {journal} {Annals of Physics}\ }\textbf {\bibinfo
  {volume} {324}},\ \bibinfo {pages} {1547--1572} (\bibinfo {year} {2009})},\
  \bibinfo {note} {{J}uly 2009 Special Issue}\BibitemShut {NoStop}%
\bibitem [{\citenamefont {Fendley}\ \emph {et~al.}(2007)\citenamefont
  {Fendley}, \citenamefont {Fisher},\ and\ \citenamefont
  {Nayak}}]{Fendley2007}%
  \BibitemOpen
  \bibfield  {author} {\bibinfo {author} {\bibfnamefont {Paul}\ \bibnamefont
  {Fendley}}, \bibinfo {author} {\bibfnamefont {Matthew P.~A.}\ \bibnamefont
  {Fisher}}, and\ \bibinfo {author} {\bibfnamefont {Chetan}\ \bibnamefont
  {Nayak}},\ }\bibfield  {title} {\enquote {\bibinfo {title} {Edge states and
  tunneling of non-{A}belian quasiparticles in the $5/2$ quantum {H}all state
  and $p+ip$ superconductors},}\ }\href {\doibase 10.1103/PhysRevB.75.045317}
  {\bibfield  {journal} {\bibinfo  {journal} {Phys. Rev. B}\ }\textbf {\bibinfo
  {volume} {75}},\ \bibinfo {pages} {045317} (\bibinfo {year}
  {2007})}\BibitemShut {NoStop}%
\bibitem [{\citenamefont {Bishara}\ and\ \citenamefont
  {Nayak}(2008)}]{Bishara2008}%
  \BibitemOpen
  \bibfield  {author} {\bibinfo {author} {\bibfnamefont {Waheb}\ \bibnamefont
  {Bishara}}\ and\ \bibinfo {author} {\bibfnamefont {Chetan}\ \bibnamefont
  {Nayak}},\ }\bibfield  {title} {\enquote {\bibinfo {title} {Edge states and
  interferometers in the {P}faffian and anti-{P}faffian states of the
  $\ensuremath{\nu}=\frac{5}{2}$ quantum {H}all system},}\ }\href {\doibase%
  10.1103/PhysRevB.77.165302} {\bibfield  {journal} {\bibinfo  {journal} {Phys.
  Rev. B}\ }\textbf {\bibinfo {volume} {77}},\ \bibinfo {pages} {165302}
  (\bibinfo {year} {2008})}\BibitemShut {NoStop}%
\bibitem [{\citenamefont {Bonderson}\ \emph {et~al.}(2010)\citenamefont
  {Bonderson}, \citenamefont {Clarke}, \citenamefont {Nayak},\ and\
  \citenamefont {Shtengel}}]{Bonderson2010}%
  \BibitemOpen
  \bibfield  {author} {\bibinfo {author} {\bibfnamefont {Parsa}\ \bibnamefont
  {Bonderson}}, \bibinfo {author} {\bibfnamefont {David~J.}\ \bibnamefont
  {Clarke}}, \bibinfo {author} {\bibfnamefont {Chetan}\ \bibnamefont {Nayak}},
  and\ \bibinfo {author} {\bibfnamefont {Kirill}\ \bibnamefont {Shtengel}},\
  }\bibfield  {title} {\enquote {\bibinfo {title} {Implementing arbitrary phase
  gates with {I}sing anyons},}\ }\href {\doibase%
  10.1103/PhysRevLett.104.180505} {\bibfield  {journal} {\bibinfo  {journal}
  {Phys. Rev. Lett.}\ }\textbf {\bibinfo {volume} {104}},\ \bibinfo {pages}
  {180505} (\bibinfo {year} {2010})}\BibitemShut {NoStop}%
\bibitem [{\citenamefont {Nilsson}\ and\ \citenamefont
  {Akhmerov}(2010)}]{Nilsson2010}%
  \BibitemOpen
  \bibfield  {author} {\bibinfo {author} {\bibfnamefont {Johan}\ \bibnamefont
  {Nilsson}}\ and\ \bibinfo {author} {\bibfnamefont {A.~R.}\ \bibnamefont
  {Akhmerov}},\ }\bibfield  {title} {\enquote {\bibinfo {title} {Theory of
  non-{A}belian {F}abry-{P}erot interferometry in topological insulators},}\
  }\href {\doibase 10.1103/PhysRevB.81.205110} {\bibfield  {journal} {\bibinfo
  {journal} {Phys. Rev. B}\ }\textbf {\bibinfo {volume} {81}},\ \bibinfo
  {pages} {205110} (\bibinfo {year} {2010})}\BibitemShut {NoStop}%
\bibitem [{Note10()}]{Note10}%
  \BibitemOpen
  \bibinfo {note} {For a discussion of energy partitioning in the
  Luttinger-liquid context, see Ref.~\protect \rev@citealp
  {LLpartitioning}.}\BibitemShut {Stop}%
\bibitem [{Note11()}]{Note11}%
  \BibitemOpen
  \bibinfo {note} {One might have naively guessed that Ising-anyon tunneling
  instead admits a perturbative treatment provided the incoming fermion
  momentum $k_e$ is sufficiently large that $t_\sigma /(v_e k_e^{7/8}) \ll 1$.
  When this inequality holds, it would appear that one is probing the system at
  high energies for which $t_\sigma $ has not yet flowed to strong coupling.
  The fallacy in this argument stems from energy partitioning. Regardless of
  the magnitude of $k_e$, at the constriction the incident fermion splinters
  into Ising anyons that share the incident energy in all permissible ways. In
  particular, the allowed partitionings include cases where an Ising anyon
  tunneling across the constriction carries arbitrarily small momentum, and for
  such events $t_\sigma $ can not be regarded as weak. Consequently, finite
  length $L_a$ is required to define a perturbative regime, corresponding to
  the quoted inequality $t_\sigma L_a^{7/8}/v_e \ll 1$.}\BibitemShut {Stop}%
\bibitem [{Note12()}]{Note12}%
  \BibitemOpen
  \bibinfo {note} {Actually, in the full microscopic model $\delta \epsilon
  (k)$ reduces the overall bandwidth even though the velocity remains fixed.
  This effect likely yields a slightly smaller critical $U$ compared to what
  would occur if the bandwidth was also fixed.}\BibitemShut {Stop}%
\bibitem [{Note13()}]{Note13}%
  \BibitemOpen
  \bibinfo {note} {For a chiral CFT, one can transform between the `position
  slice' quantization and the `time slice' quantization via a Wick rotation
  (real time to imaginary time), a $90^\circ $ Euclidean rotation (exchange
  time and space), and then another Wick rotation (imaginary time back to real
  time).}\BibitemShut {Stop}%
\bibitem [{\citenamefont {Nayak}\ and\ \citenamefont
  {Wilczek}(1996)}]{NayakWilczek:2nStatesQHPf:96}%
  \BibitemOpen
  \bibfield  {author} {\bibinfo {author} {\bibfnamefont {Chetan}\ \bibnamefont
  {Nayak}}\ and\ \bibinfo {author} {\bibfnamefont {Frank}\ \bibnamefont
  {Wilczek}},\ }\bibfield  {title} {\enquote {\bibinfo {title} {{$2n$-quasihole
  states realize $2^{n-1}$-dimensional spinor braiding statistics in paired
  quantum {H}all states}},}\ }\href {\doibase 10.1016/0550-3213(96)00430-0}
  {\bibfield  {journal} {\bibinfo  {journal} {Nucl. Phys. B}\ }\textbf
  {\bibinfo {volume} {479}},\ \bibinfo {pages} {529--553} (\bibinfo {year}
  {1996})}\BibitemShut {NoStop}%
\bibitem [{\citenamefont {{Wolfram
  Research}}(2001{\natexlab{a}})}]{EllipticK:08.02.04.0009.01}%
  \BibitemOpen
  \bibfield  {author} {\bibinfo {author} {\bibnamefont {{Wolfram Research}}},\
  }\href {http://functions.wolfram.com/08.02.04.0009.01} {\enquote {\bibinfo
  {title} {Complete elliptic integral of the first kind: formula
  08.02.04.0009},}\ } (\bibinfo {year} {2001}{\natexlab{a}}),\ \bibinfo {note}
  {accessed 2019}\BibitemShut {NoStop}%
\bibitem [{\citenamefont {{Wolfram
  Research}}(2001{\natexlab{b}})}]{EllipticK:08.02.17.0001.01}%
  \BibitemOpen
  \bibfield  {author} {\bibinfo {author} {\bibnamefont {{Wolfram Research}}},\
  }\href {http://functions.wolfram.com/08.02.17.0001.01} {\enquote {\bibinfo
  {title} {Complete elliptic integral of the first kind: formula
  08.02.17.0001},}\ } (\bibinfo {year} {2001}{\natexlab{b}}),\ \bibinfo {note}
  {accessed 2019}\BibitemShut {NoStop}%
\bibitem [{\citenamefont {{Wolfram
  Research}}(2001{\natexlab{c}})}]{EllipticE:08.01.04.0009.01}%
  \BibitemOpen
  \bibfield  {author} {\bibinfo {author} {\bibnamefont {{Wolfram Research}}},\
  }\href {http://functions.wolfram.com/08.01.04.0009.01} {\enquote {\bibinfo
  {title} {Complete elliptic integral of the second kind: formula
  08.01.04.0009},}\ } (\bibinfo {year} {2001}{\natexlab{c}}),\ \bibinfo {note}
  {accessed 2019}\BibitemShut {NoStop}%
\bibitem [{\citenamefont {{Wolfram
  Research}}(2001{\natexlab{d}})}]{EllipticE:08.01.17.0002.01}%
  \BibitemOpen
  \bibfield  {author} {\bibinfo {author} {\bibnamefont {{Wolfram Research}}},\
  }\href {http://functions.wolfram.com/08.01.17.0002.01} {\enquote {\bibinfo
  {title} {Complete elliptic integral of the second kind: formula
  08.01.17.0002},}\ } (\bibinfo {year} {2001}{\natexlab{d}}),\ \bibinfo {note}
  {accessed 2019}\BibitemShut {NoStop}%
\bibitem [{\citenamefont {Karzig}\ \emph {et~al.}(2011)\citenamefont {Karzig},
  \citenamefont {Refael}, \citenamefont {Glazman},\ and\ \citenamefont {von
  Oppen}}]{LLpartitioning}%
  \BibitemOpen
  \bibfield  {author} {\bibinfo {author} {\bibfnamefont {Torsten}\ \bibnamefont
  {Karzig}}, \bibinfo {author} {\bibfnamefont {Gil}\ \bibnamefont {Refael}},
  \bibinfo {author} {\bibfnamefont {Leonid~I.}\ \bibnamefont {Glazman}}, and\
  \bibinfo {author} {\bibfnamefont {Felix}\ \bibnamefont {von Oppen}},\
  }\bibfield  {title} {\enquote {\bibinfo {title} {Energy partitioning of
  tunneling currents into {L}uttinger liquids},}\ }\href {\doibase%
  10.1103/PhysRevLett.107.176403} {\bibfield  {journal} {\bibinfo  {journal}
  {Phys. Rev. Lett.}\ }\textbf {\bibinfo {volume} {107}},\ \bibinfo {pages}
  {176403} (\bibinfo {year} {2011})}\BibitemShut {NoStop}%
\end{thebibliography}%

\end{document}